\documentclass[fleqn,usenatbib,useAMS]{mnras}
\usepackage{graphicx}	
\usepackage{amsmath}	
\usepackage{amssymb}	
\usepackage{multicol}   
\usepackage{multirow}
\usepackage{bm}		
\usepackage[T1]{fontenc}
\usepackage{ae,aecompl}
\usepackage{longtable}
\usepackage{url} 
\usepackage{threeparttable} 
\usepackage{subfig}      
\usepackage{booktabs}
\usepackage[T1]{fontenc}
\usepackage{multicol}
\usepackage[utf8x]{inputenc}
\usepackage{hyperref}
\hypersetup{
    colorlinks=true,
    linkcolor=blue,
    filecolor=magenta,      
    urlcolor=cyan,
    pdftitle={Overleaf Example},
    pdfpagemode=FullScreen,
    }

\newcommand{\sqcm}{cm$^{-2}$\,}  
\newcommand{\kms}{$\rm km\, s^{-1}$\,} 
\newcommand{\lya}{Ly$\alpha$\,}

\newcommand{\HI}{\mbox{H\,{\sc i}}}

\newcommand{\OVI}{\mbox{O\,{\sc vi}}}

\newcommand{\CIV}{\mbox{C\,{\sc iv}}} 
\newcommand{\CV}{\mbox{C\,{\sc v}}}

\newcommand{\SiII}{\mbox{Si\,{\sc ii}}}
\newcommand{\SiIII}{\mbox{Si\,{\sc iii}}}
\newcommand{\SiIV}{\mbox{Si\,{\sc iv}}}

\newcommand{\MgII}{\mbox{Mg\,{\sc ii}}}

\newcommand{\zmin}{$z_{\rm min}$}
\newcommand{\zmax}{$z_{\rm max}$}

\newcommand{\Rvir}{$R_{\rm vir}$\,}
\newcommand{\ew}{$\rm EW_0$\,}
\newcommand{\logN}{$\log_{\rm 10} N/{\rm cm^{-2}}$\,} 
\newcommand{\dop}{Doppler parameter\,}
\newcommand{\colm}{column density\,}
\newcommand{\fc}{covering fraction\,}
\newcommand{\pms}{$\pm$}
\newcommand{\Msun}{M$_{\odot}$\,}
\newcommand{\ang}{\mbox{\normalfont\AA}}
\newcommand{\Rev}[1]{\textcolor{black}{#1}}

\DeclareRobustCommand{\VAN}[3]{#2}
\let\VANthebibliography\thebibliography
\def\thebibliography{\DeclareRobustCommand{\VAN}[3]{##3}\VANthebibliography}

\title[The relation between LAEs and \CIV\ absorbers] {MUSEQuBES: The relation between \lya\ emitters and \CIV\ absorbers at $z \approx 3.3$} 

\author[E. Banerjee et al.]{
Eshita Banerjee $^{1}$\thanks{E-mail: eshitaban18@iucaa.in},
Sowgat Muzahid $^{1}$,
Joop Schaye $^{2}$,
Sean D. Johnson $^{3}$,
and Sebastiano Cantalupo $^{4}$
\\ $^{1}$ IUCAA, Post Bag 04, Ganeshkhind, Pune-411007, India\\
$^{2}$ Leiden Observatory, Leiden University, P.O. Box 9513, NL-2300 AA Leiden, the Netherlands\\
${^3}$ Department of Astronomy, University of Michigan, 1085 S. University, Ann Arbor, MI 48109, USA\\
${^4}$ Department of Physics, University of Milan Bicocca, Piazza della Scienza 3, I-20126 Milano, Italy
}

\date{Accepted XXX. Received YYY; in original form ZZZ}

\pubyear{2022}

\begin{document}
\label{firstpage}
\pagerange{\pageref{firstpage}--\pageref{lastpage}}
\maketitle

\begin{abstract}

\noindent 
We present a detailed study of the column density and covering fraction profiles of \CIV\ absorption around 86 redshift $z \approx 3.3$ Ly$\alpha$ emitters (LAEs) detected in 8 Multi-Unit Spectroscopic Explorer (MUSE) fields of $1'\times 1'$ centered on 8 bright background quasars as part of the MUSEQuBES survey. Using Voigt profile fitting of all the \CIV\ absorbers detected along these 8 sightlines, we generated a ``blind'' absorber catalog consisting of 489 \CIV\ absorption components. We cross-matched this blind \CIV\ catalog with the MUSE-detected LAE catalog and found a significant enhancement of \CIV\ components within $\approx \pm$400 \kms of the systemic redshifts of the LAEs. Neither the \CIV\  column density ($N$) nor the Doppler parameter ($b$) of individual \CIV\ components shows any significant anti-correlation with impact parameter ($\rho$) of the LAEs in the 68 percentile range of $90\leq \rho \leq 230$ physical kpc (pkpc). We find a covering fraction of $\approx60$\% for a threshold $N(\CIV)$ of $10^{12.5}$~\sqcm, which is roughly twice as high as in random regions. The \CIV\ covering fraction remains constant at $\approx50\%$ for impact parameters in the range 150--250~pkpc ($\approx 3-6 R_{200}$). Using the covering fraction profile, we constrain the LAE--\CIV\ absorber two-point correlation function, and obtain $r_0 = 3.4^{+1.1}_{-1.0}$~comoving Mpc (cMpc) and $\gamma = 1.2^{+0.2}_{-0.3}$ for a threshold $N(\CIV)$ of $10^{13.0}$~\sqcm. The \CIV\ covering fraction is found to be enhanced for the LAEs that are part of a ``pair/group'' compared to the isolated ones.

\end{abstract}




\section{INTRODUCTION}

Based on the current consensus, the inward and outward flows of baryons driven by accretion and galactic winds, known as the ``cosmic baryon cycle'', dictate the formation and evolution of galaxies. The imprints of this baryon cycle can be found in the circumgalactic medium (CGM), the gaseous medium in the neighbourhood of galaxies which serves as the interface between the interstellar medium (ISM) and the intergalactic medium (IGM). The gas in the CGM is diffuse, metal enriched, multiphase, and kinematically complex \citep[see][for a review]{Tumlinson2017}. The gas and metal masses in the CGM are comparable to those in the galaxies and can potentially explain the ``halo missing baryons''\citep[]{Mcgaugh_2010, Werk_2014} and ``missing metals'' problems \citep[]{Peeples_2014}. Therefore, probing the CGM of galaxies at different cosmic epochs, and in different environments is crucial.

Owing to its very low density ($n_{\rm H} \sim 10^{-3} ~\rm  cm^{-3}$), it is challenging to probe the CGM in emission, particularly at large galactocentric distances. However, there is a growing number of studies reporting extended emission out to a few hundreds of kpc from active galaxies \citep[e.g.,][]{Cantalupo2014,Borisova_2016} and a few tens of kpc from normal galaxies \citep[e.g.,][]{Wisotzki2018,Zabl_2021,Rdutta2023} including the LAEs (upto $100-1000$ ckpc) \citep[e.g.,][]{Kikuchihara_2022}, mostly using strong resonant lines such as \lya\ and \MgII\ \citep[but see][]{Johnson_2022}. Due to scattering, these resonant lines are not reliable tracers of gas kinematics. Moreover, the lack of multiple transitions restricts the ability to constrain the physical and chemical properties of the CGM. On the other hand, owing to the linear dependence of optical depth on density, absorption is a more sensitive tracer of the low-density gas than emission. The spectra of bright background sources, such as quasars, are generally used to probe the diffuse gas in the CGM \citep[e.g.,][]{Chen_1998, Adelberger_2003, Turner_2014}. However, bright background galaxies can also be used for similar studies \citep[e.g.,][]{Steidel_2010, Peroux2018,rubin2018galaxies}.

Different ionic or atomic transitions seen in quasar spectra probe different phases of the absorbing gas. The neutral or low ionization transitions such as those from \HI\ and \MgII\ trace the cool ($T \sim 10^4$ K), relatively dense gas phase, whereas highly ionized transitions such as those from \CIV\ and \OVI\ can probe the low density photoionized and/or collisionally ionized warm-hot ($T\sim 10^{5}-10^{6}$~K) gas phase. However, at high redshift ($z>2$), \CIV\ absorbers are predominantly photoionized due to the higher intensity of the extragalactic UV background radiation and the relatively lower ionization potential ($\rm 47.8~eV$) \cite[e.g.,][]{joop_2003, muzahid2012high}. The higher oscillator strengths of the \CIV\ transitions and high cosmic abundance of carbon make it a useful tool to probe the CGM. Moreover, the rest-frame wavelengths of the \CIV\ doublet ($\lambda\lambda 1548, 1550$) are suitable to identify them in the red part of the quasar's \lya\ emission using the doublet matching technique.

At low redshifts ($z<1$), the connection between \CIV\ absorbers and galaxies is well-studied \citep[see e.g.,][]{Chen_2001,Borthakur_2013,bordoloi2014cos,Liang_Chen2014, Schroetter_2021,Rdutta2021}. These studies suggest that, unlike \HI, \CIV\ is closely tied to galaxies with virtually no absorbers detected beyond the virial radius (\Rvir). The COS-Dwarfs survey with $43$ sub-$L^*$ galaxies with stellar masses  $M_* \sim 10^9$~\Msun concluded that the CGM of low-mass galaxies contains more carbon within $\approx0.5$\Rvir than that present in the stars of those galaxies \citep[]{bordoloi2014cos}. They also found that for star-forming galaxies the \CIV\ covering fraction at a threshold equivalent width of  $100$~m\ang\ is $f_c = 60-80\%$ within $\approx0.5$~\Rvir. \citet{Burchett_2016} obtained a \CIV\ covering fraction of $\approx 33\%$ within the \Rvir for galaxies of redshift $0.0015 < z < 0.015$. They found that the high-mass galaxies ($>10^{9.5}$~\Msun) show an environmental dependence in which the galaxies with fewer companions have a larger \CIV\ covering fraction. They claimed that when galaxy number density increases \CV\ becomes the dominant ionization state, leading to lower covering fractions for \CIV. The low-mass galaxies in their sample ($<10^{9.5}$~\Msun) rarely exhibit \CIV\ absorption ($f_c \approx 9$\%). \citet{Sean_2017} found for a sample of 18 star-forming dwarf galaxies of stellar mass $\sim 10^{8}-10^{9}$~\Msun at $z\approx0.2$ that \OVI\ is considerably more common than \CIV\ within $\approx$~3\Rvir of the low-mass galaxies.

CGM studies at high $z$ are more limited because it is challenging to find galaxies at higher redshifts. The Keck Baryonic Structure Survey (KBSS) is the first major CGM survey at $2<z<3$ that probed the radial distribution and kinematics of gas (\HI) and metals (e.g., \CIV\, \OVI) around galaxies \citep[e.g.,][]{Adelberger_2003, Adelberger_2005, Steidel_2010, Rakic_2012, Turner_2014, Rudie_2012, Rudie_2019}. A strong correlation between \HI\ and \CIV\ absorbers and Lyman Break Galaxies (LBGs) is observed in the KBSS \citep[]{Adelberger_2003, Adelberger_2005, Turner_2014}. \citet{Steidel_2010}, for the first time,  observed the CGM of star-forming LBGs using background galaxy spectra and measured the distributions of various metals (e.g., \CIV\, \SiII, and \SiIV) within an impact parameter of $\approx125$~pkpc. They also showed the presence of broad P-Cygni profiles in higher ionization transitions which suggested strong outflows caused by the stellar winds from massive stars. \cite{Turner_2014} with their $z\approx2.4$ star-forming KBSS galaxies showed that \CIV\ absorption is extended up to 2 pMpc in the transverse direction. Using Voigt profile fitting of various metal lines around 8 KBSS galaxies for which the quasar sightlines probe the regions within \Rvir, \citet{Rudie_2019} confirmed the multiphase nature of the CGM at high redshifts. They measured a covering fraction of $\approx90$\% for \CIV\ within \Rvir\ for a threshold column density of $10^{12.5}$~\sqcm.

None of these high $z$ studies have investigated the connections between galaxy properties (e.g., SFR, environment) and CGM properties. Additionally, because of the narrow/broad-band color selections, KBSS galaxies are biased towards high stellar masses ($M_*\sim10^{10}-10^{11}$~\Msun) and high star-formation rates (SFRs $\gtrsim10$~\Msun~$\rm yr^{-1}$). Since low-mass galaxies at high redshifts are too faint to be detected in continuum emission, CGM investigations at high redshifts have hitherto been limited to such massive, bright galaxies.

With the availability of state-of-the-art integral field spectrographs (IFS) such as VLT/MUSE \cite[]{bacon2010muse} and Keck/KCWI \cite[]{Morrissey_2018}, it is now possible to detect high-redshift galaxies purely via emission lines (primarily via the \lya\ emission) that are too faint to be detected in continuum emission. The MUSE Quasar-field Blind Emitters Survey (MUSEQuBES) is the first survey focused on studying the CGM around high-$z$ ($3<z<4$), low-mass ($M_* \sim 10^{9}$ \Msun), mildly star-forming (SFR $\sim 1$~\Msun~$\rm yr^{-1}$) \lya\ emitters (LAEs) using background quasars \citep[see][]{Muzahid_2020, Muzahid_2021}. The MUSEQuBES survey presented a sample of 96 LAEs, detected around 8 high-$z$, UV-bright ($V<19$) quasars within the $1' \times 1'$  field of view (FOV) of MUSE. Using spectral stacking analysis, \cite{Muzahid_2021} reported significant excess \HI\ and \CIV\ absorption near LAEs out to $\approx250$~pkpc in the transverse and $\pm500$~\kms\ along the line of sight directions. Moreover, the \HI\ and \CIV\ equivalent widths were measured to be larger around pair/group LAEs than near isolated ones. A strong correlation between \HI\ absorption and the SFR of the LAEs was also seen, which was not present for \CIV. The recent MUSE Analysis of Gas around Galaxies (MAGG) survey also found a higher \HI\ covering fraction near pair/group LAEs than around isolated LAEs \citep[][]{Lofthouse_2023}. They searched for LAEs around known strong \HI\ absorbers ($\log_{10} N(\HI)/\rm cm^{-2} > 16.5$) and confirmed that high column density absorbers and LAEs basically trace each other. Using the MAGG sample,  \cite{Galbiati_2023} reported the \CIV\ covering fraction of $\approx40$\% out to $\approx250$~pkpc for a threshold of equivalent width of $0.1$~\AA. The \CIV\ covering fraction in `group' is found to be $\approx3$ times higher compared to isolated LAEs. They further showed that the luminosity function of the LAEs associated with \CIV\ absorbers having a factor of $\approx2.5$ higher normalization factor compared to the field LAEs.

\begin{table*}
\begin{center} 
\begin{threeparttable}[b] 
\caption{Summary of the quasar fields surveyed in this work.}
\begin{tabular}{lcccccccc}
    \hline
    Quasar$^{1}$ & $\rm RA_{qso}^{2}$ & $\rm Dec_{qso}^{3}$ & $z_{\rm qso}^{4}$ & $N_{\rm gal}^{5}$ & Median & Limiting & $z_{\rm min}^8$ & $z_{\rm max}^9$\\
    &&&&&SNR$^{6}$ & ${\rm log_{10}}N$(\CIV)$^7$&&\\
    \hline
    BRI1108-07& 11:11:13.6     & $-$08:04:02&  3.922 & 22(17)  & 25  & 12.2 & 2.9 & 3.872\\
    J0124+0044 & 01:24:03.0    & $+$00:44:32&  3.834 & 4(3)    & 72  & 11.7 & 2.9 & 3.786\\
    PKS1937-101& 19:39:57.3    & $-$10:02:41&  3.787 & 2(1)    & 89  & 11.6 & 2.9 & 3.521\\
    QB2000-330& 20:03:24.0     & $-$32:51:44&  3.773 & 14(14)  & 80  & 11.7 & 2.9 & 3.725\\ 
    Q1621-0042& 16:21:16.9     & $-$00:42:50&  3.709 & 12(11)  & 105 & 11.6 & 2.9 & 3.662\\
    Q1317-0507& 13:20:30.0     & $-$05:23:35&  3.700 & 22(20)  & 97  & 11.6 & 2.9 & 3.653\\
    Q0055-269& 00:57:58.1      & $-$26:43:14&  3.655 & 12(12)  & 82  & 11.7 & 2.9 & 3.608\\
    Q1422+23& 14:24:38.1       & $+$22:56:01&  3.620 & 8(8)    & 217 & 11.2 & 2.9 & 3.574\\
    \hline
    \end{tabular}
    \label{qso_list}
    \vskip0.2cm 
Notes -- 1: Name of the quasar sightline; 2 \& 3: RA and Dec of the quasar (J2000); 4: redshift of the quasar; 5: Total number of LAEs detected in MUSE. The number in parentheses indicates the number of LAEs for which we have \CIV\ coverage outside the LAF region and free from  significant skyline contamination; 6: Median SNR of the quasar spectrum at 6657 \pms100~\ang; 7: Limiting \CIV\ \colm\ (in log-scale) at the median SNR of the respective sightlines for a typical gas temperature of $\approx 10^4$K; 8: The minimum redshift probed; 9: The maximum redshift of interest, $ z_{\rm max} = z_{\rm qso}- (1+z_{\rm qso})\times \frac{3000}{c}$. \\
\end{threeparttable}  
\end{center}
\end{table*} 

Here we continue with the analysis of the MUSEQuBES data. We blindly searched for the \CIV\ doublet along the 8 MUSEQuBES quasars and created a ``blind'' \CIV\ catalog using Voigt profile decomposition of all the  blindly detected \CIV\ absorbers. We cross-matched this catalog with the MUSE-detected LAE catalog to explore possible relations between the \CIV\ absorbers and LAEs at $z\approx3.3$. In order to correct for the resonant scattering effects of the ``\lya\ redshifts'' \footnote{The redshift determined from the peak of the \lya\ emission line is often redshifted with respect to the systemic redshift owing to the resonant scattering of \lya\ photons  \citep[e.g.,][]{verhamme20063d, Steidel_2010}.} we adopted the empirical relation obtained for the MUSEQuBES galaxies in \citet{Muzahid_2020}. The LAE redshifts lie between $2.9 - 3.8$, near the peak of the cosmic star formation rate density \citep[]{Madau2014review}. The lower bound on the redshift is set by the minimum wavelength covered by MUSE, whereas the upper bound is determined by the redshift of the most distant quasar in our sample. Quasars at $z>4$ were avoided deliberately in  MUSEQuBES since the \lya\ forest (LAF) regions become too crowded to obtain accurate \HI\ column densities.

This paper is organized as follows. In Section~\ref{sec2} we briefly describe our \CIV\ absorber and LAE catalogs. In Section~\ref{sec3} we present our analysis and results. In Section~\ref{sec4}, we discuss our results and compare them with the literature, followed by the conclusions in Section~\ref{sec5}. Throughout this study, we adopted a flat $\Lambda$CDM cosmology with $H_0$ = 70 km $\rm s^{-1}\, Mpc^{-1}$, $\Omega_{\rm M}$ = 0.3 and $\Omega_\Lambda$ = 0.7, and all distances given are in physical units unless specified otherwise.

\section{DATA} 
\label{sec2}

\subsection{Quasar spectra and the ``blind'' \texorpdfstring{\CIV\ }\ absorber catalog}

The quasar spectra used in this study are the same as presented in \citet[][see Section~2.2]{Muzahid_2021}. Briefly, for all 8 quasars, we have excellent quality optical spectra obtained with the VLT/UVES and/or Keck/HIRES spectrographs with a resolving power of $R \approx 45,000$. All but one spectrum have a signal-to-noise ratio (SNR) of $> 70$ per pixel in the region red-wards of the quasars' \lya\ emission (see Table~\ref{qso_list}). The corresponding limiting rest-frame equivalent width (REW) is 0.0015 \ang\ (the limiting \CIV\ \colm\ is ${\rm log_{10}}~ N(\CIV)/{\rm cm^{-2}}$ = 11.6) for a gas temperature of $\sim 10^{4}$~K.

We searched for \CIV\ absorbers in these 8 spectra manually using the so-called doublet matching technique, which relies on the fact that the observed wavelength ratio of the two members of the \CIV\ doublet ($\lambda\lambda1548,1550$) is equal to the ratio of their rest-frame wavelengths. We started by shifting the wavelength array of a given spectrum by the ratio of the rest-frame wavelengths of the \CIV\ doublet and plotting the shifted spectrum on top of the original spectrum. We then inspected all the coincidences of lines between the original and shifted spectra redward of the \lya\ forest.

While searching for \CIV\ absorbers using this method, we verified that (i) the overall shape of the \CIV~$\lambda1548$ line is similar to the \CIV~$\lambda1550$ line, and (ii) in the optically thin regime the \CIV~$\lambda1548$ line will be $\approx2$ times stronger than the \CIV~$\lambda1550$ line, provided that the lines are free from significant contamination (which is true for most of the \CIV\ absorbers in our sample). However, whenever possible, we determined the identities of the interlopers if present.

Since our ultimate goal is to cross-match the blind \CIV\ catalog with the catalog of LAEs detected in these fields \citep{Muzahid_2020,Muzahid_2021}, we did not include the \CIV\ absorbers at $z<2.9$, since no LAEs can be detected with MUSE below $z=2.9$. Moreover, we excluded any absorbers detected within the quasar's proximate zones, i.e., $3000$~\kms\ blue-ward of the quasars' redshifts, from our analysis.

Once the search is complete, an initial catalog of \CIV\ absorbers was created. Next, we fit the Voigt profile of every \CIV\ absorber (both the doublets) using the {\sc VPFIT} software \citep[v12.2;][]{2014ascl.soft08015C}. {\sc VPFIT}\footnote{\url{https://people.ast.cam.ac.uk/~rfc/}} returns the best-fitting redshift ($z$), column density ($N$), and Doppler parameter ($b$) of each \CIV\ component of a given absorber, estimated using an iterative $\chi ^2$ minimization technique.

\begin{figure}
\includegraphics[width=0.98\linewidth]{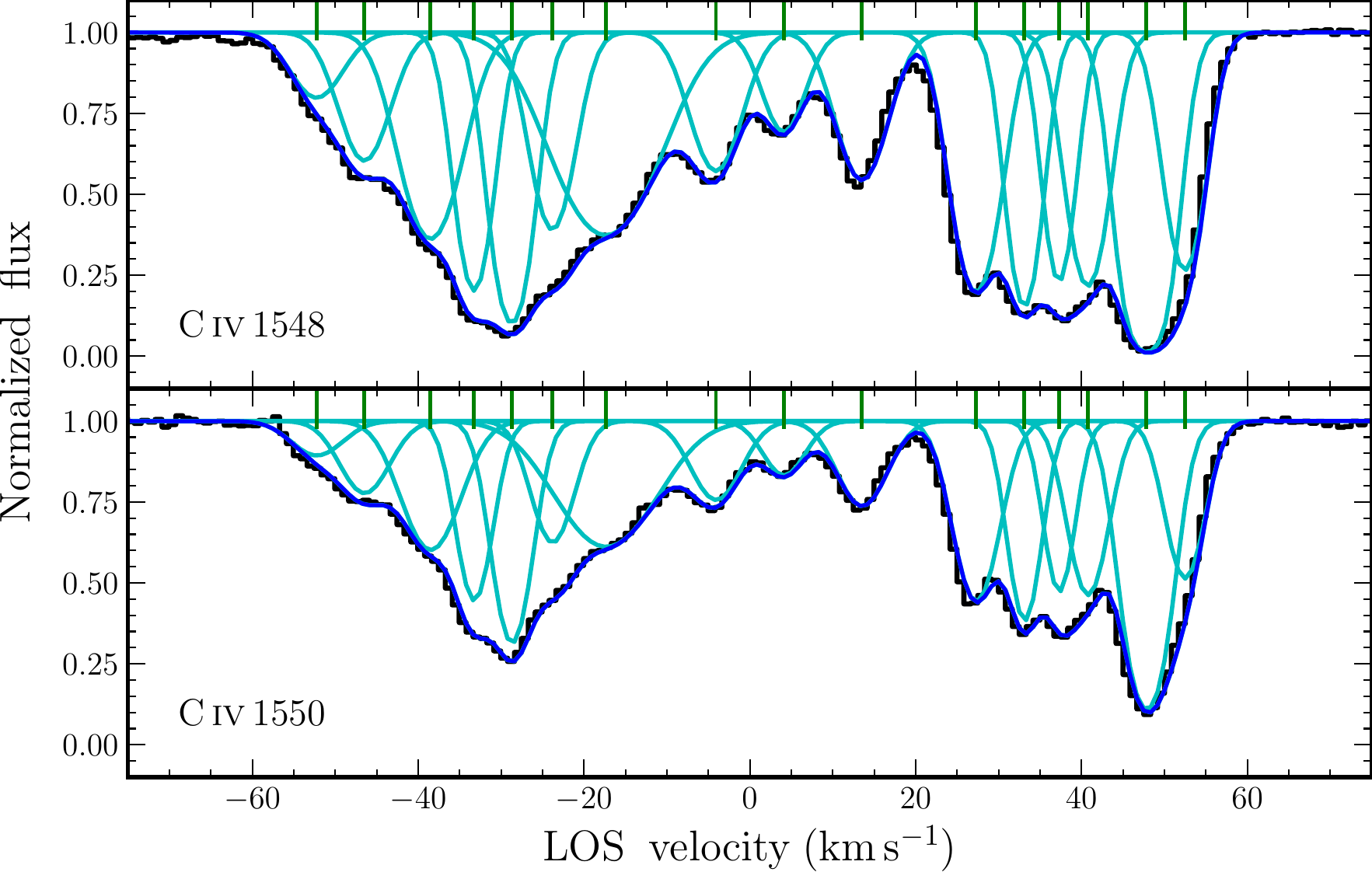}
\caption{An example of Voigt profile decomposition using {\sc vpfit}. The \CIV\ system is detected towards the sightline Q~1621--0042. The {\tt top} and {\tt bottom} panels show the \CIV~$\lambda1548$ and \CIV~$\lambda1550$ lines, respectively. The individual \CIV\ components are shown in cyan and the resultant model profile is shown in blue. The green vertical ticks indicate the centroids of the individual \CIV\ components. No contamination is present in this system.}  
\label{fig:vpfit}
\end{figure}

It is known that the number of components used in the Voigt profile decomposition of an absorber is subjective. We generally used the minimum number of components required to obtain a reduced $\chi^2$ close to unity. For each absorber, we repeated the fitting procedure iteratively to obtain the best-fit parameters. In the case of a blended \CIV\ absorber, we measured the absorption parameters from the uncontaminated doublet member. Our final \CIV\ absorber catalog comprises $\approx$ 152  \CIV\ systems comprising of 489 Voigt components in the redshift range $2.900 \leq z \leq 3.835$ (median $z = 3.260$). ``Absorption systems'' are defined by sorting all the component redshifts in a given spectrum obtained from the Voigt profile, and then grouping the components within $\pm300$~\kms to each other starting from the lowest redshift component. The best-fit parameters for these components are presented in Table~\ref{tab:abs_cat}. Fig.~\ref{fig:vpfit} shows an example of our Voigt profile decomposition. The profiles in cyan represent individual \CIV\ components. Their centroids are shown by the  vertical green ticks. The resultant model \CIV\ profile is shown in blue.

\subsection{LAE catalog}
\label{sec:lyz}

The 96 LAEs in the redshift range $2.9 < z < 3.8$ used in this study are detected via their \lya\ emission as part of our MUSEQuBES survey (P.I. Joop Schaye) \citep[see][]{Muzahid_2020,Muzahid_2021}.\footnote{MUSEQuBES is a dual survey. For the low-$z$ counterpart, see \citet{SDutta2023}.}  MUSEQuBES consists of 8 MUSE pointings centered on the 8 UV-bright quasars with $\rm 50h$ of guaranteed time observations. The individual fields are observed for $\rm 2h-10h$ with excellent seeing conditions (typically $<0.8''$). The details of the MUSE observations and data reduction procedure are described in Sections~$2.1$ and $3.1$ of \citet{Muzahid_2021}.

Line emitters in the 8 MUSE cubes are identified using {\sc CubEx} \citep[v1.6;][]{Cantalupo2018}, and subsequently classified manually with the help of the 1D spectra and pseudo-narrow-band images. In total 96 objects were classified as LAEs. There are $9$ LAEs ($5$ towards BRI1108--07, $2$ towards Q~1317--0507, 1 each towards Q~1621--0042 and J~0124+0044) for which the corresponding \CIV\ wavelengths are heavily affected by strong skylines. \Rev{There is no \CIV\ coverage for one of the LAEs detected towards PKS1937--101.} We thus excluded these $10$ LAEs from our analysis. The median redshift for the rest of the $86$ LAEs is $z = 3.3$. The properties of these LAEs are summarized in Table~\ref{tab:lae}. 

 
It is well known that the Ly$\alpha$ emission peak is generally  redshifted by a few hundred of \kms\ with respect to the systemic redshift of a galaxy \citep[see,][]{Shibuya_2014, Verhamme+18}. The \lya\ peak redshifts of the 96 LAEs are corrected using the empirical relation $V_{\rm offset} = 0.89\times {\rm FWHM} -58$~\kms\ obtained in \citet{Muzahid_2020} for our sample, where $V_{\rm offset}$ is the velocity offset between the \lya\ emission peak and the systemic redshift and $\rm FWHM$ is the   full width at half maximum of the \lya\ line.

We estimated the halo mass ($M_{\rm vir}$) and subsequently the \Rvir\ for only those LAEs ($35$ out of $86$) for which we could measure the star formation rate (SFR) from the UV continuum flux density \citep[see][]{Muzahid_2020}. Note that, these SFR values are dust-uncorrected and are calculated from the measured UV luminosity values using the local calibration relation of \cite{Kennicutt_1998} corrected to the \cite{Chabrier_2003} initial mass function. The halo masses (\Rvir) range from $10^{10.97}-10^{11.72}$~\Msun\ ($32.8-58.6$~pkpc) with a median value of $10^{11.29}$~\Msun\ ($42.4$~pkpc).


\begin{table}
\begin{center}
\begin{threeparttable}[b] 
\caption{Properties of the 86 LAEs in our sample.}
\begin{tabular}{lrrrc} 
    \hline
    Parameter & Min.$^1$ & Max.$^2$ & Med.$^3$ & $N_{\rm gal}$\\
    \hline
    Redshift  & 2.918 & 3.814 & 3.306 & 86\\
    Impact parameter (pkpc) & 16.1 & 315.2 & 163.7 & 86\\
    log$_{10}$(SFR/ \Msun $\rm yr^{-1}$) & $-$0.47 & 0.85 & 0.11 & 35\\
    log$_{10}\, {L\rm (Ly \alpha)}/{\rm erg \,s^{-1}}$ $^a$ & 41.29 & 42.91 & 42.00 & 86\\
    \ew ($\ang$) $^b$ & 9.1 & 113.4 & 48.7 & 35\\
    log$_{10}\, {M_{*}}/\rm M_{\odot}$ $^c$ & 8.33 & 9.65 & 8.91 & 35\\
    \Rvir (pkpc)$^{d}$ & 32.8 & 58.6 & 42.3 & 35 \\
    \hline
    \end{tabular}
    \label{tab:lae}
Notes-- a: \lya\ line luminosity; b: Rest-frame equivalent width of the \lya\ emission line; c: Stellar mass; d: Virial radius. The values for SFR, \ew, stellar masses and virial radii are calculated using only the 35 out of 86 LAEs with robust SFR estimates (see text). $^1$Minimum value, $^2$maximum value, and $^3$median value of the parameter.   
\end{threeparttable}  
\end{center} 
\end{table}


\section{RESULTS}
\label{sec3}

\begin{figure*}
    \includegraphics[width=0.8\textwidth]{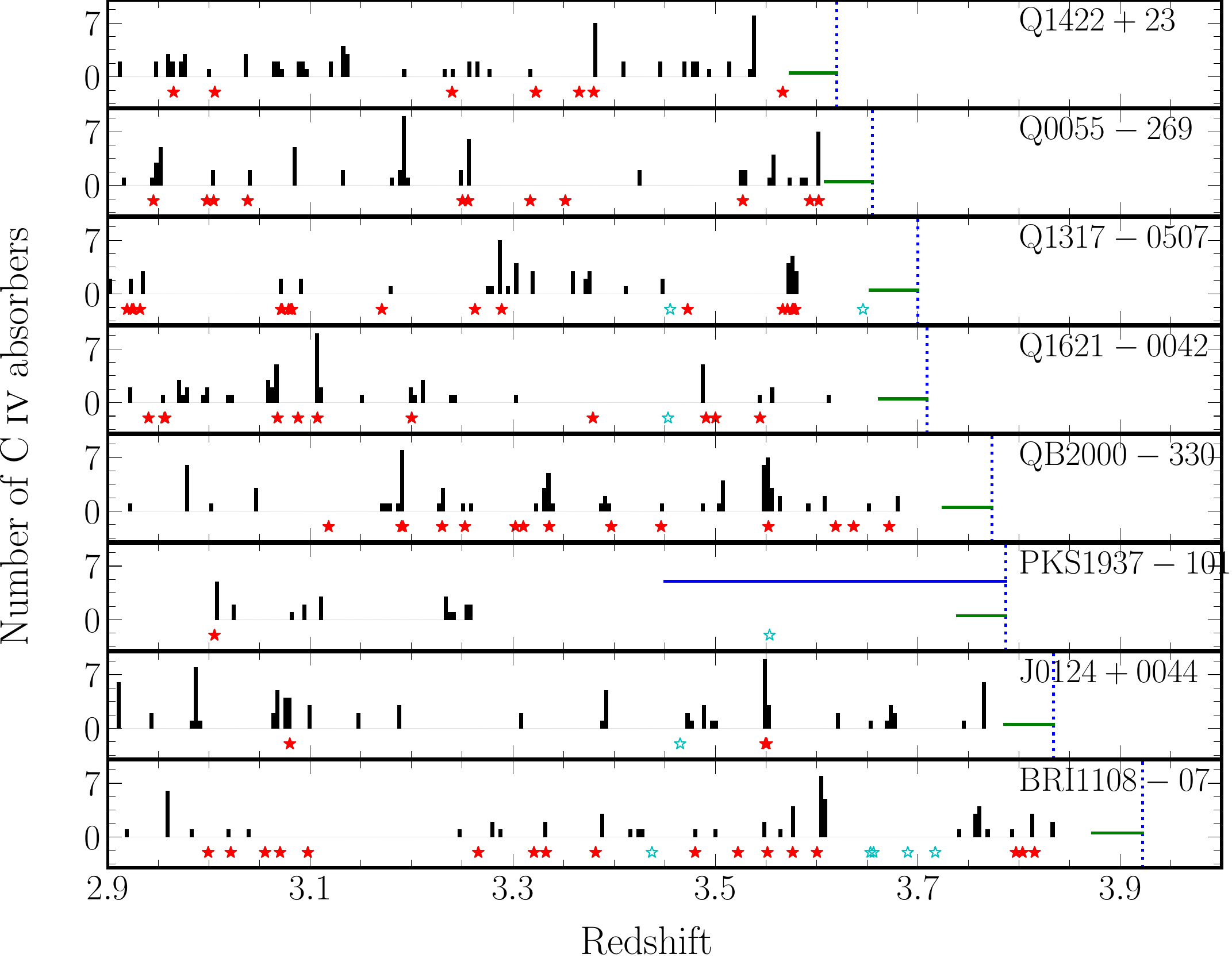}
    \caption{Histograms of the redshifts for the $489$ \CIV\ components obtained from Voigt profile fitting in bins of $\Delta z = 0.004$ (corresponding to $300$~\kms at $z\approx3$). The quasar sightlines are sorted in ascending order of the quasar redshifts indicated by the blue dotted lines from top to bottom. The 86 LAE redshifts in the different quasar sightlines are represented by the red star symbols, while the empty cyan stars represent the LAEs for which the \CIV\ wavelengths are either severely affected by skylines or not covered. The green horizontal lines are the excluded $3000$~\kms ranges blueward of the quasar redshifts. The blue horizontal line for PKS1937-101 indicates the lack of spectral coverage. Note that there are some LAEs without associated \CIV\ absorbers. Similarly, for some of the absorbers, there are no nearby LAE counterparts.}
    \label{fig:survey_summary} 
\end{figure*}

Our blind-absorber catalog of \CIV\ absorption in the 8 high-resolution, high SNR quasar spectra contains 489 Voigt components (152 systems) in the redshift range $2.900 \leq z \leq 3.835$ (median $z = 3.260$). The redshift distribution of the components along each sightline is shown by the black histograms in Fig.~\ref{fig:survey_summary}. The sightlines are sorted in ascending order of the quasar redshifts, shown by the vertical dotted blue lines, from top to bottom. We excluded any absorbers within $3000$~\kms of the quasars' emission redshifts (horizontal green lines) in order to avoid the quasars' proximity zones. The Q1422+23 sightline exhibits the highest number of \CIV\ systems (25) likely because of the excellent SNR of the spectrum (Table~\ref{qso_list}). PKS1937-101 has the lowest number of systems (7) possibly due to the lack of sufficient spectral coverage shown by the blue horizontal line in Fig.~\ref{fig:survey_summary}.

The red star symbols in Fig.~\ref{fig:survey_summary} indicate the redshifts of the 86 LAEs for which we have \CIV\ coverage. The open cyan star symbols represent the 10 LAEs for which the data are heavily contaminated by skylines (no \CIV\ coverage for one LAE towards PKS~1937--101). Note that there are some LAEs with no associated \CIV\ absorbers. Similarly, for some of the absorbers, there are no nearby LAE counterparts.

\subsection{Analyses of the ``blind''  \texorpdfstring{\CIV\ }\ absorbers}   

\subsubsection{Statistics of the \CIV\ absorption components} 
\label{sec:colm_limit} 

\begin{figure*}
\hbox{
	\includegraphics[width=0.5\textwidth]{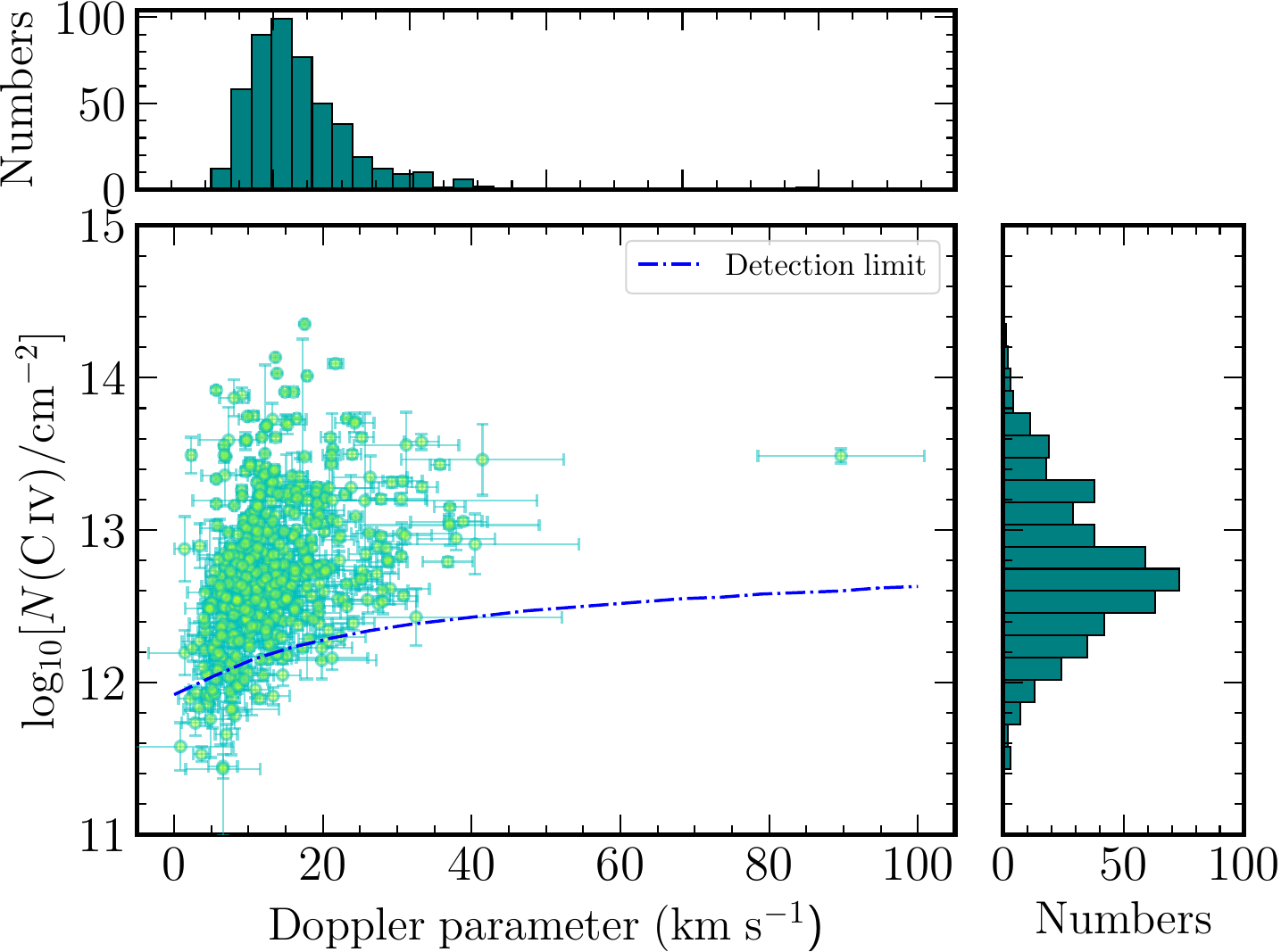}
	\includegraphics[width=0.5\textwidth]{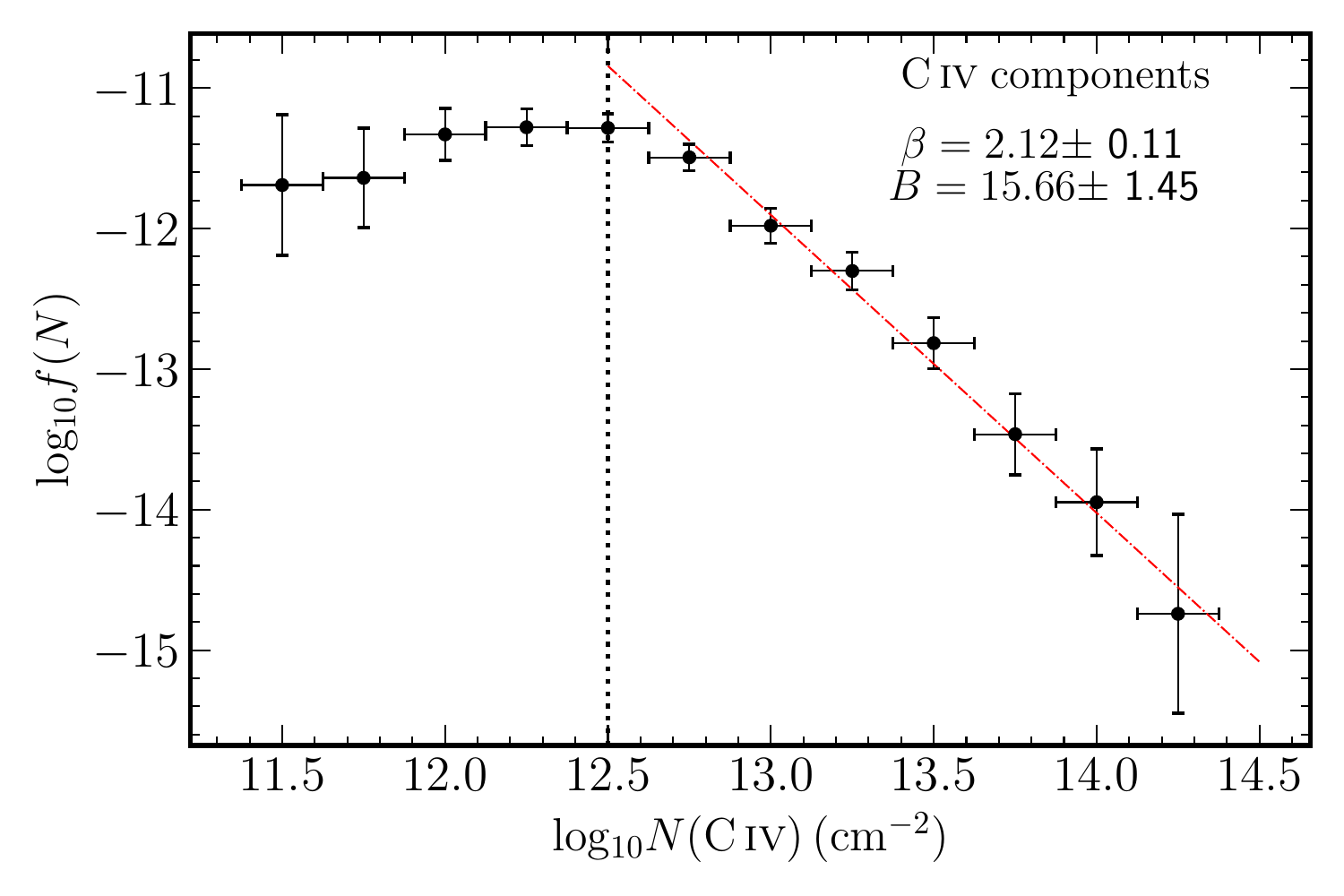}
	} 
    \caption{{\tt Left:} Best-fit values of \colm\ and \dop\ of all the \CIV\ absorption components. The distributions of the respective parameters are shown in the side panels. The blue dashed line indicates the maximum $N(\CIV)$ that can be hidden within the average $3\sigma$ noise of the spectra for a given \dop\ (see the text for details). The median \dop\ of individual \CIV\ components is $12.0$~\kms\ and median \colm\ is $\log_{10} N/\rm cm^{-2}$ = $12.7$. The median value of \colm\ of \CIV\ systems is $\log_{10} N/\rm cm^{-2}$ = $13.0$, where the linking velocity used to define a system is 300 \kms. A correlation is found between the $N(\CIV)$ and $b(\CIV)$. {\tt Right:} The CDDF for individual \CIV\ components. The error bars along the y-axis indicate the $1\sigma$ Poisson errors. Along the x-axis, the error bars represent the bin size (0.25 dex). The best-fit power-law relation, shown by the red dot-dashed line, is obtained using non-linear least squares  fitting for the data points rightwards of the vertical dotted line at $\log N/\rm cm^{-2} = 12.5$, below which our survey suffers from incompleteness.}
    \label{fig:bN_scatter_cddf}
\end{figure*}

In this section, we discuss the distributions and statistical properties of the Doppler parameter ($b$) and column density ($N$) of the \CIV\ components obtained from the blind search (i.e., identified without any prior knowledge of the LAE catalog). The left panel of Fig.~\ref{fig:bN_scatter_cddf} shows the scatter plot of $N$(\CIV) versus $b(\CIV)$ for all the \CIV\ components in our sample. The distributions of the parameters are also shown on the side panels. The column density of the ``blind'' \CIV\ ``components'' shows a range of $\log_{10} N(\CIV)/\rm cm^{-2} = 11.4 - 14.3$ with a mean of $12.7\pm0.5$ (median $12.7$). The mean and standard deviation of the \colm\ distribution for the \CIV\ ``systems'' is $\log_{10} N(\CIV)/\rm cm^{-2}$ = $13.1\pm0.6$ (median $13.0$) for our sample.

The $b$ parameters range from $0.87-41.4$~\kms, except for the one component with a very large $b$ value (89 \kms) detected towards Q0055-269. We believe this outlier is owing to the poor continuum placement at the wavelength of the absorber. The mean value of the \dop\ for the complete sample is $13.6\pm8.0$~\kms (median 12.0 \kms). Considering pure thermal broadening, the median $b(\CIV)$ value corresponds to an upper limit on the gas temperature of $1.1 \times 10^5$~K. Since this is close to the temperature at which \CIV\ ionization fraction peaks under collisional ionization equilibrium, it is likely that the bulk of the \CIV\ absorbers in our sample are photoionized \citep[see also][]{muzahid2012high}.

The blue dot-dashed line in Fig.~\ref{fig:bN_scatter_cddf} represents the $3\sigma$ detection limit corresponding to the median SNR of 50 per pixel. It indicates the maximum \CIV\ \colm\ that can be hidden within the $3\sigma$ noise of the spectrum for a given \dop. The \colm\ upper limits are estimated from the limiting equivalent width ($W_{\rm lim}$) for the \CIV~$\lambda1550$ line assuming the linear part of the  curve-of-growth relation.\footnote{Note that we used the weaker member of the \CIV\ doublet in this calculation which provides a conservative upper limit. This is because of the fact that  our visual identification technique relies on the detection of both the \CIV\ doublet members. However, in case of contamination, we used the $\lambda1548$ line and scaled the $W_{\rm lim}$ measurement to the one corresponding to the $\lambda1550$ line.} The $W_{\rm lim}$ is calculated using the following equation from \cite{hellsten1998observability}: 

\begin{equation}
\label{eqn:colm_lim}
W_{\rm lim} = \frac{N_{\sigma}~ \sqrt{N_{\rm pix}}  ~\Delta \lambda_{\rm pix}}{{\rm SNR}~ (1 + z_{\rm abs})}~\AA,  
\end{equation}
where, $N_{\sigma}~(= 3)$ is the detection threshold, $\Delta \lambda_{\rm pix}$ is the pixel size in \AA, SNR is the signal-to-noise ratio per pixel, and $N_{\rm pix}$ is the number of pixels contributing to the unresolved absorption. Note that $N_{\rm pix}$ depends on the assumed $b$-parameter. For a given $b$ value, we first calculate the $\sigma$ of the Line Spread Function (LSF) convolved line  (i.e., $\sigma^2 = \sigma_{\rm lsf}^2 + \sigma_{\rm line}^2$).  The $N_{\rm pix}$ is then determined by counting the number of pixels within the FWTM (full width at tenth maximum).

A correlation is seen between the column density ($N$) and Doppler parameter ($b$) of the \CIV\ components for the full sample (with Spearman rank correlation coefficient $\rho_s = 0.41$, $p$-value$= 4.2\times 10^{-21}$) and for the \CIV\ components above the detection limit curve ($\rho_s = 0.37$, $p$-value$=6.5\times 10^{-16}$). \Rev{We also found a positive correlation between $N$ and $b$ of all the \CIV\ components with \logN$>12.5$, which is the overall completeness limit for this survey (see the next section), with a $\rho_s = 0.19$ and $p$-value$=0.00058$.}

\subsubsection{Column Density Distribution Function}
\label{sec:cddf} 

The column density distribution function (CDDF), $f(N)$, is defined as the  number of absorbers (either individual components or systems) per unit \colm\ per unit redshift range. The CDDF can be parameterized as $f(N)dN = BN^{-\beta}dN,$ where $\beta$ and $B$ are the power-law index and normalization, respectively.

The {\em right} panel of Fig.~\ref{fig:bN_scatter_cddf} depicts the CDDF for individual \CIV\ components for our survey with a total redshift path length of $\Delta z = 6.20$. The plot also shows the best-fit power-law model.\footnote{We used the {\sc curve-fit} package of {\sc Python} which returns the best-fit parameters based on the non-linear least-squares method.} We obtained best-fit values of $\beta= 2.12\pm0.11$ and $B = 15.66\pm1.45$. The fitting is performed for $\log_{10} N(\CIV)/\rm cm^{-2} > 12.5$ below which our survey suffers from incompleteness. The $\beta$ value obtained for the \CIV\ systems is $1.77\pm0.17$ (plot not shown). The shallower power-law slope for systems is expected as multiple low-column density components are added to construct a system. Finally, we note that our $\beta$ values are consistent with earlier estimations at $z\gtrsim2$ in the literature \citep[e.g.,][]{songaila2001minimum, Ellison_2000, 10.1111/j.1365-2966.2009.15856.x,muzahid2012high}.


\subsection{Linking the LAEs and the \texorpdfstring{\CIV\ }\ absorbers}  
\label{sec:abs_distribution}

We have so far discussed the properties of the \CIV\ absorbers detected via a blind search. Now we will focus on the properties of the \CIV\ absorbers associated with LAEs.

\begin{figure}
\includegraphics[width=0.48\textwidth]{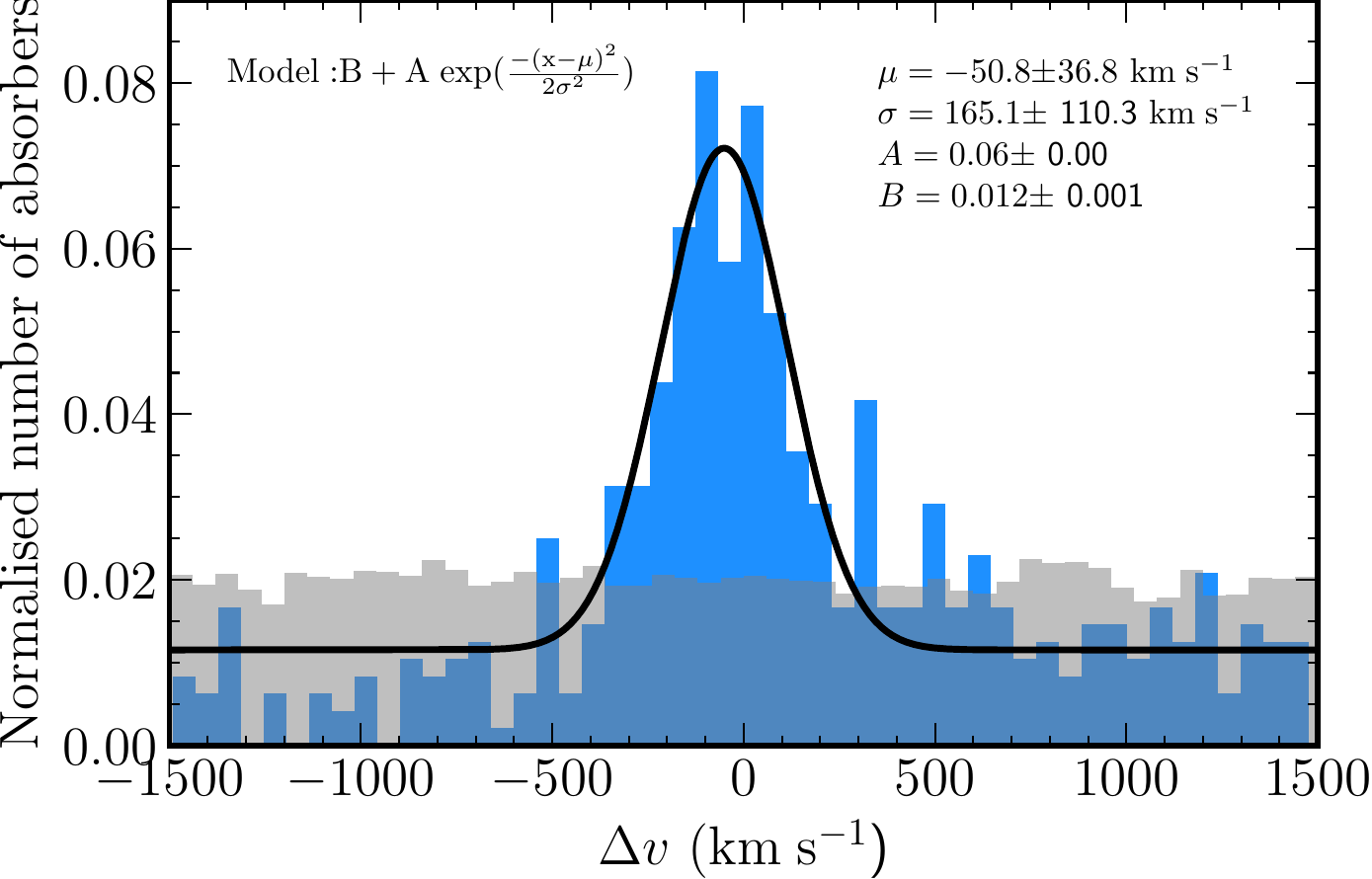}
\caption{The LOS velocity distribution of all the \CIV\ components within \pms1500~\kms of the LAE redshifts (corresponding to 0 \kms) is shown by the blue histogram.  The grey shaded histogram shows the same but for the random sample (see text). The number of \CIV\ components is enhanced near the $0$~\kms. We fit the distribution of the entire sample (black solid line) using a Gaussian+Linear model. The best-fit parameters of the model indicate an excess of \CIV\ components compared to a random region within 165~\kms, which corresponds to an FWHM of $\approx 400$~\kms. }
\label{fig:abs_distribution}
\end{figure}

The blue histogram in Fig.~\ref{fig:abs_distribution} shows the line of sight (LOS) velocity distribution of the individual \CIV\ components detected within \pms1500 \kms of the LAEs. The zero velocity ($\Delta v = 0$~\kms) corresponds to the redshift of the LAEs. It is evident from the figure that the number of \CIV\ components is enhanced near $\approx 0$~\kms. The best-fit Gaussian + baseline model (black solid line) returns a $\sigma$ of $\approx165 \pm 110$~\kms, corresponding to an FWHM of $\approx 390$~\kms.\footnote{The errors on the $\mu$ and $\sigma$ 
 values indicated in the plot are estimated using 1000 bootstrap realizations.} This indicates a velocity-correlation between the LAEs and \CIV\ absorbers within $\approx$ \pms400~\kms. The grey-shaded histogram in Fig.~\ref{fig:abs_distribution} shows the LOS velocity distribution of ``random'' \CIV\ absorbers. The ``random'' sample is constructed by generating 1000 uniform random numbers along each sightline within the corresponding \zmin\ and \zmax\ values listed in Table~\ref{qso_list}. We will refer to these 8000 redshifts as the ``random  sample'' throughout this paper. As expected, no clustering is seen for the LOS velocity distribution for this random sample.

Next, we explored that the high column density absorbers (i.e., \logN$> 13.0$) are more tightly correlated with the LAEs in velocity with a $\sigma$ value of $\approx136$~\kms ($\rm FWHM\approx320$~\kms; see Fig.~\ref{fig:abs_distribution_appendix} {\tt left} panel). We also inspected the $N$(\CIV)-weighted LOS velocity distribution of the \CIV\ absorbers within $\pm1500$~\kms of the LAEs. A similar Gaussian + baseline fit returned a significantly smaller $\sigma$ value of $\approx 62$~\kms. This indicates that the higher column density \CIV\ absorbers are more tightly correlated to the LAEs. We confirmed that these results are insensitive to the velocity range (i.e. $\pm1500$~\kms) we considered here. Hence, analyzing the absorbers within \pms500~\kms of the LAEs would be sufficient to study the interplay between metal-rich gas and low-mass  galaxies at high-$z$.

Out of the 489 \CIV\ components 202 lie within $\pm500$~\kms of the 86 LAE redshifts. We found that the column density and Doppler parameter distributions of these 202 \CIV\ components are not statistically different compared to the remaining 287 components ($p$-values from two-sided KS-tests $>0.05$). The median \CIV\ column density (${\rm \log_{10}} N(\CIV)/{\rm cm^{-2}} \approx 12.7$) and $b$-parameter ($\approx12.0$~\kms) of the two samples are also consistent with each other. These facts suggest two possibilities: ({\tt i}) a fraction of the absorbers that are unrelated to the detected LAEs are related to fainter galaxies below our detection thresholds and/or ({\tt ii}) not all the associated \CIV\ components (i.e., within \pms500~\kms of the LAEs) are actually physically related to the LAEs.

\begin{figure}
    \includegraphics[width=0.48\textwidth]{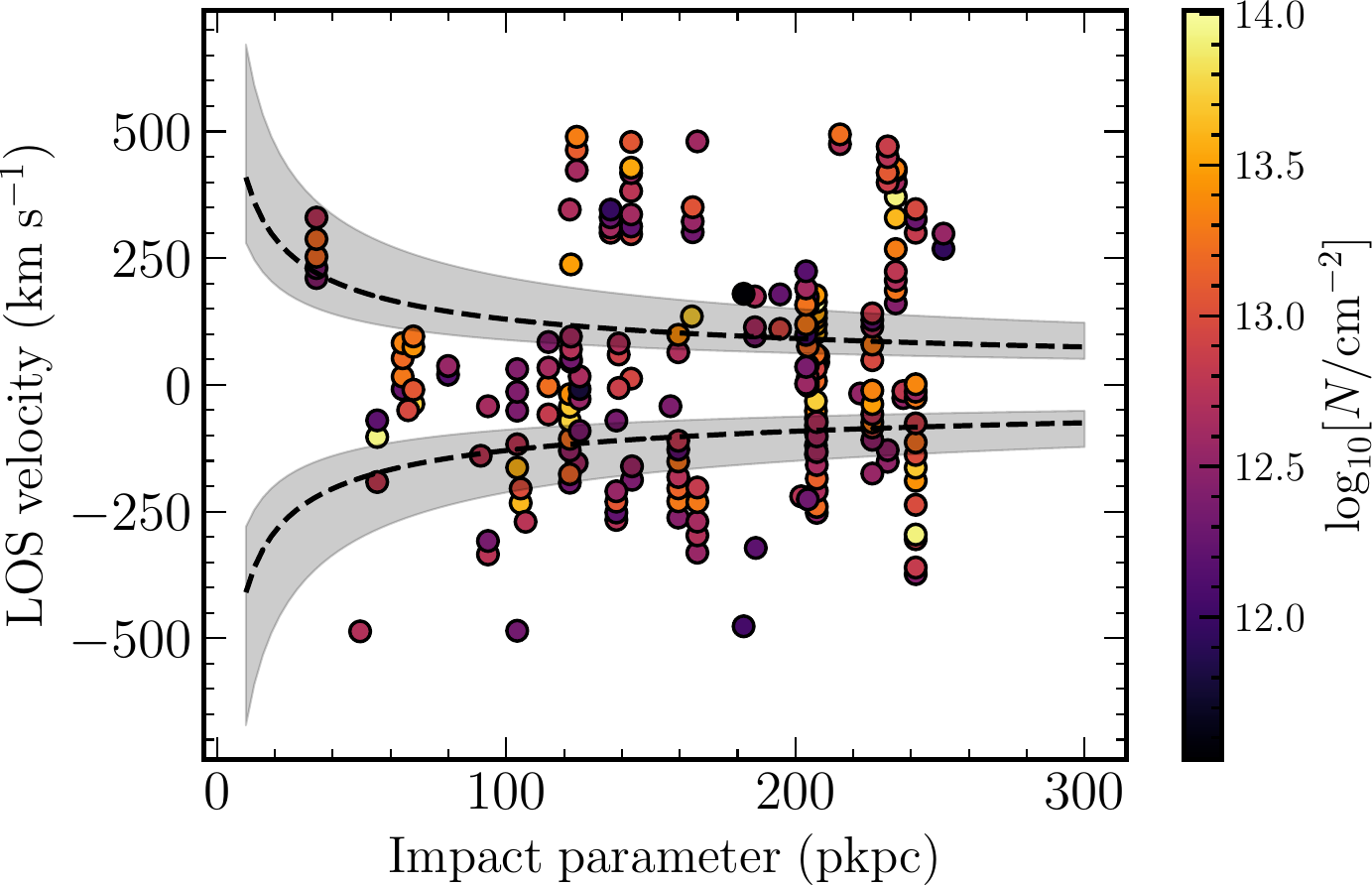}
    \caption{The LOS velocity of the \CIV\ components residing within \pms500~\kms of the LAEs as a function of the impact parameter of the respective LAEs. The points are color-coded by their respective column densities. The escape velocity as a function of distance for a halo mass range encompassed by our LAE sample i.e. $10.96 \leq {\rm log}_{10}\;M_{h}/{\rm M}_{\odot} \leq 11.72$, is indicated by the black shaded region. The black dashed line indicates the same for the median halo mass of our sample (${\rm log}_{10}\;M_{h}/{\rm M}_{\odot} = 11.29$). Roughly $45\%$ of the \CIV\ absorbers residing within \pms500~\kms of the LAEs are not bound.} 
    \label{fig:escape_velo}
\end{figure}

Fig.~\ref{fig:escape_velo}~shows the LOS velocity of the \CIV\ components detected within \pms500~\kms of the LAEs as a function of the impact parameter of the respective LAEs. A Spearman rank correlation test suggests no specific trend between these two parameters ($\rho_s = 0.08$ and $p$-value = $0.25$). The points are color-coded by their  column densities. The black shaded region shows the escape velocity at different distances for the halo mass range covered by our LAE sample\footnote{The method for deriving the halo masses of these LAEs is mentioned in section~\ref{Rvir}.} , i.e., $10.96 \leq \log_{10} M_{h}/\rm M_{\odot} \leq 11.72$. The black dashed line indicates the same for the median halo mass of our sample ($\log_{10} M_{h}/\rm M_{\odot} = 11.29$). To draw these curves, we used the relation, $v_{\rm esc} = \sqrt{\frac{\rm 2G M_{h}}{R}}$, and assumed that the impact parameter is the same as the 3D distance (R) from the center of the halo.

Out of the 202 \CIV\ components detected within \pms500~\kms of the LAE redshifts 111 ($\approx 55\%$) fall inside this shaded region. It is safe to say that the absorbers falling beyond the shaded region are not bound to the LAEs. However, interpreting the nature of the absorbers within the $v_{\rm esc}$ bounds is not straightforward since the impact parameter and LOS velocity of an absorber are merely lower limits on the actual 3D distance and velocity, respectively. Nevertheless, from now on, we will refer to these components as ``bound absorbers''.

We did not find any significant difference between the \colm\ of ``bound absorbers'' (111 components) and the remaining ``unbound absorbers'' (378 components) ($D_{KS}= 0.14$ and $p$-value = $0.25$). The same holds for the \dop\ distributions ($D_{KS} = 0.16$ and $p$-value = $0.12$). The median \CIV\ column densities for the bound and unbound components are \logN = $12.8$ and $12.7$ respectively, and the median Doppler parameters are $12.4$ and $10.9$~\kms.

We take into consideration the fact that the LOS velocity depends on the systemic redshifts of the LAEs. The systemic redshifts are calculated from the \lya\ redshifts using the empirical relation provided in \citet{Muzahid_2020}. We confirmed that there is no correlation between the LOS velocities of various \CIV\ components and the velocity offsets applied to correct the \lya\ redshifts. This ensures that there is no systematic uncertainty, related to the scatter in the empirical relation, in determining  the nature of the absorbers (i.e., unbound vs. bound).

Finally, we reconstructed Fig.~\ref{fig:abs_distribution} using only the bound components (see {\tt right} panel of Fig.~\ref{fig:abs_distribution_appendix}). The line centroid of the best-fit model ($\mu = -35.7\pm33.3$~\kms) is consistent with $\Delta v \approx 0$~\kms\ while in Fig.~\ref{fig:abs_distribution} we noted a $\approx 1.4\sigma$ offset ($\mu = -50.8\pm36.8$~\kms) with respect to the systemic redshift. Note that, \citet{Muzahid_2021} also reported an offset of $-30\pm22$~\kms\ between their stacked \CIV\ profile and the systemic redshifts of LAEs.  The best-fit $\sigma$ of 105~\kms\ obtained from the plot  corresponds to an FWHM of $\approx 247$~\kms. As a result, for the majority of our analysis, we focused on a velocity window of \pms250~\kms. \Rev{The distribution of \CIV\ absorbers at velocities $\gtrsim|300|$~\kms, however, is not symmetric, as seen in  Fig.~\ref{fig:abs_distribution} and Fig.~\ref{fig:abs_distribution_appendix}, with the number of absorbers being somewhat higher on the red-ward (i.e., $+ve$ velocities). We discuss the possible reason(s) for this in Section~\ref{sec4}.}


\begin{figure}
\includegraphics[width=0.47\textwidth]{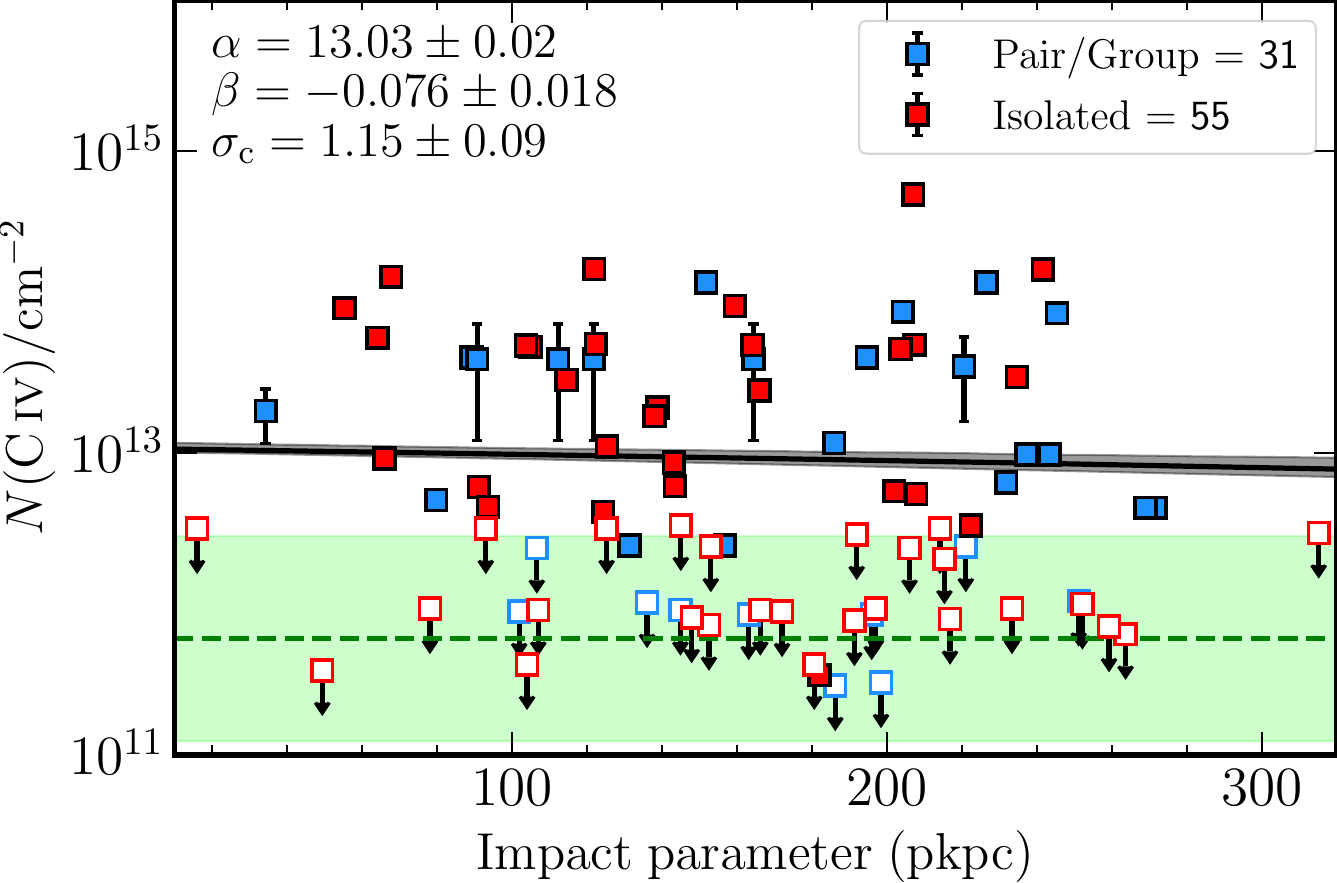}
    \caption{The total \CIV\ \colm\ as a function of impact parameter. The solid squares represent the total $N(\CIV)$ measured within \pms250~\kms\ of the LAEs. The open squares with downward arrows indicate $3\sigma$ \colm\ upper limits for the LAEs without detectable \CIV\ absorption within \pms250~\kms. The black solid line shows the best-fit log-linear model, and the grey-shaded region shows the $1\sigma$ uncertainty of the fitted curve. This uncertainty does not consider the intrinsic scatter of the plot. No significant correlation exists between total $N({\mbox{C\,{\sc iv}}})$ and $\rho$. The green dashed line and the green shaded region represent the mean $N(\CIV)$ including the contribution from upper limits using the Kaplan-Meier estimator and $1\sigma$ scatter obtained for the random sample. The $N(\CIV)$ values measured near the LAEs are significantly higher compared to the random sample.}  
    \label{fig:n_profile}
\end{figure}

\subsubsection{\CIV\ column density profile}
\label{sec:N_profile}

One of the main objectives of this work is to probe the radial variations of the \CIV\ line parameters in the CGM of the high-$z$, low-mass galaxies. In this section, we will investigate the variation of \CIV\ column density as a function of the impact parameter ($\rho$). In the previous section, we observed an enhancement of \CIV\ components around the systemic redshifts of the LAEs out to a few hundred \kms (see Fig.~\ref{fig:abs_distribution}), indicating a connection between the galaxies and the metal-enriched gas around them. In Fig.~\ref{fig:n_profile} we present the distribution of total \CIV\ column density within a $\pm$250~\kms\ velocity window against the impact parameter for the 86 LAEs in our sample. In cases where no \CIV\ component is found, we obtained $3\sigma$ upper limits on $N$(\CIV) using the method described in section \ref{sec:colm_limit} for the SNR measured near the \CIV~$1550$~\ang\ line. When the $1550$~\ang\ line is contaminated, we used the $1548$~\ang\ line to calculate the upper limit of equivalent width and then calculated the column density upper limit corresponding to the $1550$~\ang\ line using the relation of the linear part of the curve-of-growth. We assumed that the undetected line has a Doppler parameter of $12.0$~\kms (median $b$ of our ``blind'' components) in order to calculate the $N_{\rm pix}$ term in Eq.~\ref{eqn:colm_lim}. The upper limits on the \CIV\ column density are shown as downward arrows in the plot.

\begin{table}
\centering
\caption{Best-fit parameters for log-linear fit to the \CIV\ column density profile} 
\begin{tabular}{cccc} 
    \hline
    $dv$ (\kms) & $\alpha$ & $\beta$ & $\sigma$\\
    \hline
    $\pm500$  & 13.04\pms0.03 & $-$0.068\pms0.023 & 1.06\pms0.10\\
    $\pm300$  & 13.03\pms0.02 & $-$0.074\pms0.019 & 1.14\pms0.10\\
    $\pm250$  & 13.03\pms0.02 & $-$0.076\pms0.018 & 1.15\pms0.09\\
    $\pm150$  & 13.02\pms0.02 & $-$0.082\pms0.013 & 1.26\pms0.13\\
    \hline
    \end{tabular}
    \label{tab:log_linear}    
\end{table}

The horizontal green dashed line in Fig.~\ref{fig:n_profile} shows the median column density measured for the random sample.  The green-shaded region shows the corresponding $1\sigma$ scatter. For each random redshift, we determined the total \CIV\ column density using our ``blind'' catalog within $\pm250$~\kms. In the absence of any detected \CIV\ component, we estimated the $3\sigma$ \colm\ upper limit. Using the empirical cumulative distribution function (ECDF) for censored data (in the presence of upper limits) using the Kaplan-Meier estimator\footnote{We used the {\sc cenfit} function under the {\sc nada} package in {\sc R}. However, in this instance, rather than being randomly distributed, the censored data points are almost always lower  compared to the detected values.}, we calculated the mean \CIV\ \colm\ for the random sample as ${\rm \log_{10}} N(\CIV)/\rm cm^{-2} = 11.77$ with a $1\sigma$ scatter of $0.68$ dex. The vast majority of the \CIV\ components within the $\pm250$~\kms\ of the LAE redshifts show considerably higher column densities compared to the median \CIV\ column density observed for the random sample. A two-sided $\rm logrank$ test\footnote{Performed using the {\sc survival} package in {\sc R}.} suggests that the observed \CIV\ column density distribution around the LAEs is significantly different than the random regions with a $p$-value of $<10^{-7}$.

While the \CIV\ column densities observed near the LAE redshifts are significantly higher than random regions, no significant trend is seen between the \CIV\ column density and impact parameter. A Kendall's $\tau$ rank-order correlation  test, which takes care of the upper limits, suggests no significant anti-correlation between the total \colm\ within $\pm250$~\kms\ of the LAEs and impact parameter with a $\tau_{\rm K} = -0.08$ and a $p$-value of $0.25$\footnote{We used the {\sc cenken} function under the {\sc nada} package in {\sc R}.}. This is also true for the absorbers within $\pm500$~\kms\ of the LAEs ($\tau_{\rm K} = -0.09$ and $p$-value $=0.21$).

Following \citet[see also \citet{nielsen2013magiicat}]{chen2010empirical} we characterized the dependence between \colm\ and $\rho$ (in pkpc) assuming a log-linear model described as:  
\begin{equation}
\label{eqn:colm_profile_model}
{\rm log_{10}}~ N({\mbox{C\,{\sc iv}}}) = \alpha + \beta~ (\frac{\rho}{\rm 160 \, pkpc})~.  
\end{equation}
To determine the values of the model parameters ($\alpha$, $\beta$), we used the following maximum-likelihood function that takes care of both $n$ number of measurements and $m$ number of upper limits: \\

\begin{math}
\label{likelihood}
    \mathcal{L}(N) = \left( \prod _{i=1}^{n} \frac{1}{\sqrt{2 \pi \sigma_i ^2}} {\rm exp} \left\{ - \frac{1}{2} \left[ \frac{N_i - N(\rho_i)}{\sigma_i} \right]^2 \right\} \right) \\
            \times \left( \prod _{i=1}^{m} \int _{- \infty}^{N_i} \frac{dN^{'}}{\sqrt{2 \pi \sigma_i ^2}} {\rm exp} \left\{ - \frac{1}{2} \left[ \frac{N^{'} - N(\rho_i)}{\sigma_i} \right]^2 \right\} \right), \\
\end{math}

\noindent 
where, $N_i$ is the sum of the individual component column densities  within $\pm250$~\kms or the $3\sigma$ upper limit, and $N(\rho_i)$ is the expected value of log$_{10}N({\mbox{C\,{\sc iv}}})$ from the model (Eq.~\ref{eqn:colm_profile_model}) for individual $\rho$. The total error, $\sigma_i = \sqrt{\sigma_{c}^2 + \sigma_{mi}^2}$; with $\sigma_{mi}$ denoting the error associated with $i$th measurement and $\sigma_c$ accounting for the intrinsic scatter. The best-fit parameters of Eq.~\ref{eqn:colm_profile_model} calculated for different velocity windows are given in Table~\ref{tab:log_linear}. These parameters are obtained using the nested sampling Monte Carlo algorithm MLFriends \citep[]{Buchner_2014,Buchner_2019} using the {\tt UltraNest}\footnote{\url{https://johannesbuchner.github.io/UltraNest/}} package of Python \cite[]{Buchner_2021}. The best-fit slope (e.g., $-0.076\pm0.018$ for $\pm250$~\kms) with a large scatter ($\sigma_c =1.15$ dex) clearly rules out any strong trend between $N(\CIV)$ and the impact parameter. To cross-verify our results, we used the Python {\tt linmix} package \cite[]{Kelly_2007} \footnote{\url{https://github.com/jmeyers314/linmix}}, which employs a hierarchical Bayesian model for fitting a straight line to data. We found the best-fit values as $\alpha = 12.83 \pm 0.44$, $\beta = -0.44 \pm 0.41$, and $\sigma_c = 1.79 \pm 0.47$. Despite the fact that $\alpha$ and $\sigma_c$ are consistent with previous values, the $\beta$ value is different here. The $\beta$ value returned by {\tt linmix} is consistent with zero within $1\sigma$ significance, which further confirms the absence of any strong trend as suggested by Kendal's $\tau$ tests.

\begin{table}
\centering
\caption{Kendall's $\tau$ correlation test results between $N$(\CIV) and $\rho$}
\begin{tabular}{lcc} 
    \hline
    Sample (No. of LAEs) & $\tau_{K}$ & $p$-value\\
    \hline
    All (86)         & $-0.09$ & 0.21 \\
    Pair/Group (31) & $-0.06$ & 0.67 \\
    Isolated (55)    & $-0.14$ & 0.14 \\
    \hline
    \end{tabular}
\label{tab:colm_rho}
\end{table}

The data points in Fig. \ref{fig:n_profile} are color-coded by the environments (``isolated'' or ``pair/group'') of the LAEs. To define the environment, we used the procedure followed by \citet{Muzahid_2021}. \Rev{We defined a galaxy pair/group in such a way that each member of a pair/group has at least one companion within \pms500~\kms\ and within the MUSE FoV ($\approx 300$~pkpc on a side)}. These conditions are met by 31 out of our 86 galaxies. The rest are classified as ``isolated'' galaxies. \citet{Muzahid_2021}  argued that detecting 2 LAEs within \pms500~\kms\ within the MUSE FoV corresponds to a galaxy overdensity of $\approx 10$. The blue squares represent the $N(\CIV)$ for the pair/group LAEs whereas the red squares are for the isolated LAEs. It is readily apparent that the bulk of the upper limits on $N$(\CIV) (non-detections) arises from the isolated systems. Specifically, $\approx44$\% of the isolated LAEs (24 out of 55) show upper limits, whereas only $\approx17$\% (5 out of 29) of the pair/group LAEs exhibit \CIV\ non-detection. This hints at a higher \CIV\ covering fraction for the pair/group galaxies compared to the isolated LAEs. We further investigate this in Section~\ref{Rvir}

Next, we fit the log-linear relation (Eq.~\ref{eqn:colm_profile_model}) separately for isolated and pair/group galaxies. The slopes, intercepts, and scatter for the isolated and pair/group subsamples are consistent with each other and with the full sample. The lack of a strong trend for these two subsamples is also apparent from the Kendall's $\tau$ results summarised in Table~\ref{tab:colm_rho} (for column density measured within $\pm 500$~\kms). On a passing note, we also looked for any correlation between column density and LOS velocity from the LAE and did not find anything significant (Spearman correlation test result: $\rho_s= -0.06$, $p$-value=0.4).

Finally, we note that the fraction of pair/group LAEs increases with the impact parameter. For example, from 0--100 pkpc, the fraction is $\frac{4}{14}$ (pair/group-4, isolated-10); from 100--200 pkpc, the ratio becomes $\frac{16}{44}$ (pair/group-16, isolated-28); and from 200--300 pkpc, it is $\frac{11}{28}$ (pair/group-11, isolated-17). This can be attributed to the ``edge effect'' owing to the limited FoV of MUSE. A pair/group LAE can be confused with an isolated LAE because the companion LAE is just outside the MUSE FoV.


\subsubsection{\CIV\ \dop\ profile}

\begin{figure}
	\includegraphics[width=0.48\textwidth]{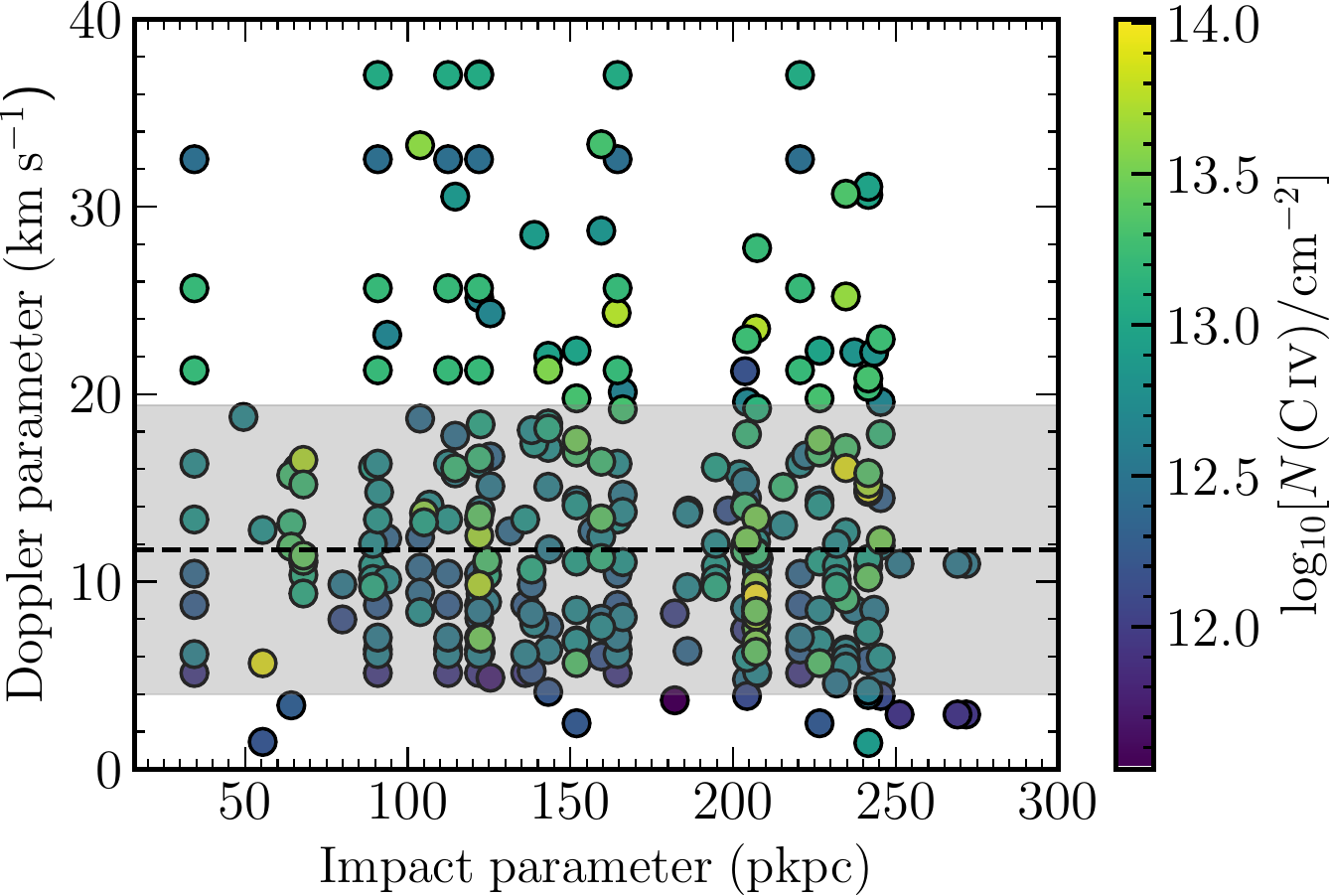}
    \caption{\dop\ of the \CIV\ components within \pms500~\kms\ from the LAEs against impact parameter. The data points are color-coded by their column densities as indicated by the color bar. The black dashed line and gray shaded region are  the median value of the Doppler parameter and its $1\sigma$ scatter for the random sample. No significant correlation exists between $b(\CIV)$ and $\rho$.} 
    \label{fig:b_profile}
\end{figure}

Fig.~\ref{fig:b_profile} shows the $b$-parameters of individual \CIV\ components associated with the LAEs as a function of the impact parameter, color-coded by the \colm, detected within $\pm$500~\kms\ from the LAEs. It seems there is no  significant trend between $b$(\CIV) and $\rho$. This visual impression is also supported by the Spearman rank correlation test results summarized in Table~\ref{tab:brho}. Since we observed a mild correlation between $N(\CIV)$ and $b$ in Section~\ref{sec:colm_limit}, we split the sample into three different \colm\ bins and performed the Spearman tests  between $b(\CIV)$ and $\rho$ for each of these three \colm\ bins (see Table \ref{tab:brho}). The table also includes the correlation test results for the pair/group and isolated LAEs separately. The overall trend suggests the $b(\CIV)$ and impact parameters of \CIV\ components are not significantly correlated for any of the subsamples except maybe for the lowest column density bin ($p$-value = $0.03$). Similarly, no significant correlation was found between $b$ and the LOS velocity of the \CIV\ components with respect to the LAEs (Spearman correlation test result: $\rho_s= -0.02$, $p$-value$=0.7$).

We also generated a random sample of Doppler parameters considering all the \CIV\ components within \pms500~\kms of our random redshifts described in section \ref{sec:abs_distribution}. In case of a non-detection, we considered the $b$ value of the nearest detected \CIV\ component to that random redshift. We obtained a median of 11.7~\kms\ and a standard deviation of 7.7~\kms\ for the random sample which are  shown by the black dashed line and the grey-shaded region, respectively, in Fig.~\ref{fig:b_profile}. A KS test between the $b$-distributions of the CGM sample and the random sample gives $D_{KS} = 0.07$ and $p$-value~$ = 0.12$. This suggests these two populations are not statistically different.

\begin{table}
\begin{center}
\begin{threeparttable}[b] 
\renewcommand{\arraystretch}{1.2}
\caption{Spearman rank correlation test results between $b$ and impact parameter for the \CIV\ components}
\begin{tabular}{lrcc}
    \hline
    Sample$^1$ & Number$^2$ & $\rho_{s}^3$ & $p$-value$^4$\\
    \hline
    \hline
    All                 & 303 &  $-0.08$ & $0.15$ \\ 
    $11.5 \leq$ \logN $<12.5$ & 80  & $-0.24$ & $0.03$ \\
    $12.5 \leq$ \logN $<13.0$ & 127 & $-0.13$ & $0.15$ \\
    $13.0 \leq$ \logN $<14.5$ & 96  & $+0.04$ & $0.67$ \\
   Related to Pair/Group LAEs & 151  & $-0.09$ & $0.27$ \\
    Related to Isolated LAEs & 152  & $-0.08$ & $0.32$ \\
    \hline
    \end{tabular}
\label{tab:brho}
Notes--1: \CIV\ sample, based on \colm\ and environment; 2: Number of \CIV\ components contributing to the sample. Note that, the total number of \CIV\ components in sample ``All" is greater than the number of components present within $\pm500$~\kms of the LAEs (202) because closely spaced LAEs will share some of the \CIV\ components with each other, so some of the components will contribute multiple times; 3: Spearman rank correlation coefficient; 4: Probability that there is no correlation.
\end{threeparttable}
\end{center}
\end{table}

\subsubsection{\CIV\ covering fraction }

Covering fractions of different atomic and ionic species in the CGM of galaxies are a measure of the patchiness of the medium and metal distribution, and provide stringent constraints on galaxy evolution models. In this section, we determine the \CIV\ covering fraction in the neighbourhood of the LAEs in our sample. For a given threshold column density, $N'$, the covering fraction ($f_{\rm c}$) is defined as: 
$$f_c = \frac{n_{\rm Hit} (N \geq N')}{n_{\rm Total}},$$ 
where, $n_{\rm Hit}$ is the number of LAEs with a total column density $N \geq N'$, measured within a specific LOS velocity window around the LAEs, and $ n_{\rm Total}$ is the total number of LAEs for which the spectra were sensitive enough to detect absorption down to the threshold column density of $N'$.

\begin{figure}
	\includegraphics[width=0.48\textwidth]{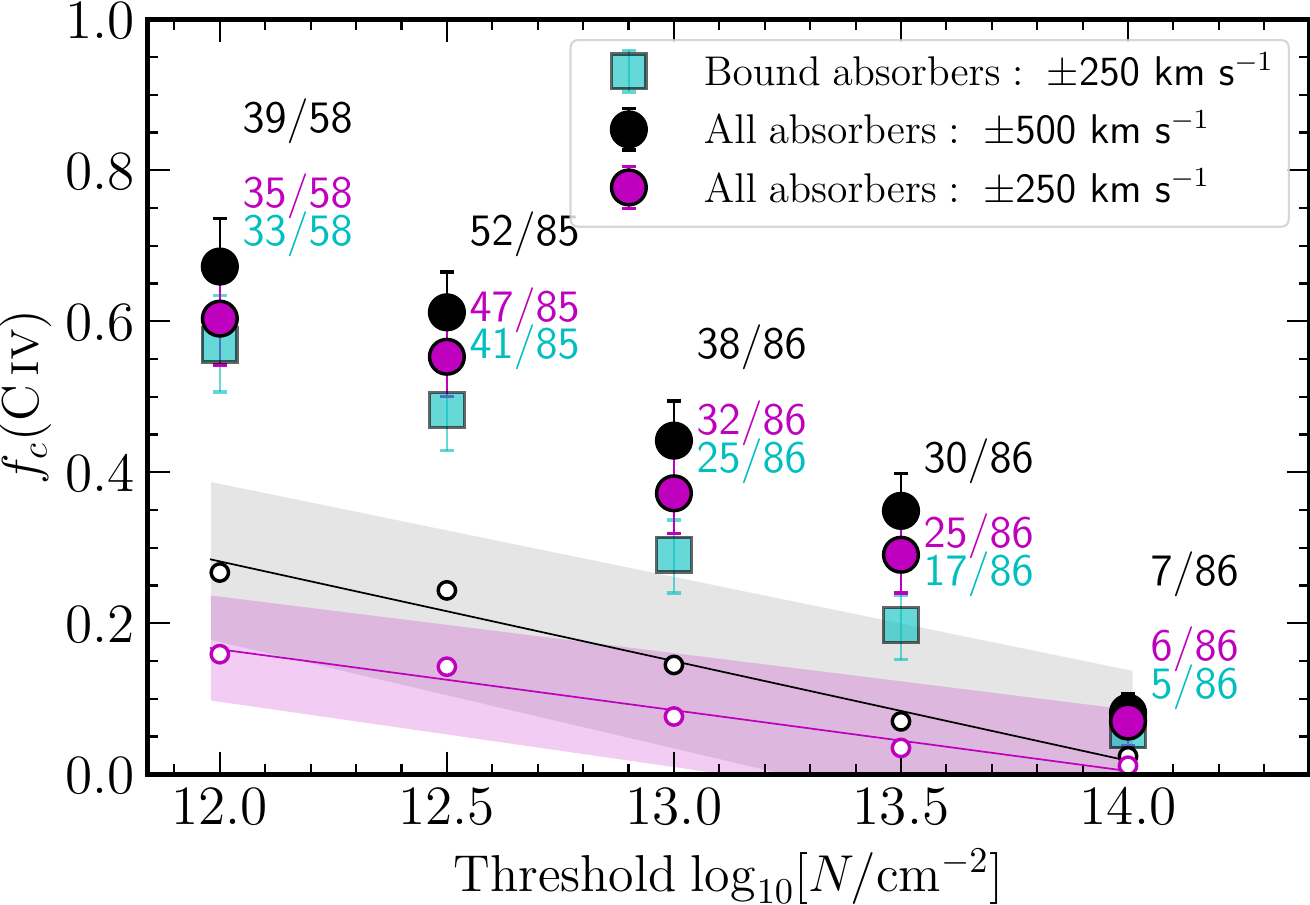}
    \caption{The \CIV\ \fc\ ($f_{\rm c}$) within 16--315~pkpc of the LAEs for different threshold column densities. The filled circles indicate the covering fraction calculated using all the absorbers within the LOS velocity separations from the LAEs indicated by the legends. The cyan squares represent the covering fractions estimated using the bound absorbers only. The error bars in $f_{\rm c}$ are calculated using the binomial proportion confidence interval (BPCI) Wilson Score method with $68\%$ confidence level. The $n_{\rm Hit}$ and $n_{\rm Total}$ for each measurement are indicated by the numbers in the numerator and denominator of similar colors adjacent to the points. The black and magenta small open circles represent the \CIV\ covering fractions measured for the random sample for $\pm500$ and $\pm250$~\kms velocity windows, respectively. The solid lines represent the corresponding best-fit linear relations between the threshold \colm\ and the $f_c$ for the random sample, while the shaded regions represent the $1\sigma$ uncertainties.} 
    \label{fig:fc}
\end{figure}

\renewcommand{\arraystretch}{1.8}
\begin{table*}
	\centering
	\caption{Summary of \CIV\ covering fraction measurements}
	\label{tab:fc}
	\begin{tabular}{|c|ccccc|}
        \hline
	\multicolumn{1}{c}{\centering Sample} &  \multicolumn{5}{c}{\centering Threshold \logN} \\
        \cline{2-6}
        & $12.0$ & $12.5$ & $13.0$ & $13.5$ & $14.0$\\
        \hline
        \multicolumn{1}{p{2cm}}{\centering All absorbers: $\pm500$}
        & \multicolumn{1}{p{2cm}}{\centering $0.77^{+0.11}_{-0.16}$}
        & \multicolumn{1}{p{2cm}}{\centering $0.64^{+0.05}_{-0.06}$}
        & \multicolumn{1}{p{2cm}}{\centering $0.44^{+0.05}_{-0.05}$}
        & \multicolumn{1}{p{2cm}}{\centering $0.35^{+0.05}_{-0.05}$}
        & \multicolumn{1}{p{2cm}}{\centering $0.08^{+0.03}_{-0.02}$}\\
        
        \multicolumn{1}{p{2cm}}{\centering All absorbers: $\pm250$}
        & \multicolumn{1}{p{2cm}}{\centering $0.60^{+0.06}_{-0.06}$}
        & \multicolumn{1}{p{2cm}}{\centering $0.55^{+0.05}_{-0.05}$}
        & \multicolumn{1}{p{2cm}}{\centering $0.37^{+0.05}_{-0.05}$}
        & \multicolumn{1}{p{2cm}}{\centering $0.29^{+0.05}_{-0.05}$}
        & \multicolumn{1}{p{2cm}}{\centering $0.07^{+0.03}_{-0.02}$}\\

        \multicolumn{1}{p{2cm}}{\centering All absorbers: $\pm125$}
        & \multicolumn{1}{p{2cm}}{\centering $0.52^{+0.06}_{-0.06}$}
        & \multicolumn{1}{p{2cm}}{\centering $0.41^{+0.05}_{-0.05}$}
        & \multicolumn{1}{p{2cm}}{\centering $0.28^{+0.04}_{-0.05}$}
        & \multicolumn{1}{p{2cm}}{\centering $0.19^{+0.04}_{-0.04}$}
        & \multicolumn{1}{p{2cm}}{\centering $0.05^{+0.02}_{-0.03}$}\\

        \multicolumn{1}{p{2cm}}{\centering Bound absorbers:\\$\pm250$}
        & \multicolumn{1}{p{2cm}}{\centering $0.59^{+0.06}_{-0.06}$}
        & \multicolumn{1}{p{2cm}}{\centering $0.51^{+0.05}_{-0.05}$}
        & \multicolumn{1}{p{2cm}}{\centering $0.33^{+0.05}_{-0.05}$}
        & \multicolumn{1}{p{2cm}}{\centering $0.22^{+0.05}_{-0.04}$}
        & \multicolumn{1}{p{2cm}}{\centering $0.06^{+0.03}_{-0.02}$}\\
        \hline

        \multicolumn{1}{p{2.5cm}}{\centering Pair/Group LAEs:\\$d_{\rm cov}=\pm250$\\$d_{\rm link}=\pm500$}
        & \multicolumn{1}{p{1.5cm}}{\centering $--$}
        & \multicolumn{1}{p{2cm}}{\centering $0.57^{+0.09}_{-0.09}$}
        & \multicolumn{1}{p{2cm}}{\centering $0.39^{+0.09}_{-0.09}$}
        & \multicolumn{1}{p{2cm}}{\centering $0.36^{+0.09}_{-0.08}$}
        & \multicolumn{1}{p{2cm}}{\centering $--$}\\

        \multicolumn{1}{p{2.5cm}}{\centering Isolated LAEs:\\$d_{\rm cov}=\pm250$\\$d_{\rm link}=\pm500$}
        & \multicolumn{1}{p{1.5cm}}{\centering $--$}
        & \multicolumn{1}{p{2cm}}{\centering $0.37^{+0.09}_{-0.08}$}
        & \multicolumn{1}{p{2cm}}{\centering $0.17^{+0.08}_{-0.06}$}
        & \multicolumn{1}{p{2cm}}{\centering $0.14^{+0.08}_{-0.05}$}
        & \multicolumn{1}{p{2cm}}{\centering $--$}\\

        \multicolumn{1}{p{2.5cm}}{\centering Pair/Group LAEs:\\$d_{\rm cov}=\pm250$\\$d_{\rm link}=\pm250$}
        & \multicolumn{1}{p{1.5cm}}{\centering $--$}
        & \multicolumn{1}{p{2cm}}{\centering $0.68^{+0.09}_{-0.10}$}
        & \multicolumn{1}{p{2cm}}{\centering $0.50^{+0.10}_{-0.10}$}
        & \multicolumn{1}{p{2cm}}{\centering $0.50^{+0.10}_{-0.10}$}
        & \multicolumn{1}{p{2cm}}{\centering $--$}\\

        \multicolumn{1}{p{2.5cm}}{\centering Isolated LAEs:\\$d_{\rm cov}=\pm250$\\$d_{\rm link}=\pm250$}
        & \multicolumn{1}{p{1.5cm}}{\centering $--$}
        & \multicolumn{1}{p{2cm}}{\centering $0.38^{+0.10}_{-0.09}$}
        & \multicolumn{1}{p{2cm}}{\centering $0.23^{+0.10}_{-0.07}$}
        & \multicolumn{1}{p{2cm}}{\centering $0.14^{+0.09}_{-0.05}$}
        & \multicolumn{1}{p{2cm}}{\centering $--$}\\
        \hline
        \end{tabular}
        \begin{tabular}{|l|l|l|l|}
        \multicolumn{1}{c}{\centering Variable} & 
        \multicolumn{1}{c}{\centering $dv$} &
        \multicolumn{1}{c}{\centering Type of LAEs} &
        \multicolumn{1}{c}{\centering Covering fraction$^a$} \\

        \hline

        \multicolumn{1}{p{2.5cm}}{\centering Redshift, $z<3.3$}&
        \multicolumn{1}{p{2.5cm}}{\centering $\pm250$}&
        \multicolumn{1}{p{2.5cm}}{\centering All}
        & \multicolumn{1}{p{5cm}}{\centering $0.55^{+0.07}_{-0.08}$~ ($0.30^{+0.07}_{-0.06}$)}\\

        \multicolumn{1}{p{2.5cm}}{}&
        \multicolumn{1}{p{2.5cm}}{}&
        \multicolumn{1}{p{2.5cm}}{\centering Isolated}
        & \multicolumn{1}{p{5cm}}{\centering $0.46^{+0.09}_{-0.09}$~ ($0.30^{+0.09}_{-0.08}$)}\\

        \multicolumn{1}{p{2.5cm}}{}&
        \multicolumn{1}{p{2.5cm}}{}&
        \multicolumn{1}{p{2.5cm}}{\centering Controlled $\rho$}
        & \multicolumn{1}{p{5cm}}{\centering $0.53^{+0.09}_{-0.09}$~ ($0.31^{+0.09}_{-0.08}$)}\\

        \multicolumn{1}{p{2.5cm}}{\centering Redshift, $z>3.3$}&
        \multicolumn{1}{p{2.5cm}}{\centering $\pm250$}&
        \multicolumn{1}{p{2.5cm}}{\centering All}
        & \multicolumn{1}{p{5cm}}{\centering $0.56^{+0.07}_{-0.07}$~ ($0.44^{+0.07}_{-0.07}$)}\\

        \multicolumn{1}{p{2.5cm}}{}&
        \multicolumn{1}{p{2.5cm}}{}&
        \multicolumn{1}{p{2.5cm}}{\centering Isolated}
        & \multicolumn{1}{p{5cm}}{\centering $0.57^{+0.09}_{-0.09}$~ ($0.39^{+0.09}_{-0.09}$)}\\

        \multicolumn{1}{p{2.5cm}}{}&
        \multicolumn{1}{p{2.5cm}}{}&
        \multicolumn{1}{p{2.5cm}}{\centering Controlled $\rho$}
        & \multicolumn{1}{p{5cm}}{\centering $0.55^{+0.09}_{-0.09}$~  ($0.45^{+0.09}_{-0.09}$)}\\
        \hline

        \multicolumn{1}{p{2.5cm}}{$\rho: 16--100$ pkpc}&
        \multicolumn{1}{p{2.5cm}}{\centering $\pm250$}&
        \multicolumn{1}{p{2.5cm}}{\centering All}
        & \multicolumn{1}{p{5cm}}{\centering $0.71^{+0.10}_{-0.13}$~ ($0.43^{+0.13}_{-0.12}$)}\\

        \multicolumn{1}{p{2.5cm}}{}&
        \multicolumn{1}{p{2.5cm}}{}&
        \multicolumn{1}{p{2.5cm}}{\centering Isolated}
        & \multicolumn{1}{p{5cm}}{\centering $0.60^{+0.14}_{-0.16}$~ ($0.30^{+0.15}_{-0.12}$)}\\

        \multicolumn{1}{p{2.5cm}}{$\rho: 100--200$ pkpc}&
        \multicolumn{1}{p{2.5cm}}{\centering $\pm250$}&
        \multicolumn{1}{p{2.5cm}}{\centering All}
        & \multicolumn{1}{p{5cm}}{\centering $0.46^{+0.07}_{-0.07}$~ ($0.39^{+0.07}_{-0.07}$)}\\

        \multicolumn{1}{p{2.5cm}}{}&
        \multicolumn{1}{p{2.5cm}}{}&
        \multicolumn{1}{p{2.5cm}}{\centering Isolated}
        & \multicolumn{1}{p{5cm}}{\centering $0.52^{+0.09}_{-0.09}$~ ($0.39^{+0.09}_{-0.08}$)}\\

        \multicolumn{1}{p{2.5cm}}{$\rho: 200--315$ pkpc}&
        \multicolumn{1}{p{2.5cm}}{\centering $\pm250$}&
        \multicolumn{1}{p{2.5cm}}{\centering All}
        & \multicolumn{1}{p{5cm}}{\centering $0.60^{+0.08}_{-0.09}$~ ($0.32^{+0.09}_{-0.08}$)}\\

        \multicolumn{1}{p{2.5cm}}{}&
        \multicolumn{1}{p{2.5cm}}{}&
        \multicolumn{1}{p{2.5cm}}{\centering Isolated}
        & \multicolumn{1}{p{5cm}}{\centering $0.47^{+0.12}_{-0.12}$~ ($0.29^{+0.12}_{-0.09}$)}\\

        \hline
	\end{tabular}\\
   Note-- The errors in \fc\ are calculated using the binomial proportion confidence interval (BPCI) Wilson Score method with $68\%$ confidence level.     
   $^{a}$These values are calculated at threshold \logN $= 12.5\; (13.0)$. 
\end{table*}

To determine the \CIV\ covering fraction, we first calculated the $3\sigma$ limiting column density for each LAE in our sample using the SNR of the  corresponding  quasar spectra near the expected \CIV\ wavelengths (preferably at $1550$~\ang) following the procedure described in Section~\ref{sec:N_profile}. Fig.~\ref{fig:fc} shows the covering fraction of \CIV\ as a function of threshold column density, for two different LOS velocity windows around the LAEs. The black (magenta) dots correspond to the measurements within \pms500~\kms (\pms250~\kms) of the LAEs. The \CIV\ covering fraction calculated for the entire impact parameter range of $16-320$~pkpc of our LAEs is close to $60\%$ for a threshold column density of $10^{12.5}$~\sqcm,  but declines to $\approx10$\% for strong \CIV\ absorbers with $N(\CIV)>10^{14.0}$~\sqcm. The cyan squares in Fig.~\ref{fig:fc} show the covering fraction estimated using the bound absorbers for a velocity window of $\pm250$~\kms. The \CIV\ covering fraction for the bound absorbers alone is as high as $\approx50$\% for a threshold column density of $10^{12.5}$~\sqcm. We have also estimated the same for different absorber and galaxy sub-samples using different velocity windows in the following sections. All the measurements are summarized in Table~\ref{tab:fc}.

We also estimated \CIV\ covering fraction at random locations using the random sample. In Fig.~\ref{fig:fc}, the colored empty dots are the covering fraction for this random sample calculated within the LOS velocity ranges mentioned in legends. The solid lines are the best-fit straight lines while the shaded regions represent the $1\sigma$ uncertainties. It is worth noting that the probability of finding a \CIV\ absorber is noticeably higher (more than twice) in the neighbourhood of the LAEs than in random locations for almost the entire range of column density probed here.


\subsubsection{Redshift evolution of \CIV\ \fc} 
\label{sec:f_redshift}
\begin{figure}
    \includegraphics[width=0.48\textwidth]{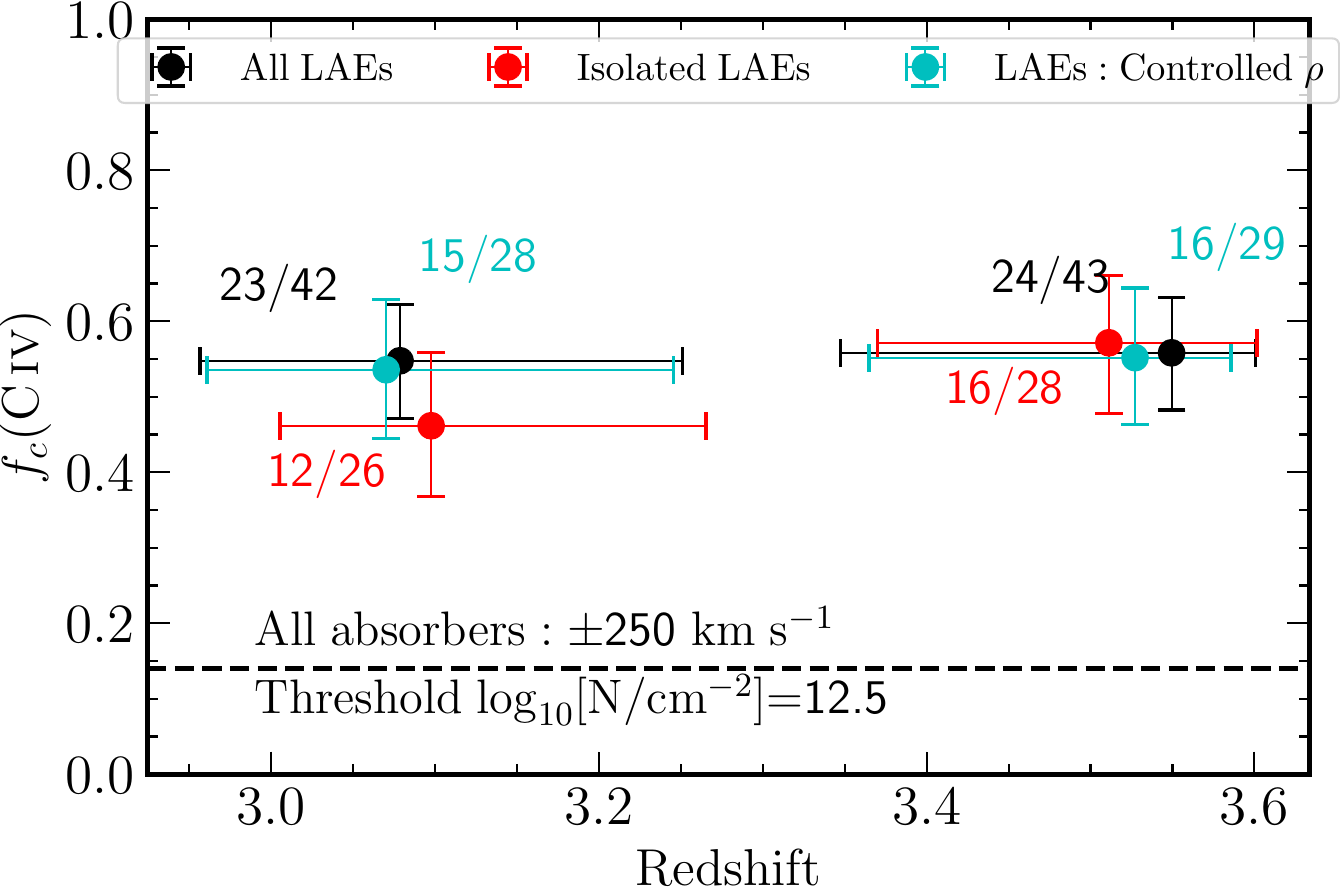}
    \caption{The redshift dependence of \CIV\ \fc estimated using the absorbers within \pms250~\kms of all LAEs (black), only for the isolated LAEs (red) and for the impact parameter controlled LAE sample (cyan). We have divided the LAE populations into two redshift bins based on their median redshifts. The calculations are done at a threshold \logN = 12.5. The x-position and error bar reflects the median value and the 16-84 percentile range. In all the cases, the values of $f_{\rm c}$ are consistent within 1-$\sigma$, indicating no significant redshift evolution within our sample. The black dashed line indicates the \fc for a random sample at this threshold \colm.} 
    \label{fig:f_redshift}
\end{figure}

In this section, we will examine whether the \CIV\ \fc\ evolves with redshift. The cosmic time span of our sample is $\Delta t = 0.6$ Gyr for $z = 2.92$ to $3.82$ and $\Delta z$/(1+$z_{\rm median}$) is small ($\approx0.2$), hence, we do not expect a strong redshift evolution. Fig.~\ref{fig:f_redshift} shows the  \CIV\ \fc\ for the \CIV\ absorber-sample within \pms250~\kms\ of the LAEs for two redshift bins partitioned by the median redshift ($z\approx3.3$). Here, we have chosen a threshold \colm\ of \logN = 12.5.

To eliminate any possible environmental  effects, we further calculated the \fc\ for only isolated LAEs, shown by the red points in Fig.~\ref{fig:f_redshift}. Similarly, the impact parameter controlled $f_c$ measurements shown in cyan eliminate any possible effects due to the difference in the $\rho$ distributions. For controlling the impact parameter, first, we split the LAEs into two redshift bins based on their median redshift, and then we chose the LAEs in such a way that for each LAE in the lower bin, there was a LAE in the upper bin with an impact parameter consistent within $\pm10$ pkpc. In all three cases (i.e., full sample, isolated sample, and impact-parameter controlled sample), the covering fractions for the high-$z$ and low-$z$ bins are consistent within the $1\sigma$ allowed uncertainty, indicating no redshift evolution. Finally, no redshift evolution is seen even when we included all the associated components detected within $\pm500$~\kms of an LAE redshift.

\subsubsection{Environmental dependence of \CIV\ covering fraction}
\label{sec:f_env}
 
\begin{figure*}
	\includegraphics[width=0.7\textwidth]{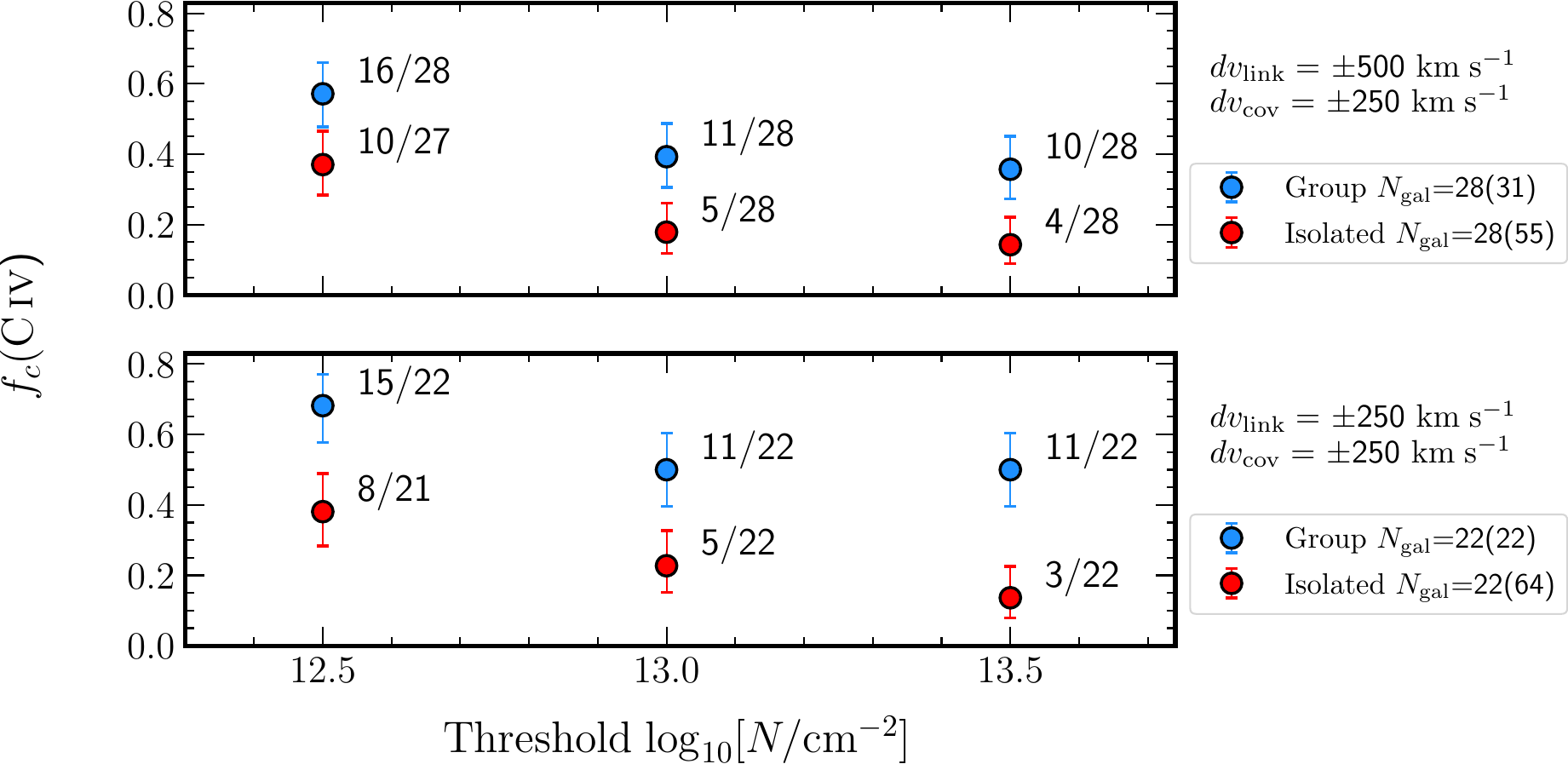}
    \caption{{\em Top:} Same as Fig~\ref{fig:fc} but for the ``isolated'' and ``pair/group'' LAEs separately. We defined a galaxy-pair/group in such a way that each member of a pair/group has at least one companion within $\pm500$~\kms\ ($dv_{\rm link}$) within the MUSE FoV. The impact parameters of the two subsamples are controlled so that for each threshold column density, the $\rho$ distributions are indistinguishable. Here we used all the \CIV\ absorbers within the LOS velocity window of \pms250~\kms ($dv_{\rm cov}$). The numbers next to the figure indicate the total number of pair/group or isolated LAEs in the controlled sample; the total number of pair/group LAEs is provided in parenthesis. ``Pair/group'' LAEs exhibit consistently higher covering fractions than the ``isolated'' ones.  {\em Bottom:} Similar to the {\tt top}, but for a $dv_{\rm link}$ of $\pm250$~\kms\ instead of  $\pm500$~\kms\ for defining a pair/group. The environmental dependence appears to be stronger in this case.}
    \label{fig:f_env}
\end{figure*}

In this section, we will investigate the effect of the galactic environment on the covering fraction of metal-rich gas traced by \CIV. The top panel of Fig.~\ref{fig:f_env} shows the \CIV\ covering fraction for the pair/group and isolated LAEs for absorbers within \pms250~\kms of the LAEs. Here we controlled for the impact parameter of the two subsamples in such a way that for each LAE in the isolated subsample, there is a galaxy in the pair/group subsample with an impact parameter consistent within $\pm10$~pkpc. In the previous section, we already found no significant redshift evolution of the \CIV\ \fc. Additionally, it is unlikely that the SFR will influence the \fc\ as it covers a small dynamical range in our sample. Note that, for the threshold \colm\, \logN = 12.5, the covering fraction for pair/group is $0.57^{+0.09}_{-0.09}$ whereas that of isolated LAEs is $0.37^{+0.09}_{-0.08}$. \Rev{Moreover, we note that for the higher threshold of \logN = 13.0 (13.5), the \CIV\ covering fraction for the pair/group galaxies is $0.39^{+0.09}_{-0.09}$ ($0.36^{+0.09}_{-0.08}$), which is almost twice of isolated galaxies, $0.17^{+0.08}_{-0.06}$ ($0.14^{+0.08}_{-0.05}$).}


In the top panel, we defined a pair/group in such a way that each member of a pair/group has at least one companion within a linking velocity, $dv_{\rm link}$ of $\pm500$~\kms. If we use a $dv_{\rm link}$ of  \pms250~\kms\ instead of \pms500~\kms, the difference of \CIV\ covering fractions between pair/group and isolated LAEs further increases for all the threshold column densities. This is shown in the bottom panel of Fig.~\ref{fig:f_env}. Here again, we considered the \CIV\ components within $\pm250$~\kms of the LAEs (see Table~\ref{tab:fc}). The enhancement of \fc\ is probably due to the fact that employing a wider velocity range for the group definition actually includes some isolated galaxies, which diminishes the environmental effect.

We investigated the environmental dependence of covering fraction using the median redshift of the group members to calculate the covering fraction for the pair/group galaxies. However, because of the small number of pairs/groups (10 for $dv_{\rm link}=\pm500$~\kms and 9 for $dv_{\rm link}=\pm250$~\kms), the $f_c$ values of the isolated and pair/group LAEs are not significantly different.


\subsubsection{The \CIV\ covering fraction profile} 
\label{Rvir}

\begin{figure}
    \centering
\includegraphics[width=0.48\textwidth]{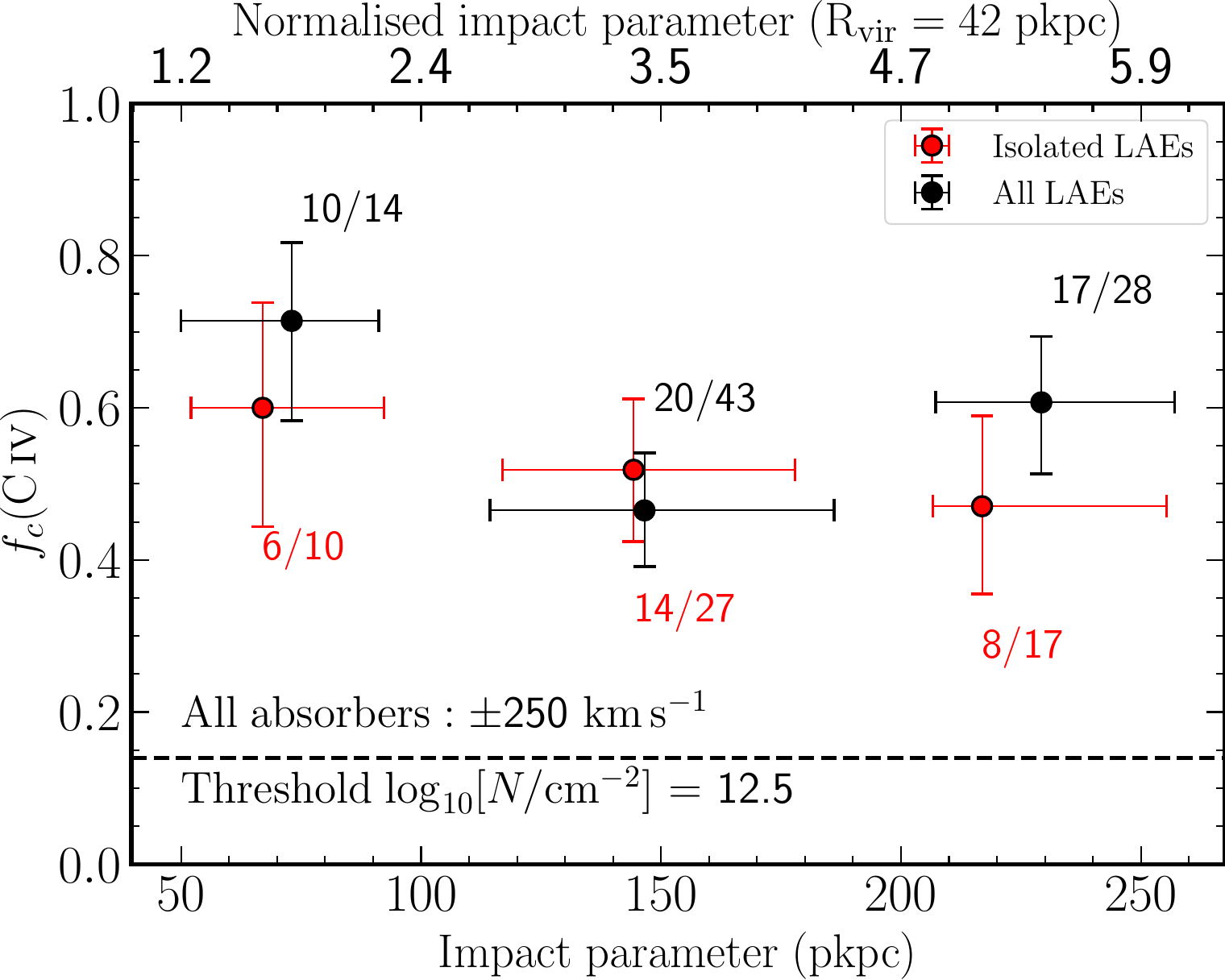}
    \caption{Covering fraction of \CIV\ absorbers for the full LAE sample (in black) and for the isolated LAEs (in red) as a function of impact parameter, $\rho$. We split the sample into three impact parameter bins: 0-100 pkpc, 100-200 pkpc, and 200-320 pkpc. The top x-axis shows the normalized impact parameter where the normalization is done using the median $R_{\rm vir}= 42$~pkpc. The calculations are performed at a threshold \logN $=12.5$ and using a LOS velocity window $\pm250$~\kms. The x-position of the dots reflects the median value in each impact parameter bin, while the error bar along the x-axis represents the $16-84$~percentile range. The black dashed line is the \fc for the random sample at this particular threshold \colm.} 
    \label{fig:f_impact}
\end{figure}

Our sample consists of galaxies with impact parameters ranging from 16.1 to 315.2 pkpc. Here we investigate the variation of the \CIV\ \fc\ with the impact parameter. The black points in Fig.~\ref{fig:f_impact} show the covering fractions calculated in three bins of the impact parameter (0--100 pkpc, 100--200 pkpc, and 200--320 pkpc) for a threshold \colm\ of \logN = 12.5 and a LOS velocity window of \pms250~\kms. The $f_c$ for the innermost 
 impact parameter bin is $\approx 71^{+0.10}_{-0.13}$\%. It declines to $\approx46^{+0.07}_{-0.07}$\% for the intermediate impact parameter bin but rises again for the outermost bin to $\approx60^{+0.08}_{-0.09}$\%. Such an upturn, however, is not seen for the isolated galaxies (red points in the plot). This suggests that the upturn is indeed owing to the underlying environmental dependence.

The impact parameter normalized to the median virial radius of $\approx42$~pkpc is shown along the top x-axis. The virial radius, $R_{\rm vir}$ or $\rm R_{200}$, for an LAE is calculated from the halo mass as: 
$$\rm R_{200}^3 = 3 M_{halo}/4 \pi \Delta_{vir} \rho_{c},$$ where $\rm \Delta_{vir} = 200$ and $\rm \rho_{c}$ is the critical density of the universe at galaxy redshift. We first obtain the stellar masses from the UV SFR assuming that the LAEs lie on the star forming main sequence relation \citep[][]{10.1093/mnras/stz1182}. The halo mass is then estimated from the halo abundance matching relation of \citet{moster2013galactic}. Note that we could only estimate the virial radius for the 35 LAEs with SFR measurements. We would like to emphasize that even for the farthest bin, with $\rho > 200$~pkpc ($\geq 5R_{\rm vir}$), the \fc of \CIV\ around the isolated LAEs is significantly high {\bf ($\approx 50\%$). }

In passing, we note that a similar analysis with the bound absorbers alone leads to similar conclusions. Finally, we did not find any significant dependence of the \CIV\ \fc\ on the SFR, \lya\ line luminosity ($L$(\lya)), or rest-frame equivalent width (\ew) within the dynamical ranges probed by our survey.

\section{DISCUSSION} 
\label{sec4}

\Rev{We studied the connection between low mass, star-forming galaxies and \CIV\ absorbers at $z\approx 3.3$ using a sample of 86 \lya\ emitting galaxies (LAEs). The Ly$\alpha$ emission peak is known to be redshifted by a few hundred of \kms\ with respect to the systemic redshift of a galaxy. \citet{Verhamme+18} provided an empirical relation to estimate the systemic redshifts of LAEs based on non-resonant rest-frame optical lines. Although, the relation is shown to over-correct LAE redshifts for $L$(\lya) $>10^{41.3}\; \rm erg\,s^{-1}$ \citep[see][]{Matthee+23}. Here we used the empirical correction obtained by \citet{Muzahid_2020} for the same LAE sample we studied here. However, we verified that we obtain consistent results if we instead use the relation given by \citet{Verhamme+18}. Here we discuss our main findings from this work.}

\subsection{Connecting the LAEs and \texorpdfstring{\CIV\ }\ absorbers}

We cross-matched the blind \CIV\ catalog with the LAE catalog and found an enhanced number of absorption components within $\pm 400$~\kms of the LAE redshifts, indicating a significant clustering between metal-enriched gas and low-mass, star-forming galaxies at $z\approx3.3$. A similar trend is seen for $z\approx2.3$ LBGs and \HI\ absorbers \citep{Rudie_2012}. We further found that the higher column density \CIV\ absorbers ($N>10^{13}$~\sqcm) are more tightly correlated with the galaxies in velocity. Using the high resolution {\sc eagle} cosmological simulation, \citet{ho2021identifying} recently showed that for low-$z$ \OVI\ gas selected via a velocity window of $\pm500$~\kms (or $\pm300$~\kms) produces higher \OVI\ column density than calculated within a fixed radius ($\approx$ 2-3 \Rvir) due to projection effects. The \OVI\ covering fraction beyond $1$~\Rvir from $M_* \sim 10^9$~\Msun\ galaxies is dominated by absorbers located at 3D distances of more than 2~\Rvir.
However, in order to be consistent with the previous studies in the literature we used a $\pm500$~\kms\ velocity window for assigning absorption to a given galaxy \citep[see e.g.,][]{Galbiati_2023}.

We found that 202 out of the 489 `blind' \CIV\ components lie within $\pm500$~\kms\ of the LAE redshifts. The column density and Doppler parameter distributions of these 202 components are not significantly different from the remaining 287 components that are not associated with any LAEs ($p$-value of KS test being $>0.05$). This could suggest two possibilities. First, the components with velocities $>500$~\kms\ from any LAEs are likely associated with galaxies that are fainter in \lya\ emission compared to the ones in our sample. Note that a fraction of LBGs does not show prominent \lya\ emission \citep[]{Kusakabe2020}. However, from the LBG luminosity function presented in \cite{steidel1999} we found that only $\approx2$ LBGs, brighter than $I$-band magnitude of 25, are expected to be present for the co-moving volume probed by our survey. On the other hand, those  components can be related to low mass and/or low SFR galaxies that are below our detection threshold. Of course, a fraction of the components can be truly intergalactic in nature, though the metals must have been produced by stars inside galaxies.

The other possibility could be that not all the \CIV\ components within $\pm500$~\kms\ of the LAE redshifts are physically related to the LAEs themselves. To investigate this we plotted the LOS velocity of all the \CIV\ components within $\pm500$~\kms\ of the LAE redshifts as a function of impact parameter and compared the LOS velocities to the escape velocity corresponding to the typical range in galaxy mass for our sample (see Fig.~\ref{fig:escape_velo}). Since the LOS velocity and impact parameter represent only lower limit on the actual 3D velocity and 3D distance from the galaxy, it is not straightforward to determine which components are actually bound to the galaxy from such a plot. However, the components outside the shaded regions in Fig.~\ref{fig:escape_velo} are undoubtedly unbound to the galaxies. We found that $45\%$ of the \CIV\ components we associated with the LAEs (i.e. 91 out of 202) are not really bound to the LAEs. Earlier \citet{Rudie_2019} also found that 5 out of 7 LBGs in their sample had some \CIV\ components with LOS velocities greater than the corresponding escape velocity. This is in stark contrast to the studies at low-$z$ relatively high-mass galaxies, where the majority of the CGM absorbers are bound to the host halos \citep[e.g.,][]{ Tumlinson_2011, bordoloi2014cos, Werk2016}. However, there are certain instances of metal-rich gas with $v > v_{\rm esc}$ that have been identified using highly ionized metal lines such as \OVI\ \citep[i.e.,][]{Tripp2011, Muzahid2015, Rosenwasser2018}.

We found that the median \colm\ of the bound (unbound) components is \logN~$= 12.8~(12.7)$ and the median \dop\ is $b = 11.8~(12.0)$~\kms. The column density and Doppler parameter distributions of the bound and unbound components do not seem to differ statistically, as suggested by KS-test results. In this context, recall that some fraction of the unbound components might not be related to the galaxies in question, but are associated with undetected lower-luminosity clustered galaxies. Fig.~\ref{fig:abs_distribution_appendix} ({\tt Right}) shows that the majority of the bound \CIV\ components are within $\pm250$~\kms\ of the LAE redshifts. The centroid of the best-fitting Gaussian ($-35.7\pm33.3$~\kms) is consistent with the systemic redshift within $1.1\sigma$ allowed uncertainty.   

\Rev{Although the velocity distribution of the \CIV\ 
 components is centered on the systemic velocity within $\approx1\sigma$, we noticed that the number of \CIV\ components is somewhat higher at $\Delta v \gtrsim 300$~\kms\ compared to $\Delta v \lesssim -300$~\kms\ (see Fig.~\ref{fig:abs_distribution} and Fig.~\ref{fig:abs_distribution_appendix}). In line with this, \citet{Momose_21} reported an anisotropic cross-correlation function (CCF) between LAEs and \lya\ forest transmission fluctuations (at $z \approx 2$) with the near side ($-ve$ velocities) CCF showing lower signals up to $r = 3-4 h^{-1}$ cMpc ($\sim 400$~\kms\ at $z=2.0$) indicating that the average \HI\ density on the near side of LAEs is lower than that on the far side. According to \citet{Momose_21}, this is due to the observational bias in which LAEs on the far side of a dense region are more difficult to detect. If \CIV\ absorbers follow a similar pattern then the asymmetry in our plot at $|\Delta v| > 300$~\kms\ could be attributed to the findings of \citet{Momose_21}. However, we note that they excluded the LOS  separation of $1 h^{-1}$~cMpc ($\approx 100$~\kms\ at $z=2.0$) from each galaxy to eliminate the influence of \HI\ in the CGM, and we do not find any significant asymmetry in the \CIV\ velocity distribution within the CGM-scale. If at all, we see a tentative negative offset, which could result from a fraction of LAEs having larger offset from the systemic redshift than the mean relation given by \cite{Muzahid_2020}. We found that this offset, however, disappears when we limit the distribution to only the bound \CIV\ absorbers
 (see Fig.~\ref{fig:abs_distribution_appendix}: {\tt Right}).}

\begin{figure}
    \includegraphics[width=0.48\textwidth]{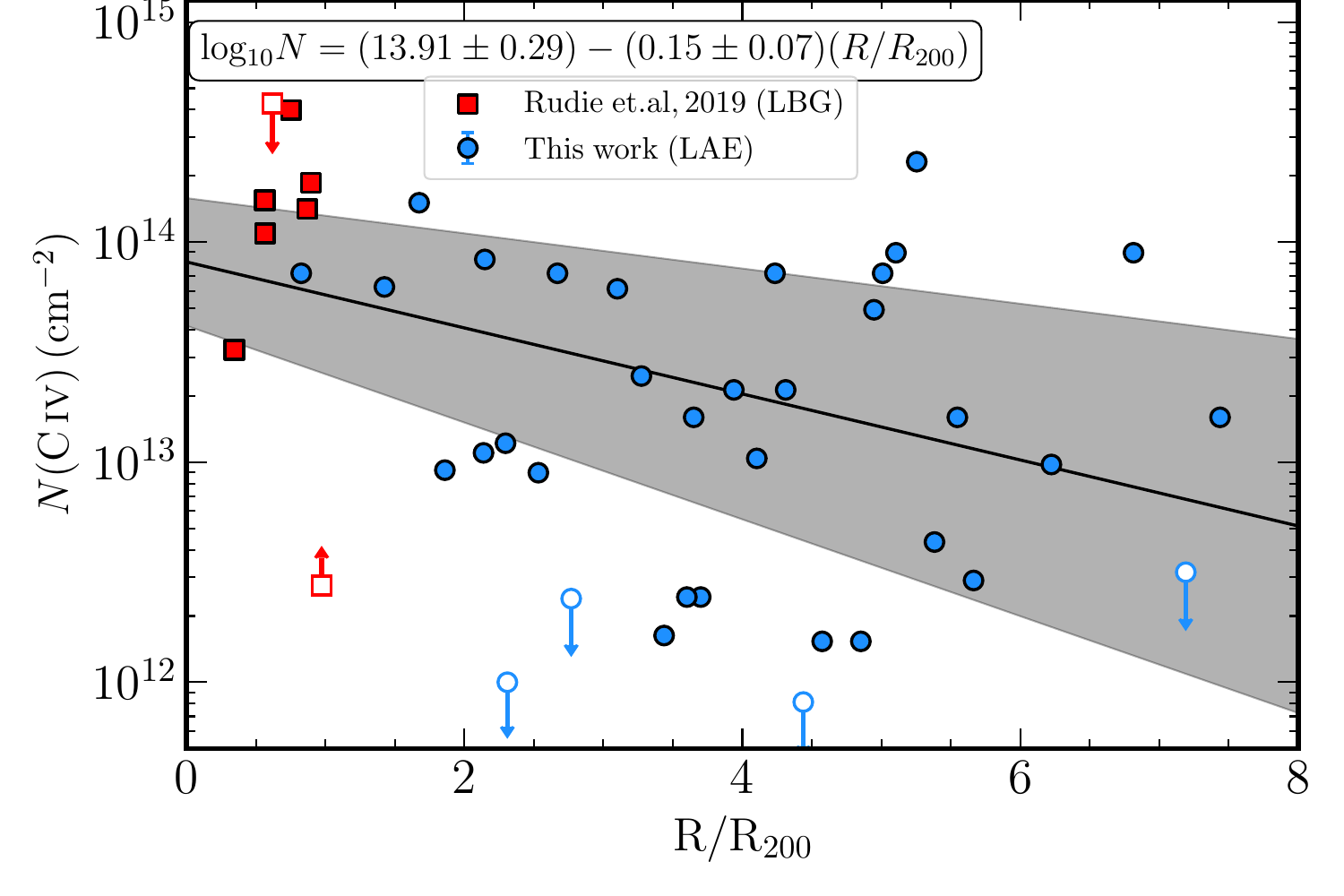}
    \caption{\CIV\ \colm\ profile compiled from the available literature. The filled squares (dots) represent \CIV\ \colm\ detections from \citet{Rudie_2019} (this work), whereas the down-arrows (up-arrow) are the $3\sigma$ upper limits (lower limit) of the respective sample. The black solid line shows the model equation written on top of the figure. The best-fit parameters are estimated from the {\sc cenken} function of {\sc NADA} package ({\sc R}) using only the high-$z$ data points. The shaded region corresponds to the $1\sigma$ uncertainty of the fitted curve derived from the bootstrapped errors on the parameters.}
    \label{fig:discussion_rudie_rvir}
\end{figure}

\subsection{The column density profile} 

We constructed the \CIV\ \colm\ profile (see Fig.~\ref{fig:n_profile}) for  the associated components (detected within \pms250~\kms\ of an LAE). We found that the $N$(\CIV) values associated with the LAEs are almost an order of magnitude higher compared to the mean $N$(\CIV) estimated for the random sample. This indeed suggests that the associated components trace overdense regions with significant metal enrichment, likely the large-scale structures in which LAEs are embedded. However, we did not detect any statistically significant (anti) correlation between the \CIV\ \colm\ and impact parameter, at least not out to $\approx250$~pkpc. This is also true when only the bound absorbers are considered. This is consistent with \citet{Muzahid_2021}, but in contrast with some low-$z$ surveys that reported an anti-correlation between the \CIV\ column density and impact parameter \citep[e.g.,][]{bordoloi2014cos, Liang_Chen2014}. \cite{Turner_2014} found a significant drop in the normalized \CIV\ optical depth within \Rvir, followed by a flat but still enhanced region beyond \Rvir\ for LBGs. Because our LAEs are of lower mass (consequently lower \Rvir) we may be probing the flat part of the radial profile. This warrants a larger sample size with impact parameters of $< R_{\rm vir}$. A wider FoV of the future BlueMUSE \citep[see][]{Richard_2019}, on the other hand, can probe the large projected distances where the absorption signal asymptotically matches the IGM.

Similar to our finding, \citet{Rudie_2019} also did not notice any significant trend between the total \CIV\ column density, calculated within $\pm1000$~\kms\ velocity window, and impact parameter ($<R_{\rm vir}$) likely because of a smaller sample size of 8 $z\approx 2.3$ LBGs. The red squares in Fig.~\ref{fig:discussion_rudie_rvir} represent their data points. The blue circles indicate the $N$(\CIV) measurements within $\pm1000$~\kms\ for the 35 MUSEQuBES LAEs for which we could estimate \Rvir. Performing a Kendall-$\tau$ correlation test on this combined high-$z$ sample (LBGs+LAEs), we found an anti-correlation between the total \colm\ and impact parameter with $\tau_K = -0.2$ and $p$-value $=0.03$, with a best-fit power-law relation: $N(\CIV) = 10^{13.91\pm0.29}~{\rm cm^{-2}} \times (R/R_{\rm vir})^{-0.15\pm0.07}$.\footnote{We used the {\sc cenken} function of {\sc NADA} package in {\sc R} which takes care of the upper limits. We excluded the only lower limit in the LBG sample from this exercise.} The gray-shaded region corresponds to the $1\sigma$ uncertainty of this fitted curve derived using 
 1000 bootstrap realizations.

\begin{figure}
	\includegraphics[width=0.48\textwidth]{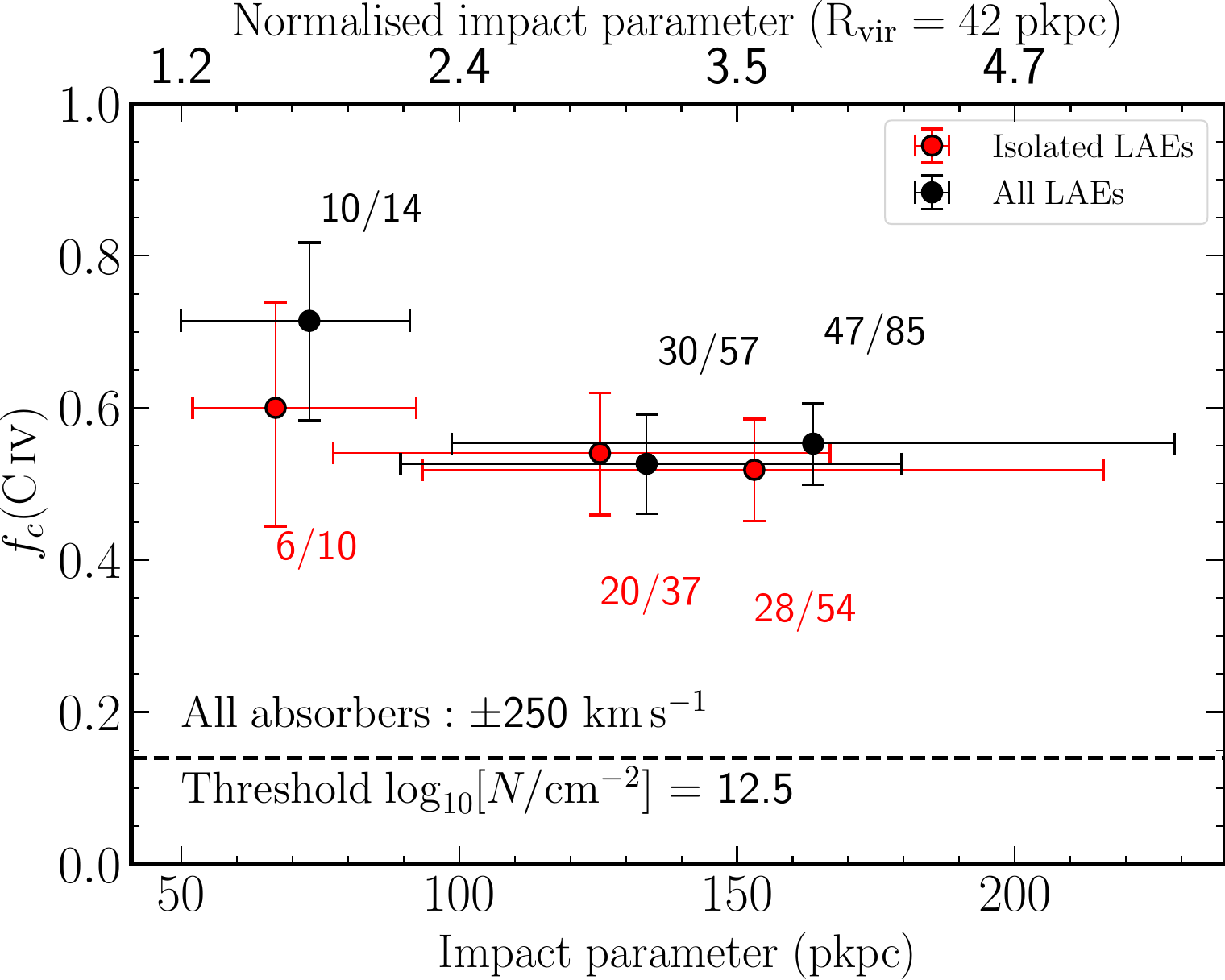}
    \caption{ Cumulative \fc of all \CIV\ absorbers within $\pm250$~\kms\ of the LAEs as a function of impact parameter. The bin boundaries are chosen as $100$, $200$, and $320$ pkpc. The rest is the same as in Fig.~\ref{fig:f_impact}. With the impact parameter, the \fc around isolated LAEs reaches a value of $\approx54\%$.}
    \label{fig:f_impact_cumulative}
\end{figure}

\subsection{The covering fraction}  

Fig.~\ref{fig:fc} shows the measured \CIV\ covering fractions for different threshold column densities. For a threshold column density of $10^{12.5}$~\sqcm, the \CIV\ covering fraction of our sample is $\approx 64$\%, measured within a velocity window of $\pm500$~\kms. These $f_c$ values are almost two times higher as compared to random regions. We did not find any significant dependence of $f_c$ on either redshift or \lya\ line luminosity in our sample. Similar to the column density profile, the covering fraction profile also does not show any appreciable anti-correlation with the impact parameter (see  Fig.~\ref{fig:f_impact}). This is also evident from Fig.~\ref{fig:f_impact_cumulative} where we show the cumulative covering fraction with impact parameter. The cumulative covering fraction remains constant for the impact parameter of $>100$~kpc. A similar trend is also seen in the recent study by  \cite{Galbiati_2023}, where the \CIV\ covering fraction appears to reach a plateau at $\sim 100-250$~pkpc with a value very similar to ours (see their figure~10). The lack of any significant trend between \CIV\ covering fraction and impact parameter in our sample owes to the fact that the largest impact parameter bin shows an equally high $f_c$ value, particularly for the full sample, leading to an upturn in the profile. This upturn is less pronounced when we only consider the isolated LAEs. This suggests that the upturn is likely due to the environmental effects and that the covering fraction profile is flatter for the pair/group galaxies compared to the isolated ones.

We found a strong environmental dependence for $f_c(\CIV)$ (Fig.~\ref{fig:f_env}). Pair/group galaxies exhibit a higher \CIV\ covering fraction compared to isolated galaxies. The difference in $f_c$ increases when we consider a narrower linking velocity ($\pm250$~\kms) for defining pairs/groups. For example, at the threshold \logN~$= 12.5\, (13.5)$, $f_{\rm c}({\mbox{C\,{\sc iv}}})$ is  $\approx68\%$ ($50\%$) for pair/group, whereas for isolated LAEs it is only $\approx38\%$ ($23\%$). \citet{Muzahid_2021} reported a similar environmental dependence in which the pair/group LAEs showed a considerably stronger \CIV\ absorption in the composite spectrum of background quasars. Recently, \cite{Galbiati_2023} also showed that the covering fraction for their group LAEs was almost three times stronger than that for the isolated ones. They calculated a \CIV\ covering fraction of $\approx 60\%$ within $\sim 200$~pkpc their group galaxies and $\approx 20\%$ around isolated LAEs. Our finding is consistent with both results. The enhanced \CIV\ covering fraction around pair/group galaxies suggests that (a) they live in significantly more metal-rich environments and/or (b) their surroundings ensure favorable ionization conditions for the triply ionized carbon and/or (c) the gas density or the number density of the clouds are higher for them. To understand this better, in future works we will explore this environmental dependence for other low- and intermediate-ionization state lines such as \SiII, \SiIII, and \SiIV. The highly ionized \OVI\ lines are unfortunately too severely blended with the LAF.

\begin{figure}
	\includegraphics[width=0.48\textwidth]{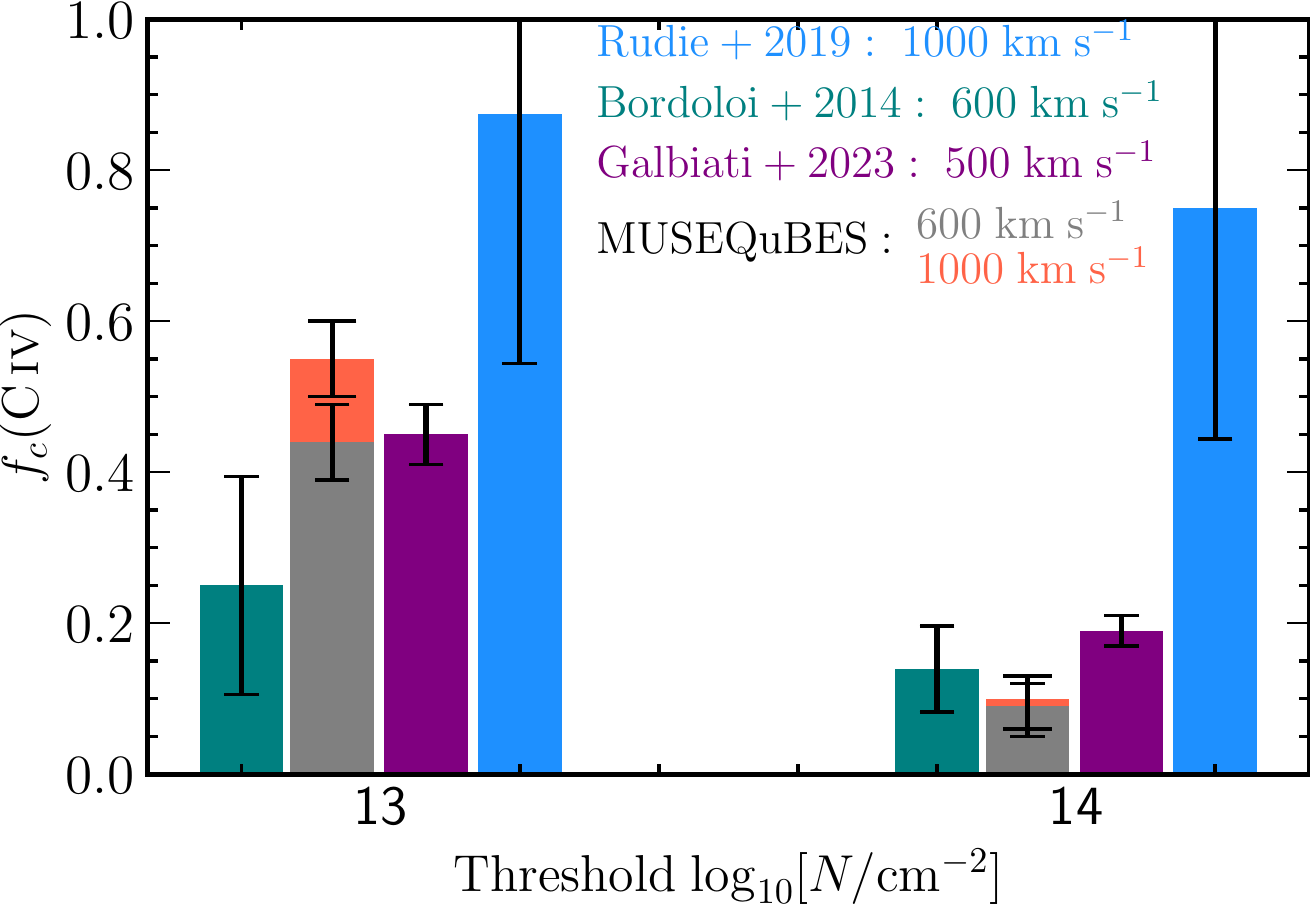}
    \caption{A comparison between the \CIV\ \fc of high-$z$ ($35 < \rho/{\rm pkpc} < 100$) LBGs from \citet[blue bar]{Rudie_2019}, low-$z$ ($14 < \rho/{\rm pkpc} < 135$) dwarf galaxies from \citet[teal bar]{bordoloi2014cos}, and $z\approx3.3$ LAEs \citet[purple bar]{Galbiati_2023} with our high-$z$ MUSEQuBES sample ($16 < \rho/{\rm pkpc} < 315$) at two different threshold column densities. The impact parameter range for \citet{Galbiati_2023} is similar to us. The legend includes the velocity window that was used in respective surveys to calculate $f_c$. To compare these findings with our results, we also estimated the $f_c$ within $\pm1000$ (orange bar) and $\pm600$~\kms (grey bar) of MUSEQuBES LAEs. While $f_c$ is calculated within the virial radii of the LBGs and low-$z$ dwarfs, the median impact parameter for both \citet{Galbiati_2023} and our sample is several times the virial radius.}
    \label{fig:discussion_rudie_f}
\end{figure}

\begin{figure*}
\centerline{\hbox{
    \hspace{4.5cm} 
    \includegraphics[width=0.60\textwidth]{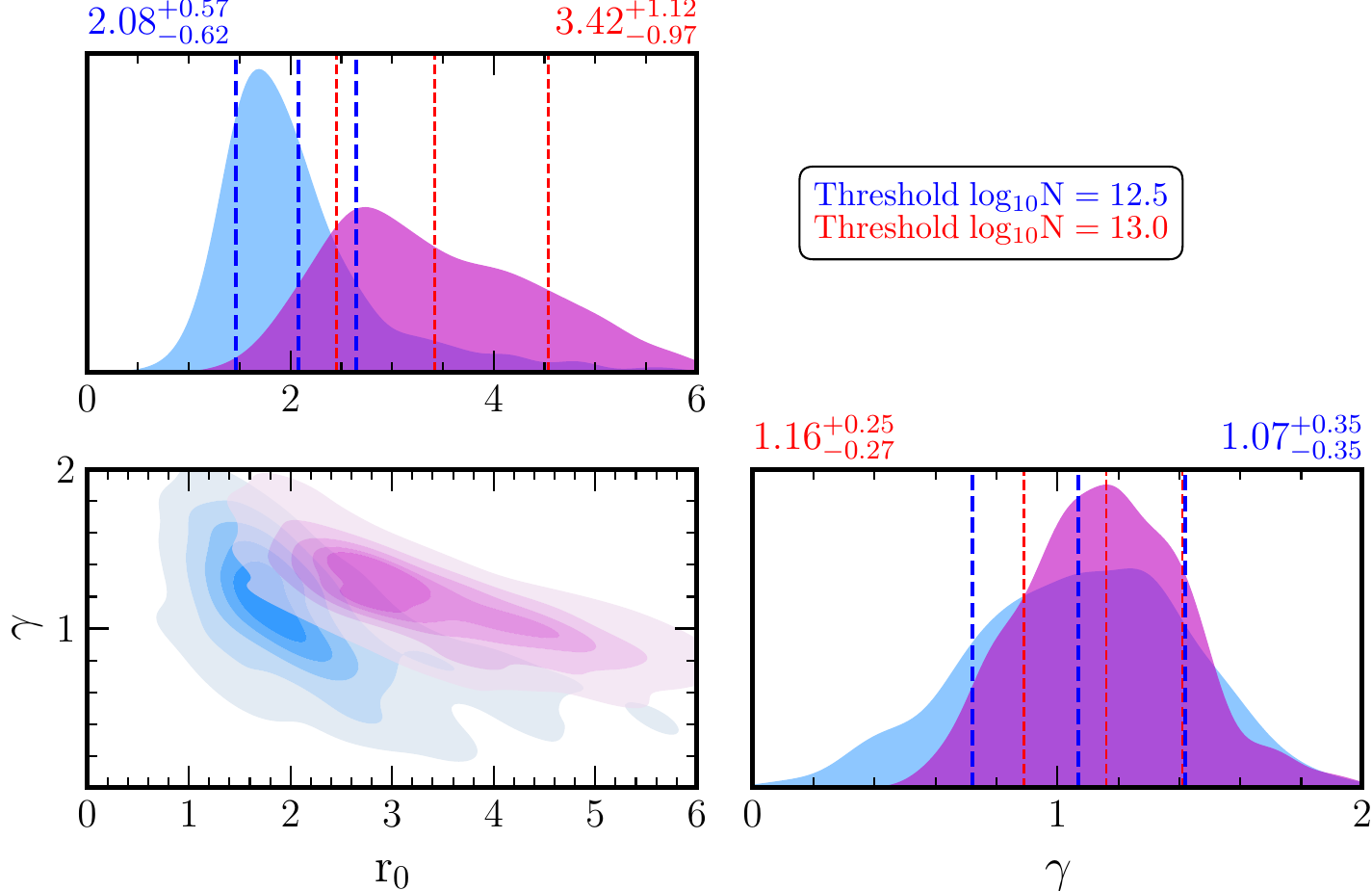}
    \hspace{-5cm} 
\centerline{\vbox{ 
    \includegraphics[width=0.40\linewidth]{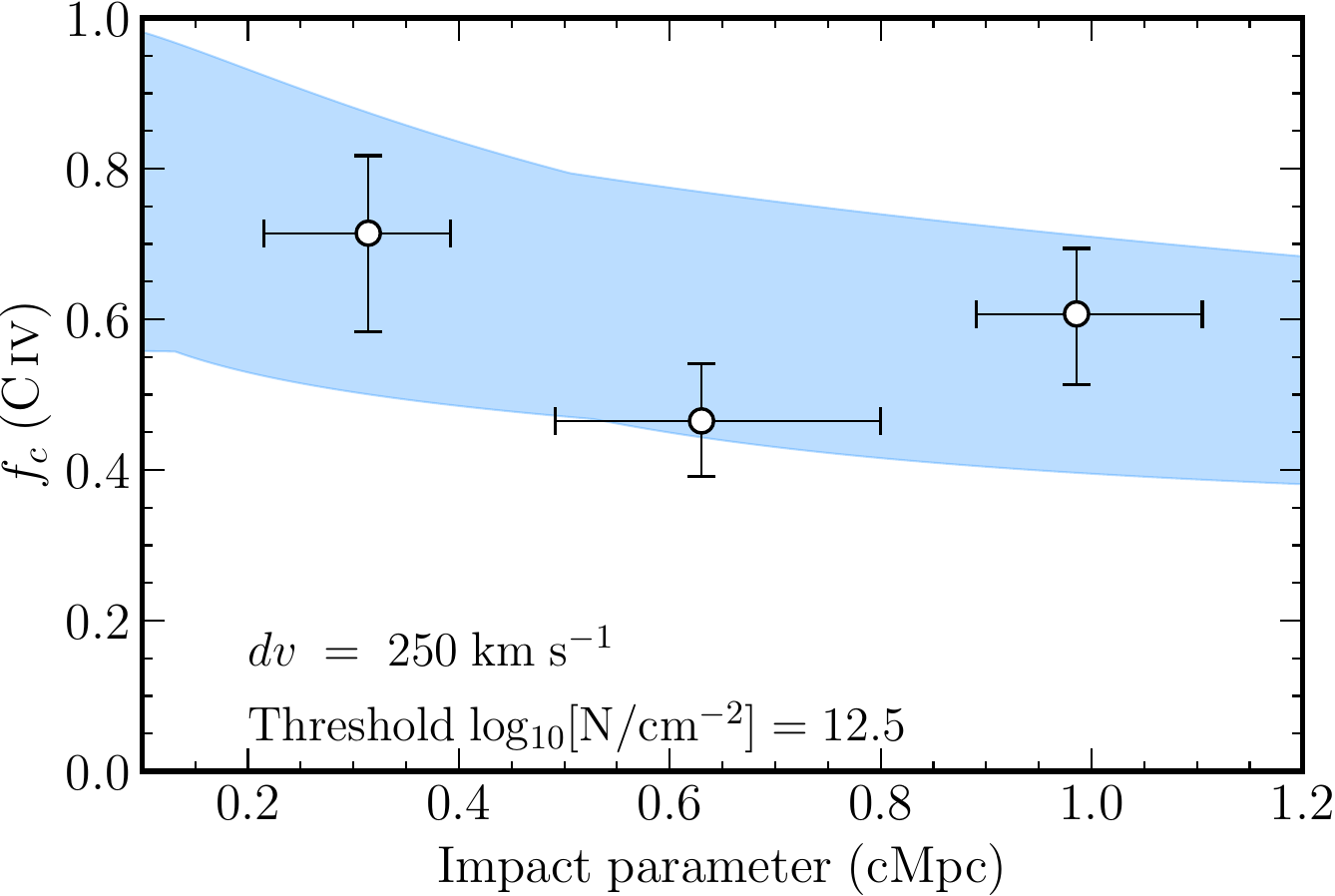}\\
    \includegraphics[width=0.40\linewidth]{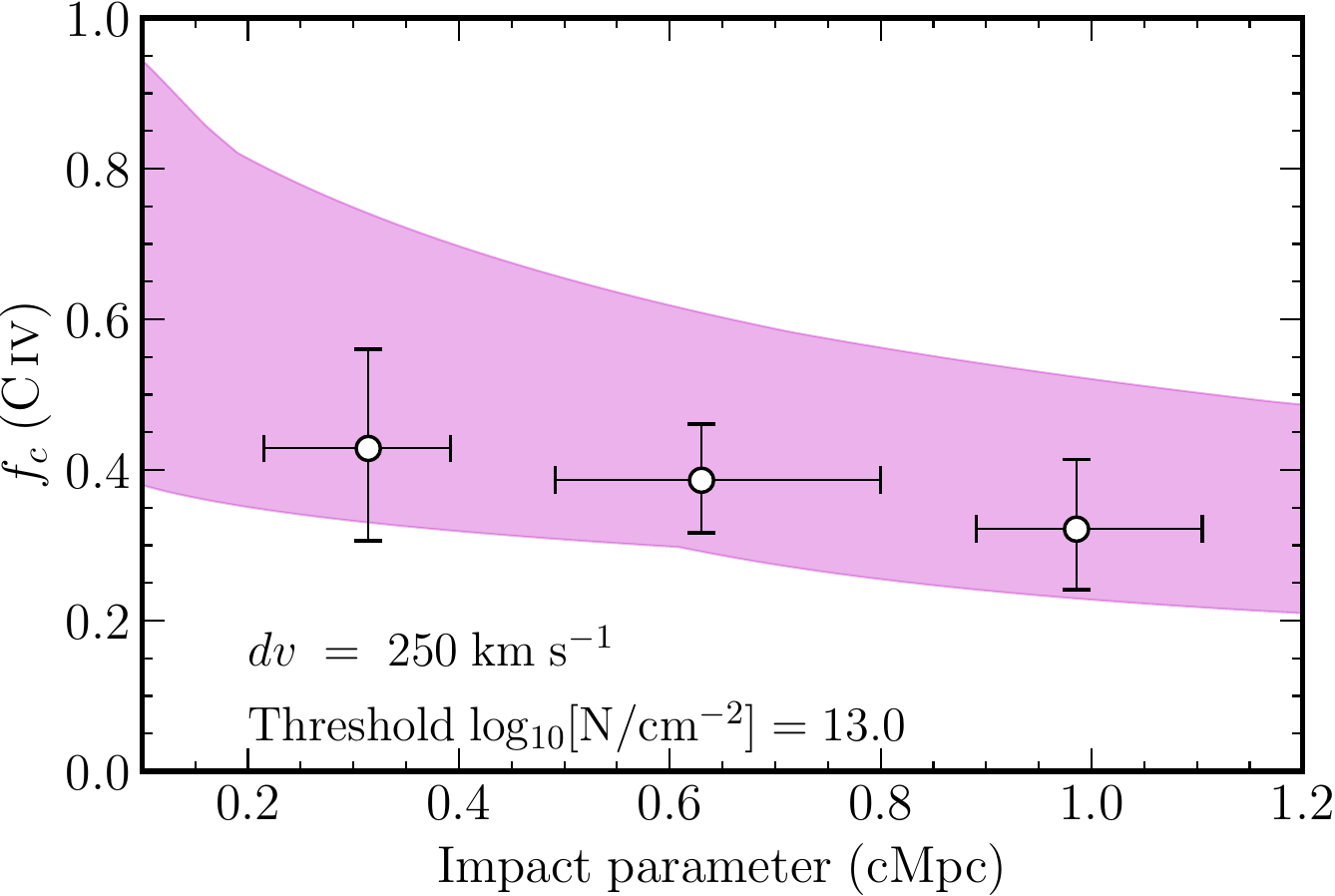}
    }}}} 
    \caption{{\tt Left:} The corner plot showing the posterior  distribution of parameters $r_0$ and $\gamma$ of the galaxy-absorber cross-correlation function ($\xi_{\rm ga}$) along with their 2D density plot for two different threshold \CIV\ column densities as indicated by the legend. Here, only the \CIV\ absorbers within $\pm250$~\kms\ of the LAEs are used. Vertical dashed lines indicate the 16--50--84 percentile range of the respective distributions. These values are also mentioned at the top of the density plots. Though the $\gamma$ values are very similar, it should be noted that the $r_0$ value is larger for higher threshold \colm. For threshold \colm\ \logN = 12.5, $r_0 =2.1^{+0.6}_{-0.6}$ cMpc and for \logN = 13.0, it is $r_0 =3.4^{+1.1}_{-1.0}$ cMpc.
    {\tt Right-top:} Covering fraction of \CIV\ as a function of impact parameter for a threshold \colm\ of \logN = 12.5 and LOS velocity window = $\pm250$~\kms (binning and error bars are same as in Fig.~\ref{fig:f_impact}). The (blue) band showing the covering fraction profile corresponding to the $r_0$ and $\gamma$ values drawn from the $99\%$ confidence level of their respective posterior sample space. {\tt Right-bottom:} Same as the {\tt top} but for a threshold column density of \logN = $13.0$.}  
\label{fig:correlation}
\end{figure*}

In Fig.~\ref{fig:discussion_rudie_f} we compared our $f_c$ measurements with that of \citet{Rudie_2019}, \citet{Galbiati_2023} and \citet{bordoloi2014cos}. The sample of \citet{Rudie_2019} contains only a handful of $z\approx2.3$ LBGs within $35 < \rho/{\rm pkpc} < 100$, whereas the sample of \citet{bordoloi2014cos} contains about 40 $z<0.1$ dwarf galaxies within $14 < \rho/{\rm pkpc} < 135$. Both the redshift ($3<z<4.5$) and impact parameter range covered by \citet{Galbiati_2023} is very similar to us. They provided \CIV\ covering fractions at different threshold equivalent widths, but for the sake of comparison, we converted them to column densities using the linear part of the COG relation. To be consistent with the velocity windows used to calculate the covering fractions in these studies, we recalculated the covering fractions for our MUSEQuBES galaxies\footnote{The $f_c$ values for MUSEQuBES sample at $dv = \pm500$~\kms\ and $\rm600$~\kms\ are identical.}. It is evident from the figure that high column density \CIV\ absorbers ($N(\CIV)>10^{14}$~\sqcm) are more frequently detected around the $z\approx2.3$ LBGs compared to the $z\approx3-4$ LAEs and $z<0.1$ dwarf galaxies. The difference in \CIV\ covering fraction around LBGs and LAEs are not significant for a lower threshold column density of $10^{13}$~\sqcm. We point out here that the quasar sightlines are always passing within the virial radii for the LBGs and low-$z$ dwarfs, whereas the median impact parameter of the high-$z$ LAEs is several times the virial radius.  
Finally, for the lower $N(\CIV)$ threshold of $10^{13}$~\sqcm\ our $f_c$ value is fully consistent with \citet{Galbiati_2023}. We confirmed that the marginal difference in $f_c$ values seen for the higher threshold column density is due to the fact that 0.3~\AA\ threshold used in \citet{Galbiati_2023} corresponds to $N(\CIV)$ of $10^{13.87}$~\sqcm\ in the linear part of the COG.

\subsection{Galaxy-absorber 2-point correlation function}

We found that the \CIV\ column densities and covering fractions are significantly enhanced near the LAE redshifts compared to random regions, indicating a strong spatial correlation between the LAEs and \CIV\ absorbers. In this section we will put constraints on the LAE and \CIV\ absorber cross-correlation function using the covering fraction profile following \citet{wilde2021} \citep[see also][]{hennawi2007}. The absorber-galaxy clustering can be defined as the excess number of absorbers around galaxies with respect to random regions, i.e.,  
$$\mbox{\Large$\chi$} = \rm \frac{N_{\rm obs} - N_{\rm rand}}{N_{\rm rand}},$$ 
where, ${\rm N_{\rm rand}}$ ($= \left< \frac{d\mathcal{N}}{dz} \right> \delta z $ ) is the average number of absorbers above a threshold \colm\ within a velocity window of $\delta z =(1+z)\frac{2\delta v}{c}$. ${\rm N_{obs}}$ is the observed number of absorbers. We will adopt the standard parametrization for the galaxy-absorber correlation function,  
\[\xi_{\rm ga}(r) = \left (\frac{r}{r_0} \right)^{-\gamma},\] 
where $r$ ($\equiv \sqrt{r_{\perp}^2 + r_{\parallel}^2}$) is the 3D distance between the galaxy and the absorber. Here, $r_{\parallel}$ is estimated using pure Hubble flow and then converted to comoving scale. This $\xi_{\rm ga}$ is related to the galaxy-absorber 3D cross-correlation function:  
\[\mbox{\Large$\chi$}(r) = \frac{1}{V} \int_{V} \xi_{\rm ga}\; dV~.\]

Say that out of $\rm \mathcal{N}$ quasar-galaxy pairs, $\rm \mathcal{N'}$ exhibit  \CIV\ absorption with total \colm\ above a threshold value ($N_{\rm{C\, \textsc{iv}}}$) within a velocity window of $2\delta v$, then one can estimate the parameters $r_0$ and $\gamma$ using the following likelihood function: 
$$\mathcal{L} = \prod_{i=0}^{\rm \mathcal{N}'} P_i^{\rm hit}(r) \prod_{j=0}^{\rm \mathcal{N}-\mathcal{N}'} P_j^{\rm miss}(r)~,$$ where, $P_{\rm hit}$ ($P_{\rm miss}$) is the probability of detecting (not detecting) a \CIV\ absorber within $\pm \delta v$ of the galaxies. The probability of non-detection can also be expressed as a Poissonian distribution of zero events and a rate of $\rm N_{obs}$, i.e.,
\[P^{\rm miss}(r) = {\rm exp}\left[ -\{1 - \mbox{\Large$\chi$}(r)\} \left< \frac{d\mathcal{N}}{dz} \right> \delta z \right], \] 
and $P^{\rm hit}(r) = 1 - P^{\rm miss}(r)$. Note that $P^{\rm hit}$ is nothing but the \fc($f_c$).

For our sample, $\left< \frac{d\mathcal{N}}{dz} \right>$ for a threshold \colm\ of \logN = 12.5 (13) is $\approx 55\, (22)$ and $\delta v = 250$~\kms. In order to estimate the parameters, we have followed a Bayesian approach. Our priors are defined such that $r_0$ obeys a flat prior that ranges between 0 and 6 cMpc and $\gamma$ follow a Gaussian prior with $\mu = 1.6$ and $\sigma = 0.5$ (provided, $\gamma > 0$). These priors were chosen based on the results from the literature \citep[see][]{wilde2021, Adelberger_2005}. Using the {\sc ultranest} package of {\sc python}, we estimated the posterior distributions of the parameters. Fig.~\ref{fig:correlation} ({\tt left}) shows the corner plots of the parameters for two different threshold column densities. The {\tt right} panel of Fig.~\ref{fig:correlation} shows the \fc profiles for two threshold column densities corresponding to the posterior distributions of both parameters. Our observed data points are plotted on top (white circles) with respective $1\sigma$ binomial confidence limits. For the threshold \logN = 12.5, we obtained  $r_0 = 2.1^{+0.6}_{-0.6}$~cMpc $\equiv 488$~pkpc and power-law slope, $\gamma = 1.1^{+0.3}_{-0.3}$. Similarly, for the threshold \logN = 13.0, $r_0 = 3.4^{+1.1}_{-1.0}$~cMpc $\equiv 791$~pkpc and $\gamma = 1.2^{+0.2}_{-0.3}$. The fact that $r_0$ increases for a higher threshold \colm\ is consistent with the findings of \citet{Adelberger_2005}. The $r_0$ value for our LAE sample and high column density (\logN = 13.0) \CIV\ absorber correlation function agrees well ($\sim 1 \sigma$) with the galaxy-galaxy correlation length scales given by \cite{Herrero_Alonso_2021}) for the LAEs of $3.3<z<6$, supporting that the high column density \CIV\ absorbers are effective tracers of galaxy halos.


\section{Conclusions} 
\label{sec5}

This work is a continuation of the MUSEQuBES survey at high-$z$ \citep[]{Muzahid_2021}, which focuses on the CGM of high-$z$ ($3< z < 4$) LAEs. MUSEQuBES presented a sample of 96 LAEs in the redshift range 2.9--3.8, detected in 8 MUSE fields centered on 8 bright background quasars. The VLT/UVES and/or Keck/HIRES spectra of these 8 quasars allow us to probe the \CIV\ absorption from the CGM of 86 of the 96 MUSEQuBES galaxies.

To carry out an unbiased study, we first prepared a ``blind'' catalog of \CIV\ absorbers irrespective of the LAEs. We searched for the \CIV~$\lambda\lambda1548,1550$  doublet in the 8 high resolution (FWHM$\approx 6.6$~\kms), high SNR ($>30$ per pixel) quasar spectra visually using the doublet matching technique. Voigt profile decomposition of all the detected \CIV\ absorbers gave rise to a total of 489 \CIV\ absorption components in the redshift range 2.9--3.8, with column densities ranging from $10^{11.4}$ to $10^{14.3}$~\sqcm\ (median $10^{12.7}$~\sqcm) and Doppler parameter in the range $0.87-41.4$~\kms (median $= 12.0$~\kms). The CDDF constructed out of these \CIV\ components has a power-law slope of $-2.12\pm0.11$, which is consistent with the values obtained in the literature at similar redshifts with larger sample sizes. Next, we cross-matched the 'blind' \CIV\ absorber catalog and the LAE catalog to investigate the relation(s) between them. Our main findings are as follows:

\begin{itemize}
    \item There is an enhancement of \CIV\ absorbers within $\pm400$~\kms of the LAEs (Fig.~\ref{fig:abs_distribution}). The high column density absorbers (\logN > 13.0) are more tightly bound to the LAEs (within $\pm320$~\kms, Fig.~\ref{fig:abs_distribution_appendix}: {\tt left}). To be  consistent with the existing literature, we adopted a velocity window of $\pm500$~\kms\ for associating the LAEs and absorbers.

    \item We found that only $\approx55\%$ of these associated absorbers are bound to the LAEs, i.e., they are falling within the escape velocity range of these LAEs (Fig.~\ref{fig:escape_velo}). These bound absorbers are more tightly correlated to the LAEs in velocity (i.e., within $\pm250$~\kms, Fig.~\ref{fig:abs_distribution_appendix}: {\tt right}).

    \item We did not find any significant correlation between the total \CIV\ column density of the absorbers associated with the LAEs and the impact parameter in the range $16 \leq \rho \leq 315$~pkpc using either non-parametric or parametric methods. The observed column densities are, however, considerably higher compared to random regions (Fig.~\ref{fig:n_profile}).

    \item We did not find any significant trend  between the Doppler parameter of the associated \CIV\ components and $\rho$ (Fig.~\ref{fig:b_profile}). The $b(\CIV)$ distributions of the associated components and the rest are statistically indistinguishable.

    \item We found that the \CIV\ covering fractions around the LAEs, averaged over the complete impact parameter range of our sample at different threshold column densities, are $\approx2$ times larger compared to random regions (Fig.~\ref{fig:fc}).

    \item We did not see any redshift evolution of the \CIV\ covering fraction in our sample (Fig.~\ref{fig:f_redshift}).

    \item The \CIV\ covering fraction is $\approx2$ times higher for the pair/group sample compared to the isolated ones (Fig.~\ref{fig:f_env}).

    \item The \CIV\ covering fraction does not vary significantly with the impact parameter within $16 \leq \rho \leq 315$~pkpc. For the isolated LAEs, even for the farthest impact parameter bin (i.e., $\approx 200-250$~pkpc), we found a covering fraction of $0.5\pm0.1$ for a threshold \logN$=12.5$, which is nearly three times larger compared to random regions (Fig.~\ref{fig:f_impact}). 

    \item Finally, we estimated the best-fit parameters of the LAE-\CIV\ absorber cross-correlation function using the \CIV\ covering fraction profile. We obtained a scale-length, $r_0=2.1^{+0.6}_{-0.6}$~cMpc ($3.4^{+1.1}_{-1.0}$~cMpc) and a power-law slope, $\gamma = 1.1^{+0.3}_{-0.3}$ ($1.2^{+0.2}_{-0.3}$) for a threshold \colm\ \logN~$= 12.5$ ($13.0$) (Fig.~\ref{fig:correlation}).   
    
\end{itemize}

This is the third paper that presents results from the MUSEQuBES high-$z$ survey. In the future, we intend to present the connection between \HI\ and other metal lines associated with these LAEs along these 8 sightlines and determine the physical properties of the CGM surrounding these LAEs in greater detail.

\section*{Data Availability}
The data underlying this article are available in ESO (http://archive.eso.org/cms.html) and Keck (https://www2.keck.hawaii.edu/koa/public/koa.php) public archives.

\section*{ACKNOWLEDGEMENT}

This paper uses the following software: NumPy \cite[]{Harris_2020}, SciPy \cite[]{Virtanen_2020}, Matplotlib \cite[]{hunter_2007}, and AstroPy \cite[]{Astropy_2013, Astropy_2018}. EB thanks Labanya Kumar Guha and Sapna Mishra for helpful discussions. EB and SM thank Raghunathan Srianand for useful discussion. EB, SM, and JS thank Marijke Segers, Lorrie Straka, and
Monica Turner for their early contributions to the MUSEQuBES project. We would like to
thank the anonymous referee for the comments and suggestions which improved the clarity of the paper.



\bibliographystyle{mnras}
\bibliography{new_bib} 



\onecolumn
\appendix
\label{appendixa}
  
\section{} 
\begin{figure*}
\centering
\includegraphics[width=0.45\textwidth]{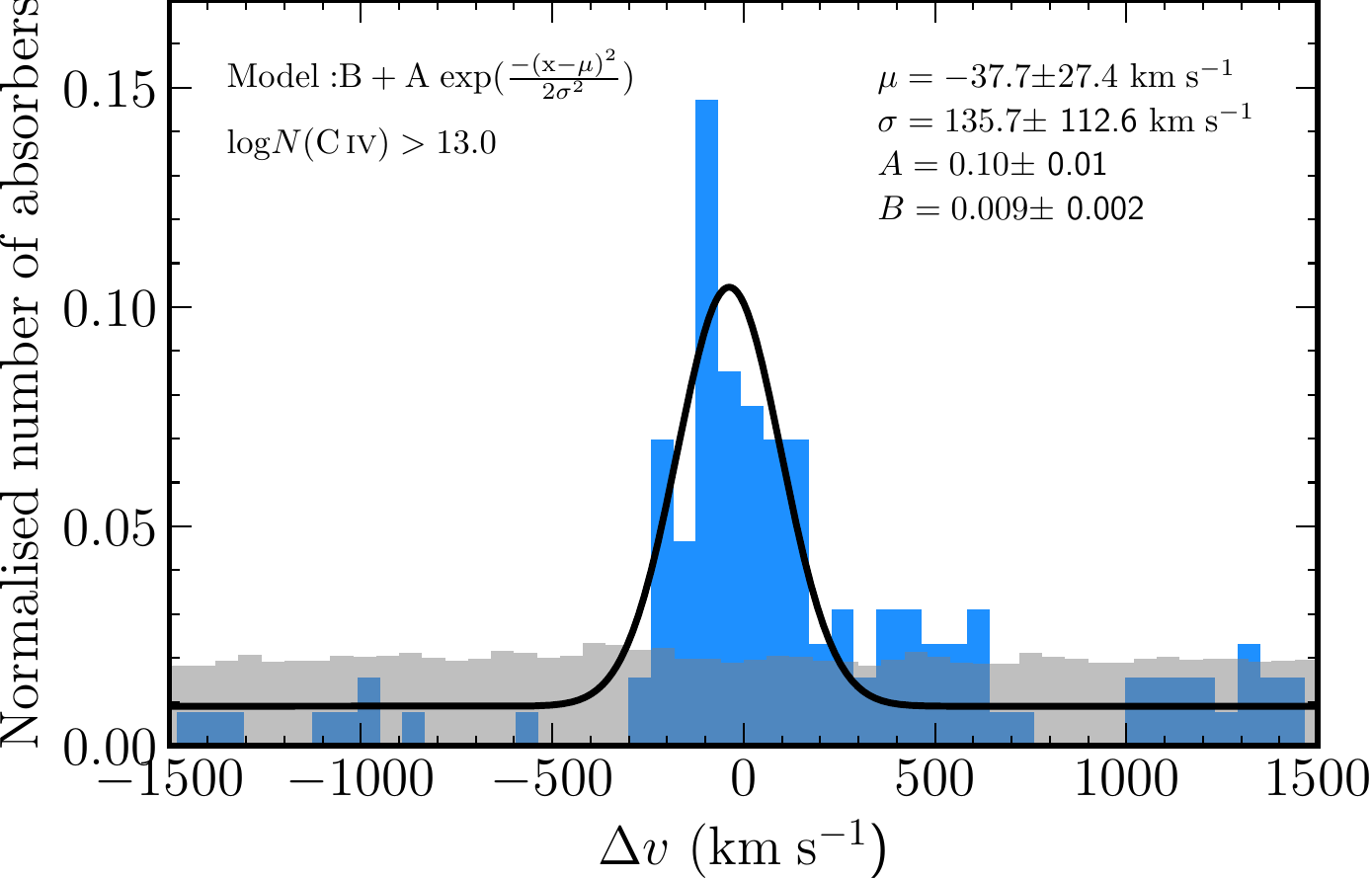}
\includegraphics[width=0.45\textwidth]{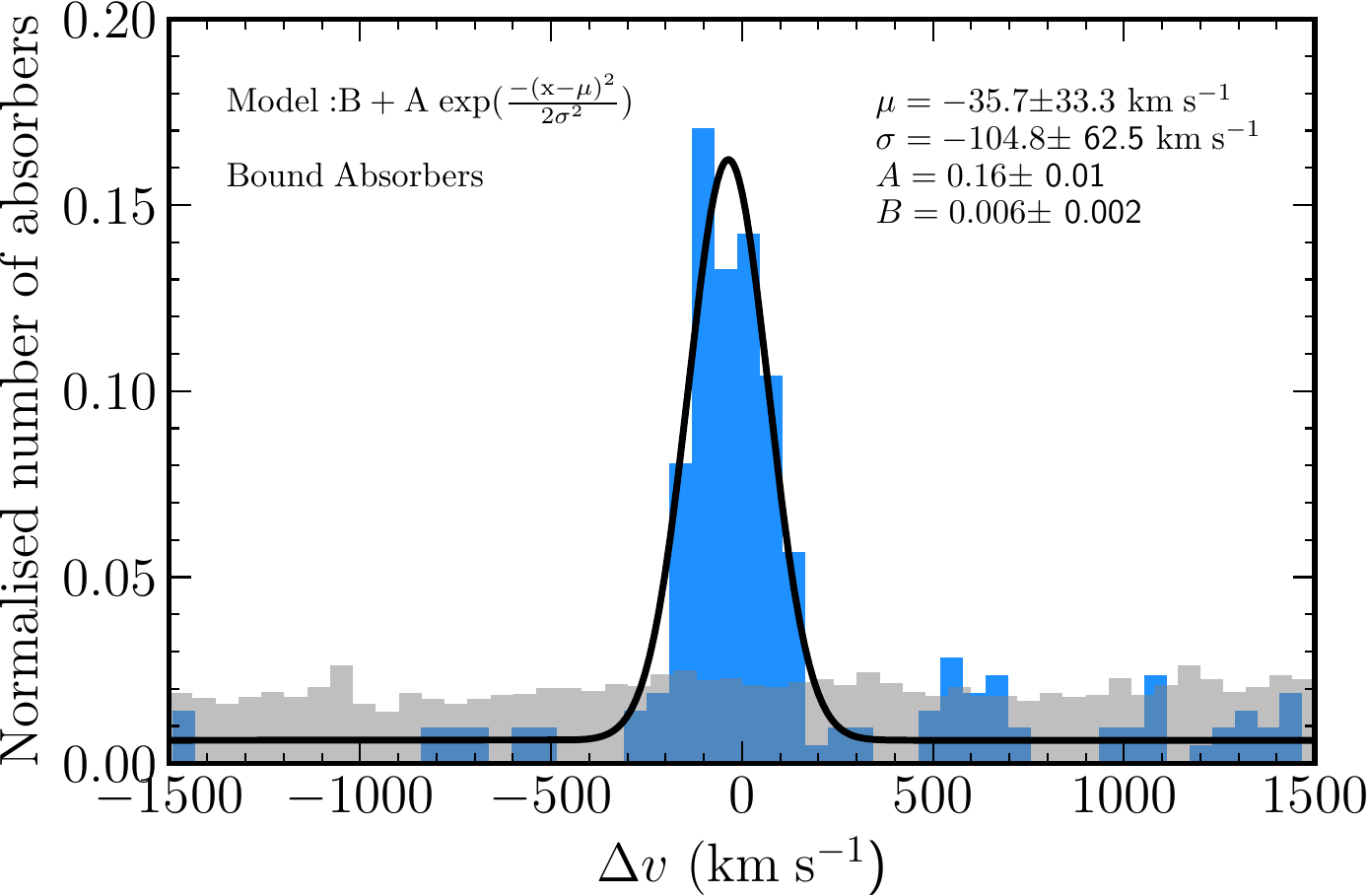}
\caption{{\tt Left:} Same as Fig.~\ref{fig:abs_distribution} but for the high column density (\logN > 13.0) absorbers only. The $\sigma$ is smaller in this case than the complete sample implying that the high column density absorbers are more tightly related to the LAEs. {\tt Right:} Same as the {\tt left} but only for the ``bound" absorbers. Note that some absorbers that are bound to some LAEs ($< \pm 500$~\kms) can be at some large velocities w.r.t. other LAEs of the same sightline. The best-fitted parameters of the model suggest an enhancement of \CIV\ absorption within \pms105~\kms corresponding to an FWHM of $\approx$ 247~\kms.}
\label{fig:abs_distribution_appendix}
\end{figure*}

\counterwithin{figure}{section}
This Appendix contains Fig.~\ref{fig:abs_distribution_appendix} and an ``online only'' Table~\ref{tab:abs_cat}. Fig.~\ref{fig:abs_distribution_appendix} shows the redshift distribution of the high-column density (\logN~$> 13.0$) \CIV\ absorbers on the {\tt Left} and of bound \CIV\ absorbers on the {\tt Right} with respect to the LAEs (see section~\ref{sec:abs_distribution} for details). Table~\ref{tab:abs_cat} presents the best-fit parameters of Voigt profiles fitted to blindly searched \CIV\ components.

\begin{longtable}{ccccccc}
\caption{The ``blind'' \CIV\ absorber catalog}
\label{tab:abs_cat}\\
\toprule
Qso Name &        $z$ &          $z$-error &         $b$ (\kms) &      $b$-error (\kms) &       \logN &      \logN-error \\
\midrule
BRI1108-07 & 2.9169454174 &   5.6285e-06 &  8.41822 &  0.66125 &   12.8155 & 0.024384 \\
BRI1108-07 &  2.959157588 &   8.4368e-06 & 21.14059 &  0.83147 &  13.43264 & 0.015268 \\
BRI1108-07 & 2.9596074097 &   7.2003e-06 & 12.43204 &  0.94801 &  13.10769 & 0.033832 \\
BRI1108-07 & 2.9600363574 &   3.8447e-06 & 11.46797 &   0.5718 & 13.383078 & 0.015505 \\
BRI1108-07 & 2.9603156541 &   8.8502e-06 &  6.75672 &  1.25923 & 12.727034 & 0.078178 \\
BRI1108-07 &   2.96058748 &  2.27194e-05 & 11.92998 &   3.0161 & 12.580692 & 0.085349 \\
BRI1108-07 &  2.960973772 &   4.0625e-06 &   7.6328 &  0.50467 & 12.897028 & 0.019229 \\
BRI1108-07 & 2.9824866348 &  1.39207e-05 & 11.71136 &  1.64549 & 12.430863 & 0.045612 \\
BRI1108-07 & 3.0185289108 &   1.4822e-05 & 15.72472 &  1.59685 & 12.746603 & 0.036159 \\
BRI1108-07 & 3.0372945814 &   6.1234e-06 & 10.63403 &  0.67735 & 12.991107 & 0.021174 \\
BRI1108-07 & 3.2464868911 &  1.32234e-05 &  9.11091 &  1.46379 & 12.625514 & 0.049883 \\
BRI1108-07 & 3.2810564231 &  4.60695e-05 & 19.42298 &  3.42687 & 13.130734 & 0.087518 \\
BRI1108-07 & 3.2814286175 &  2.24048e-05 & 11.75652 &  2.73278 & 12.943474 & 0.144878 \\
BRI1108-07 & 3.2888126674 &  2.90767e-05 & 17.36702 &  3.17348 & 12.697345 & 0.063561 \\
BRI1108-07 & 3.3299469202 &  1.21519e-05 &  7.59786 &  1.37471 & 12.363996 & 0.053985 \\
BRI1108-07 & 3.3303317431 &  1.28175e-05 & 11.72454 &  1.51328 & 12.568784 & 0.040136 \\
BRI1108-07 &  3.388731992 & 0.0001465625 & 12.99678 &  5.13952 &  12.71363 &  0.65963 \\
BRI1108-07 & 3.3890132387 &   5.8137e-05 & 15.04362 &  5.58783 & 13.217944 & 0.277319 \\
BRI1108-07 & 3.3894633487 & 0.0001244339 & 19.60389 &  9.64264 &   12.7581 &  0.27145 \\
BRI1108-07 & 3.4171119335 &  1.21083e-05 & 10.35554 &  1.32996 & 12.602007 & 0.040108 \\
BRI1108-07 & 3.4230836119 &  2.23376e-05 & 25.77128 &  2.30271 & 12.841984 &  0.03149 \\
BRI1108-07 & 3.4282285779 &    7.913e-06 &  6.43669 &  0.88295 & 12.603579 & 0.037035 \\
BRI1108-07 & 3.4820977793 &  1.09992e-05 & 24.33109 &  1.02975 & 13.715114 & 0.015974 \\
BRI1108-07 & 3.5004861158 &  2.18486e-05 & 12.88868 &  2.69492 & 12.790436 & 0.063803 \\
BRI1108-07 & 3.5478941524 &    8.612e-06 & 13.65672 &  0.68852 & 13.607976 & 0.019964 \\
BRI1108-07 & 3.5483286077 &  2.00177e-05 &  13.1603 &  1.61689 &      13.0 & 0.051599 \\
BRI1108-07 & 3.5656283892 &  2.03823e-05 & 16.07402 &  2.10604 & 12.664215 & 0.044423 \\
BRI1108-07 & 3.5762462396 &  1.53318e-05 &  3.39946 &  2.36075 & 12.267339 & 0.129615 \\
BRI1108-07 &  3.576624033 &  1.00193e-05 & 15.67167 &   1.3483 & 13.268962 &  0.02723 \\
BRI1108-07 & 3.5771759105 &   1.0305e-05 & 13.10069 &  1.35195 & 13.190222 & 0.035541 \\
BRI1108-07 & 3.5776355726 &   6.7944e-06 & 11.86062 &  0.62885 & 13.342534 & 0.018858 \\
BRI1108-07 & 3.6028311174 &  1.57584e-05 &  6.37709 &  1.77123 & 12.422924 & 0.072897 \\
BRI1108-07 &  3.603218546 &   5.6221e-06 &  9.06335 &  0.63908 & 13.262471 & 0.020255 \\
BRI1108-07 & 3.6035235853 &  1.62956e-05 &  6.27933 &  2.05163 & 12.692448 & 0.155967 \\
BRI1108-07 & 3.6037970228 &  4.28512e-05 & 12.63118 &   4.0678 & 12.782565 & 0.125579 \\
BRI1108-07 &  3.604474479 &   8.8858e-06 & 17.12057 &  0.96976 & 13.295928 & 0.018942 \\
BRI1108-07 &  3.605423934 &  1.76961e-05 & 25.20648 &  1.61766 & 13.607684 & 0.025451 \\
BRI1108-07 & 3.6060553813 &   5.0999e-06 & 16.05265 &  0.54095 & 13.906902 &  0.01374 \\
BRI1108-07 & 3.6064992921 &  1.33058e-05 &  5.88398 &  1.50312 & 12.613333 & 0.085261 \\
BRI1108-07 & 3.6068429262 &    7.973e-06 &  5.46418 &  1.13959 & 12.731315 & 0.074024 \\
BRI1108-07 & 3.6069029967 &          0.0 & 30.68376 &  2.63944 & 13.321207 & 0.028011 \\
BRI1108-07 &   3.60832794 &  2.85753e-05 & 27.42827 &  2.69359 & 12.890956 & 0.035784 \\
BRI1108-07 & 3.6094403754 &   4.3521e-06 & 11.35437 &  0.44407 & 13.228421 & 0.012591 \\
BRI1108-07 & 3.6098480553 &   8.2493e-06 &  5.92788 &  0.91944 & 12.580745 & 0.040006 \\
BRI1108-07 & 3.7424156839 &   1.5496e-05 & 15.72291 &   1.5211 & 13.193892 & 0.032714 \\
BRI1108-07 & 3.7560586951 & 0.0002598373 & 41.45069 & 10.90758 & 13.462908 & 0.233171 \\
BRI1108-07 & 3.7568289503 &  7.53063e-05 & 31.20425 &  7.04307 & 13.558572 & 0.213982 \\
BRI1108-07 & 3.7576738114 &  4.56473e-05 &  23.7913 &   2.9654 & 13.278536 & 0.082822 \\
BRI1108-07 & 3.7589457032 &  3.04498e-05 &  29.2487 &  2.72282 & 13.314114 & 0.035974 \\
BRI1108-07 & 3.7594134212 &   8.1606e-06 &  6.77224 &  1.12456 & 12.846404 &   0.0708 \\
BRI1108-07 & 3.7604079772 &   6.4357e-06 & 12.96912 &  0.61117 & 13.313706 & 0.015824 \\
BRI1108-07 & 3.7608004183 &          0.0 &  4.97388 &   1.4081 & 12.517218 & 0.068631 \\
BRI1108-07 & 3.7706050423 &  1.76249e-05 &  9.75389 &  1.67088 &  12.68681 &   0.0555 \\
BRI1108-07 & 3.7925544636 &  1.78164e-05 &  14.1048 &  1.59548 & 12.760524 & 0.040221 \\
BRI1108-07 & 3.8122200148 &  1.20139e-05 & 12.75137 &   1.4156 & 12.808177 & 0.054755 \\
BRI1108-07 & 3.8136467054 &   1.2215e-06 &  5.64586 &  0.13074 & 13.918215 & 0.018687 \\
BRI1108-07 & 3.8141739163 &   1.9807e-05 &  1.45048 &  4.88364 & 12.194455 & 0.146028 \\
BRI1108-07 & 3.8345541091 &  3.58449e-05 &  30.5483 &  3.27154 &  13.20448 &  0.03971 \\
BRI1108-07 & 3.8350861604 &  1.86327e-05 &   6.5369 &   2.2632 & 12.499411 & 0.138139 \\
J0124+0044 & 2.9088553168 &  2.08078e-05 & 10.15715 &  1.33935 & 12.904712 & 0.093632 \\
J0124+0044 & 2.9090472683 &   2.4552e-06 &  3.43677 &  0.88725 & 12.895764 & 0.145063 \\
J0124+0044 & 2.9091960165 &  3.08995e-05 & 12.23008 &  8.67081 & 13.010195 & 0.282761 \\
J0124+0044 &  2.909503963 &          0.0 & 10.87334 &   0.8106 & 13.124161 & 0.020612 \\
J0124+0044 & 2.9095335223 &   1.9436e-06 &  5.78791 &   0.3422 & 13.030647 & 0.033127 \\
J0124+0044 & 2.9113957467 &    5.624e-07 &  6.71524 &   0.0633 & 13.557463 & 0.003804 \\
J0124+0044 & 2.9420268419 &   3.0693e-06 &  7.01991 &  0.51531 & 12.559567 & 0.041892 \\
J0124+0044 &  2.942120834 &   2.2077e-05 & 22.01093 &  1.95518 & 12.547822 & 0.047192 \\
J0124+0044 & 2.9848677136 &   7.7866e-06 & 14.65674 &  0.95923 & 12.692469 &  0.02144 \\
J0124+0044 &  2.986390632 &          0.0 &   2.3009 &  0.45334 & 13.491849 & 0.118363 \\
J0124+0044 &  2.986616602 &   2.8837e-06 & 17.54242 &  0.20409 & 14.351325 & 0.007576 \\
J0124+0044 & 2.9870692555 &   4.3566e-06 &  9.58746 &  0.55653 & 13.385272 & 0.024305 \\
J0124+0044 & 2.9874633387 &   9.5325e-06 & 14.05636 &  1.40644 & 13.317219 & 0.037525 \\
J0124+0044 & 2.9876719315 &    8.754e-06 &  4.13092 &  2.07876 & 12.388823 & 0.218733 \\
J0124+0044 & 2.9882189648 &    1.815e-06 &  9.32488 &  0.36306 & 13.028174 & 0.023319 \\
J0124+0044 & 2.9888066486 &  5.05435e-05 & 36.95455 &  5.13528 & 13.053684 & 0.045168 \\
J0124+0044 & 2.9889531953 &  1.05674e-05 & 13.23301 &  1.78797 &  12.66685 & 0.120284 \\
J0124+0044 & 2.9895662893 &   7.7286e-06 & 12.58709 &  0.90116 & 12.657522 & 0.037441 \\
J0124+0044 &  3.064586192 &          0.0 &  8.80217 &  0.73688 & 12.820951 & 0.030292 \\
J0124+0044 & 3.0650482555 &   2.2211e-06 &   17.876 &  0.26213 & 14.011474 & 0.004893 \\
J0124+0044 &  3.065426192 &          0.0 &  9.12317 &  0.36206 & 13.223549 & 0.018063 \\
J0124+0044 & 3.0657839511 &  1.30854e-05 & 17.50213 &  2.89316 & 12.832504 & 0.055279 \\
J0124+0044 & 3.0660980564 &  1.30534e-05 & 10.10433 &  1.46005 & 12.648694 & 0.065904 \\
J0124+0044 & 3.0664052999 &  1.10985e-05 & 11.38854 &  1.44172 &  13.04705 & 0.079537 \\
J0124+0044 & 3.0666845358 &  1.84463e-05 & 13.61566 &  1.21344 &  13.09448 & 0.058304 \\
J0124+0044 & 3.0763240035 &   4.1844e-06 &   5.1078 &  0.73923 & 12.488993 & 0.059123 \\
J0124+0044 & 3.0764837693 &  1.86908e-05 &  27.7889 &  1.89698 & 13.203758 & 0.029401 \\
J0124+0044 & 3.0769078326 &   5.1193e-06 &  5.17306 &      0.0 & 12.665472 & 0.031577 \\
J0124+0044 &  3.077244128 &          0.0 & 19.23025 &  2.40998 & 13.130062 & 0.044222 \\
J0124+0044 & 3.0776073271 &   2.1163e-05 & 10.61341 &      0.0 & 12.635168 & 0.047786 \\
J0124+0044 & 3.0779551771 &   2.6333e-05 & 10.95246 &  4.52813 & 12.643607 & 0.088257 \\
J0124+0044 & 3.0783779214 &   6.6226e-06 &  8.50698 &  0.82653 & 12.633718 & 0.044643 \\
J0124+0044 & 3.0787609079 &   5.9662e-06 & 10.48326 &  0.86018 &  12.69786 & 0.024078 \\
J0124+0044 & 3.0980314637 &  4.02496e-05 &  11.6371 &      0.0 & 12.150027 & 0.098832 \\
J0124+0044 & 3.0983848138 &  1.80989e-05 & 14.35325 &   3.3095 & 12.569065 & 0.083826 \\
J0124+0044 & 3.0987649745 &  2.01503e-05 & 11.35259 &  1.87329 & 12.439327 & 0.074741 \\
J0124+0044 & 3.1481741759 &   6.2242e-06 &  15.3154 &  0.72663 & 12.832991 & 0.016193 \\
J0124+0044 & 3.1486446089 &  1.78212e-05 & 13.32482 &  2.06432 & 12.274797 & 0.055264 \\
J0124+0044 & 3.1876328967 &          0.0 & 10.22585 &    1.635 & 12.320567 & 0.059976 \\
J0124+0044 & 3.1879411301 &          0.0 &  11.1341 &  1.09643 & 12.610383 & 0.047349 \\
J0124+0044 & 3.1881011301 &          0.0 & 13.83601 &  0.82634 & 12.721569 & 0.031425 \\
J0124+0044 &  3.306706146 &   8.8504e-06 &  6.55355 &  1.18054 & 12.163768 & 0.077967 \\
J0124+0044 & 3.3070923429 &  4.81414e-05 &  19.8506 &  7.34394 & 12.144527 & 0.122001 \\
J0124+0044 & 3.3899686954 &   8.0989e-06 &  4.80016 &  1.03913 & 11.911439 & 0.046565 \\
J0124+0044 &  3.391141356 &          0.0 & 12.73783 &  1.75146 & 12.584715 & 0.050972 \\
J0124+0044 &  3.391441356 &          0.0 & 10.22269 &  0.93615 & 12.802791 & 0.042545 \\
J0124+0044 & 3.3917723163 &   4.8509e-06 & 13.09502 &  0.58979 & 13.257136 & 0.015126 \\
J0124+0044 & 3.3922258441 &   5.2059e-06 &  10.1836 &  0.49405 & 13.437502 & 0.021104 \\
J0124+0044 & 3.3925533965 &   4.8987e-06 & 11.78556 &   0.3525 & 13.610696 & 0.013355 \\
J0124+0044 & 3.4744329841 &          0.0 & 16.10206 &  3.70821 & 12.750119 & 0.184439 \\
J0124+0044 & 3.4745751365 &  6.20197e-05 & 11.93143 &  5.15474 &  12.85989 & 0.447882 \\
J0124+0044 & 3.4747022069 &          0.0 & 13.43278 &  3.74616 & 12.748211 & 0.407921 \\
J0124+0044 & 3.4887633737 &  1.30082e-05 & 15.43016 &   1.1514 & 13.126077 & 0.029224 \\
J0124+0044 & 3.4891768297 &  1.42136e-05 & 11.03464 &  2.05937 & 12.869979 & 0.079813 \\
J0124+0044 & 3.4895134012 &  2.50107e-05 &  9.49941 &   2.7059 & 12.571157 & 0.108333 \\
J0124+0044 &  3.498517274 &  1.39891e-05 & 19.26883 &  1.54626 & 12.758254 & 0.026626 \\
J0124+0044 & 3.5018458052 &  2.15688e-05 & 28.86161 &  2.45464 & 12.790171 & 0.028478 \\
J0124+0044 & 3.5468208395 &  1.21693e-05 & 14.33757 &      0.0 & 12.542176 & 0.023618 \\
J0124+0044 & 3.5478304885 &          0.0 &  6.86114 &  0.99962 & 12.493204 & 0.049632 \\
J0124+0044 & 3.5482004885 &          0.0 & 11.06535 &  2.47342 & 12.961629 & 0.230364 \\
J0124+0044 & 3.5483504885 &          0.0 & 16.88051 &  3.72801 & 13.288874 & 0.179076 \\
J0124+0044 & 3.5486015011 &          0.0 &  6.80625 &  1.67115 & 12.769644 & 0.223544 \\
J0124+0044 & 3.5489015011 &          0.0 &  17.5402 &  0.85966 & 13.481839 & 0.015586 \\
J0124+0044 & 3.5493089082 &          0.0 &  5.65134 &  0.19793 & 13.335191 & 0.012282 \\
J0124+0044 & 3.5502141557 &          0.0 &  22.3095 &  2.28661 & 12.960168 & 0.033862 \\
J0124+0044 & 3.5506808309 &          0.0 & 19.77268 &  1.00669 & 13.275596 & 0.016683 \\
J0124+0044 & 3.5512164853 &          0.0 &  8.46738 &  1.09965 & 12.698284 & 0.047342 \\
J0124+0044 & 3.5514164853 &          0.0 &  2.44297 &      0.0 &  12.22407 & 0.084165 \\
J0124+0044 & 3.5516164853 &          0.0 & 14.06079 &  1.89483 & 12.841188 & 0.046312 \\
J0124+0044 & 3.6217580582 &   8.6997e-06 & 10.41474 &  1.01302 &  12.73207 & 0.030941 \\
J0124+0044 &  3.622030256 &  1.13452e-05 &  5.15112 &  1.41852 & 12.234437 & 0.086571 \\
J0124+0044 & 3.6512048707 &  1.73086e-05 & 17.77363 &    2.073 & 12.713425 & 0.037734 \\
J0124+0044 &     3.669483 &          0.0 &    13.95 &      0.0 &     12.42 &      0.0 \\
J0124+0044 & 3.6735192034 &  1.04784e-05 & 16.79191 &   1.0749 & 13.350926 & 0.027247 \\
J0124+0044 & 3.6741638912 &   7.7625e-06 & 20.99755 &  1.26455 & 13.607921 & 0.021057 \\
J0124+0044 & 3.6747425239 &   3.4073e-06 & 12.25014 &  0.33179 & 13.677496 & 0.011197 \\
J0124+0044 & 3.6753436966 &    1.768e-06 & 13.82277 &  0.16899 & 14.028415 & 0.005357 \\
J0124+0044 & 3.6760261025 &   4.0686e-06 & 13.21492 &   0.3913 & 13.316987 & 0.010085 \\
J0124+0044 & 3.7465693335 &  1.15481e-05 &   9.5736 &  1.14884 & 12.629008 & 0.037528 \\
J0124+0044 & 3.7647363262 &  2.72805e-05 & 22.78427 &  3.08265 & 12.543547 & 0.044599 \\
J0124+0044 &  3.765853707 & 0.0002223312 & 40.46861 & 13.91901 & 12.907336 & 0.197389 \\
J0124+0044 & 3.7660656782 &   9.3874e-06 & 10.53095 &      0.0 & 12.855962 & 0.058679 \\
J0124+0044 & 3.7663949016 &  1.40631e-05 & 11.00803 &      0.0 &  12.72129 & 0.097898 \\
J0124+0044 & 3.7668536663 &  2.20957e-05 & 21.22726 &      0.0 & 13.068909 &  0.04851 \\
J0124+0044 & 3.7674858397 &   7.2688e-06 & 10.34264 &  0.71744 & 12.786879 &   0.0234 \\
PKS1937-10 & 3.0059594602 & 0.0001168567 & 18.38276 &  4.77784 & 13.028559 & 0.284337 \\
PKS1937-10 & 3.0060282425 &  2.02966e-05 &  6.20318 &  4.32164 & 12.308336 & 0.720141 \\
PKS1937-10 & 3.0062919775 &  3.29426e-05 &  9.34091 &  2.65874 &  12.73074 & 0.304801 \\
PKS1937-10 & 3.0066240041 &   1.6096e-05 & 13.82005 &  1.60512 &  12.62667 & 0.044753 \\
PKS1937-10 & 3.0085269381 &    6.619e-07 &  6.96317 &  0.07284 & 13.482462 & 0.003739 \\
PKS1937-10 & 3.0231541082 &    9.018e-06 & 12.62674 &  1.16307 & 12.151943 & 0.029117 \\
PKS1937-10 &  3.023553869 &   3.1707e-06 &  9.10998 &   0.3763 & 12.435804 & 0.013116 \\
PKS1937-10 &   3.08102834 &   7.1713e-06 &  8.75855 &  0.83713 & 11.980289 & 0.029349 \\
PKS1937-10 & 3.0954355402 &   3.6032e-06 &  8.95982 &  0.25523 &  12.88304 & 0.019094 \\
PKS1937-10 & 3.0956831028 &  1.44588e-05 & 11.82471 &  1.09176 & 12.498708 & 0.047187 \\
PKS1937-10 & 3.1096639002 &  1.71785e-05 & 14.87394 &   1.3135 & 12.638375 & 0.043464 \\
PKS1937-10 & 3.1098518543 &   8.4639e-06 &  4.22598 &  1.66796 & 11.922827 & 0.209162 \\
PKS1937-10 & 3.1100143699 &  2.79284e-05 & 12.33992 &  1.95192 & 12.407509 & 0.088894 \\
PKS1937-10 & 3.2349677196 &   8.6536e-06 &  5.56535 &  1.00292 & 12.020396 & 0.090856 \\
PKS1937-10 & 3.2352351849 &   6.6611e-06 & 12.94023 &  1.33419 & 12.665816 & 0.038378 \\
PKS1937-10 & 3.2356248584 &   7.0839e-06 & 12.51586 &  0.74751 & 12.679393 & 0.026262 \\
PKS1937-10 & 3.2362097722 &  1.55695e-05 &  20.7671 &  1.76629 & 12.343001 & 0.029068 \\
PKS1937-10 & 3.2417777363 &  1.15575e-05 & 14.57341 &  1.19124 & 12.155077 & 0.029167 \\
PKS1937-10 & 3.2557504304 &   6.8882e-06 &  4.53273 &  0.91127 & 11.857866 & 0.049968 \\
PKS1937-10 & 3.2560604298 &   2.2367e-06 &  11.9605 &  0.29253 & 12.850931 & 0.007521 \\
PKS1937-10 & 3.2565827816 &  1.07864e-05 & 13.94955 &  1.19187 & 12.189352 & 0.028532 \\
PKS1937-10 & 3.2603984684 &   8.5034e-06 &  7.83566 &  0.95103 & 11.902082 & 0.036199 \\
QB2000-330 & 2.9201690629 &  2.73991e-05 &  7.73526 &  3.45909 & 12.312331 & 0.128603 \\
QB2000-330 & 2.9769177584 &   5.4784e-06 &  8.35199 &  0.64632 & 12.653604 & 0.025784 \\
QB2000-330 & 2.9774250956 &          0.0 & 19.52226 &  1.07162 & 13.093465 & 0.019282 \\
QB2000-330 & 2.9776933947 &   2.1537e-06 &  8.08571 &   0.2638 &  13.16098 & 0.013061 \\
QB2000-330 & 2.9782201166 &  1.08147e-05 & 21.16592 &  1.16847 & 13.490322 & 0.022096 \\
QB2000-330 & 2.9785385681 &   4.2357e-06 &  8.58201 &  0.67934 & 12.971512 & 0.056766 \\
QB2000-330 & 2.9788095405 &          0.0 & 38.85669 &  2.53042 & 13.057131 & 0.023896 \\
QB2000-330 & 3.0022316823 &  1.15338e-05 & 11.08388 &  1.30676 & 12.367014 &  0.03895 \\
QB2000-330 & 3.0460223101 &   8.4712e-06 &  16.1297 &   0.9159 & 12.491694 & 0.020063 \\
QB2000-330 & 3.0469269532 &   8.8314e-06 & 19.80118 &  0.89362 & 12.712324 & 0.018211 \\
QB2000-330 & 3.0470091295 &    5.693e-06 &  2.12944 &  1.55236 & 11.891092 & 0.077338 \\
QB2000-330 &  3.172693149 &  1.56665e-05 & 27.80521 &  1.67711 & 12.959581 & 0.023682 \\
QB2000-330 & 3.1727914819 &   6.8044e-06 &  6.56672 &  1.08058 & 12.379101 & 0.075808 \\
QB2000-330 & 3.1793097626 &  3.34208e-05 & 27.47773 &  3.69311 & 12.513516 & 0.047527 \\
QB2000-330 & 3.1868712178 &  1.46149e-05 & 14.45369 &  1.63167 & 12.296109 & 0.037933 \\
QB2000-330 & 3.1899997545 & 0.0001165391 &  11.9037 &   5.4757 & 12.659318 & 0.390744 \\
QB2000-330 & 3.1901076739 &  2.65444e-05 &   4.7786 &  3.27719 & 12.482919 & 0.580824 \\
QB2000-330 & 3.1902405979 &   1.8965e-05 &  5.94053 &  1.41803 & 12.808087 & 0.135795 \\
QB2000-330 & 3.1905656225 &  2.99845e-05 & 19.58041 &  6.96613 & 12.645874 & 0.148936 \\
QB2000-330 &  3.191088326 &   9.9087e-06 & 17.86172 &  1.06989 & 13.195188 & 0.029756 \\
QB2000-330 & 3.1916115237 &   1.4668e-06 & 12.20251 &  0.16636 & 13.501357 & 0.005282 \\
QB2000-330 & 3.1922086227 &    6.588e-06 & 22.92278 &  0.57318 & 13.231908 & 0.010198 \\
QB2000-330 & 3.1924506126 &   6.4357e-06 &  3.89845 &  1.15835 & 12.102704 & 0.078199 \\
QB2000-330 & 3.2288828712 &  1.63192e-05 & 16.67215 &  1.68195 & 12.376465 & 0.035846 \\
QB2000-330 &  3.229773124 &  1.76092e-05 & 15.08564 &  1.73079 & 12.544327 & 0.045951 \\
QB2000-330 & 3.2300489708 &  1.89664e-05 &  4.87295 &  3.02034 & 11.759581 & 0.253765 \\
QB2000-330 & 3.2304000656 &  2.57986e-05 & 24.30942 &  3.67361 &  12.65952 &  0.05367 \\
QB2000-330 & 3.2507019314 &   5.4308e-06 & 14.76648 &  0.55685 & 12.775691 & 0.013295 \\
QB2000-330 &  3.260059386 &    1.471e-05 &   16.214 &   1.6435 & 12.475504 & 0.033943 \\
QB2000-330 & 3.3220472978 &  1.81622e-05 & 20.27832 &  1.81737 & 12.720428 & 0.032437 \\
QB2000-330 &  3.332146675 &   5.5161e-06 &  8.01623 &  0.78015 & 12.441276 & 0.049518 \\
QB2000-330 & 3.3326107153 &  2.99627e-05 &  33.3241 &  2.06043 & 13.283224 & 0.032026 \\
QB2000-330 & 3.3329811367 &   2.4618e-06 & 11.25733 &  0.46037 & 13.156346 & 0.031668 \\
QB2000-330 & 3.3333035939 &   3.7929e-06 &   7.5843 &   0.4627 & 12.729375 &  0.02763 \\
QB2000-330 & 3.3337472532 &   3.2613e-06 & 16.39261 &  0.49721 & 13.255108 & 0.010532 \\
QB2000-330 & 3.3340870853 &   8.3408e-06 &  6.01332 &  1.05543 & 12.193699 & 0.076575 \\
QB2000-330 & 3.3343385329 &          0.0 & 28.71649 &   1.4637 &   12.8028 & 0.019508 \\
QB2000-330 & 3.3368643282 &  1.00436e-05 & 12.39842 &  1.08624 & 12.684254 & 0.029154 \\
QB2000-330 & 3.3373614212 &   2.5532e-06 & 13.36767 &  0.27382 & 13.383899 & 0.006762 \\
QB2000-330 & 3.3892500819 &   2.9971e-06 &  7.27698 &   0.3306 & 12.647592 & 0.014245 \\
QB2000-330 & 3.3922883832 &   5.3238e-06 & 10.09988 &  0.53562 & 12.757138 & 0.018674 \\
QB2000-330 & 3.3926777447 &  1.72661e-05 & 12.30576 &  1.82794 & 12.344155 & 0.050435 \\
QB2000-330 &  3.396576968 &  1.36505e-05 & 23.16117 &  1.61839 &  12.64464 & 0.023248 \\
QB2000-330 & 3.4456751581 &  1.82308e-05 &  16.0585 &  2.14058 &  12.96373 & 0.044576 \\
QB2000-330 & 3.4856278416 &   6.4769e-06 &  7.30419 &  0.70693 & 12.294407 & 0.027553 \\
QB2000-330 & 3.5053953071 &    4.159e-06 & 14.78657 &  0.40151 &   12.9913 & 0.009569 \\
QB2000-330 &  3.505995264 &  1.39631e-05 &   5.4335 &  1.67764 &  11.92695 & 0.072991 \\
QB2000-330 & 3.5082177818 &  1.18304e-05 & 15.13424 &  1.35469 & 12.550616 & 0.028972 \\
QB2000-330 & 3.5087142905 &  1.04399e-05 &  9.69839 &  1.03793 & 12.410492 & 0.036133 \\
QB2000-330 & 3.5094825264 &  1.20792e-05 &  7.60101 &  1.27301 & 12.126859 &  0.04929 \\
QB2000-330 & 3.5467524668 &  1.48429e-05 &  3.94862 &      0.0 & 12.421799 & 0.097133 \\
QB2000-330 & 3.5469504548 &  1.98967e-05 & 10.84738 &      0.0 & 12.833803 & 0.043796 \\
QB2000-330 & 3.5477985575 &  1.02217e-05 &  1.38906 &  1.31225 & 12.875995 & 0.213496 \\
QB2000-330 & 3.5479377765 &   4.3097e-06 & 14.89509 &   0.3171 & 13.907153 & 0.010532 \\
QB2000-330 & 3.5488283719 &  4.85817e-05 & 30.62538 &  6.81248 & 12.976155 & 0.071391 \\
QB2000-330 & 3.5495475842 &  1.21002e-05 & 10.21595 &  0.85836 & 13.421102 & 0.054975 \\
QB2000-330 & 3.5499410897 &  1.17273e-05 & 15.22387 &  2.32308 & 13.691222 & 0.060192 \\
QB2000-330 & 3.5503226052 &  4.27136e-05 & 11.20706 &  3.54798 & 12.992337 & 0.231345 \\
QB2000-330 & 3.5506988242 &  4.17257e-05 &  20.3759 &      0.0 & 13.212498 & 0.033842 \\
QB2000-330 & 3.5512700173 &  2.42494e-05 & 14.77964 &   2.9088 & 12.876952 & 0.077252 \\
QB2000-330 & 3.5520307062 & 0.0001106067 & 20.83757 &  6.66705 & 13.277053 & 0.203506 \\
QB2000-330 & 3.5522205842 &  1.36659e-05 &  4.18004 &  2.37565 & 12.589345 & 0.225171 \\
QB2000-330 & 3.5524415803 &  4.46545e-05 & 15.78174 &  2.12181 & 13.353833 &  0.15408 \\
QB2000-330 & 3.5569795637 &    3.303e-06 &  7.32695 &  0.42128 & 12.832417 &  0.02683 \\
QB2000-330 & 3.5573765197 &   8.0608e-06 &  5.77477 &  1.51879 & 12.269029 & 0.125584 \\
QB2000-330 & 3.5576870319 &  4.18504e-05 & 31.05539 &  3.53318 & 12.963889 & 0.051462 \\
QB2000-330 & 3.5620161376 &  4.52563e-05 & 25.30735 &  2.57149 & 12.733174 & 0.061382 \\
QB2000-330 & 3.5623315132 &   6.5709e-06 &  9.71841 &  1.35297 & 12.416927 & 0.109578 \\
QB2000-330 & 3.5934015127 &   2.9534e-05 & 36.78152 &  3.35549 & 12.792591 & 0.030789 \\
QB2000-330 & 3.6070235155 &  1.71073e-05 &  7.86043 &  1.90608 &  11.83247 &  0.06923 \\
QB2000-330 & 3.6073981129 &  1.27704e-05 &  8.52637 &  1.39109 & 11.977796 & 0.048128 \\
QB2000-330 & 3.6507422424 &   8.4432e-06 & 10.11562 &  0.88145 & 12.805126 & 0.026815 \\
QB2000-330 & 3.6813321013 &  1.68912e-05 & 20.34979 &  1.78779 &  12.73497 & 0.029672 \\
QB2000-330 & 3.6819978978 &  1.14882e-05 & 12.94771 &  1.10921 & 12.628122 & 0.030629 \\
Q1621-0042 & 2.9202037093 &   8.0712e-06 & 19.87625 &  0.95644 & 13.048304 & 0.039697 \\
Q1621-0042 & 2.9202568504 &   6.9388e-06 &  8.75113 &  1.32478 & 12.538761 & 0.131255 \\
Q1621-0042 & 2.9562007455 &   8.0385e-06 &  12.6905 &  0.93519 & 12.387039 & 0.024638 \\
Q1621-0042 & 2.9710801938 &   7.0226e-06 & 16.24528 &  0.78756 & 12.699067 & 0.016954 \\
Q1621-0042 & 2.9718500025 &    8.739e-06 &  19.2464 &  0.61953 & 13.175258 & 0.019826 \\
Q1621-0042 & 2.9719376078 &   9.1185e-06 &   4.3294 &  1.75364 & 12.052836 & 0.144907 \\
Q1621-0042 & 2.9724143135 &   3.7126e-05 & 19.51034 &  3.88219 & 12.386489 & 0.078999 \\
Q1621-0042 &  2.979425732 &  2.85999e-05 & 27.26633 &  2.90624 & 12.713531 & 0.044344 \\
Q1621-0042 &  2.979777218 &    8.383e-06 &  7.99453 &    1.561 & 12.187315 & 0.110396 \\
Q1621-0042 & 2.9959220104 &   1.5311e-06 &  6.83523 &  0.18391 & 12.803731 & 0.007607 \\
Q1621-0042 & 2.9988910514 &  3.89081e-05 & 19.83642 &  4.12333 & 12.225206 & 0.078865 \\
Q1621-0042 & 2.9993581977 &  1.76324e-05 & 13.76688 &  1.93372 & 12.316708 & 0.061777 \\
Q1621-0042 & 3.0202384163 &    8.136e-06 & 10.33938 &  0.81154 & 12.419555 & 0.043293 \\
Q1621-0042 & 3.0206094678 &  1.52919e-05 & 18.46524 &  1.69351 &  12.56682 & 0.034722 \\
Q1621-0042 & 3.0587215344 &   5.0409e-06 & 12.53657 &  0.55972 & 12.988816 &  0.01548 \\
Q1621-0042 & 3.0602147712 &  1.47763e-05 & 16.21479 &  1.59503 & 13.109247 & 0.038232 \\
Q1621-0042 & 3.0605214798 &  1.83948e-05 &  8.15374 &  2.71877 & 12.506072 & 0.177261 \\
Q1621-0042 & 3.0607865086 &   1.8337e-05 &  9.64192 &  1.91905 & 12.605963 & 0.076835 \\
Q1621-0042 &       3.0611 &          0.0 & 12.36294 &  3.32545 & 12.316479 & 0.094081 \\
Q1621-0042 &  3.065468336 &  2.75103e-05 &  33.2683 &  2.32065 & 13.579396 & 0.050671 \\
Q1621-0042 & 3.0660910983 &   9.4232e-06 &  8.37271 &  1.17447 & 12.778742 & 0.057681 \\
Q1621-0042 & 3.0669925518 &  2.20156e-05 & 10.74828 &  2.71923 & 12.378219 & 0.098124 \\
Q1621-0042 & 3.0674962032 &   2.0581e-05 &  9.41455 &  2.48412 & 12.401519 & 0.085978 \\
Q1621-0042 & 3.0681030873 &  5.17318e-05 & 18.69312 &  6.74372 & 12.412712 & 0.120956 \\
Q1621-0042 & 3.1051470024 &  7.98616e-05 & 12.12798 &  3.13328 & 12.816447 & 0.318688 \\
Q1621-0042 & 3.1053840011 &  2.23207e-05 & 11.47071 &  4.35173 & 13.144879 & 0.293255 \\
Q1621-0042 & 3.1057114571 &  2.50819e-05 &  13.3836 &  4.98589 & 13.509319 & 0.189205 \\
Q1621-0042 & 3.1059269277 &  1.44935e-05 &   7.5372 &  1.73768 & 13.500672 & 0.220066 \\
Q1621-0042 & 3.1061139012 &   9.1266e-06 &  8.13542 &  1.74911 & 13.683675 & 0.128902 \\
Q1621-0042 & 3.1063150094 &   3.1993e-05 &  9.48914 &  2.41272 & 13.344427 & 0.230681 \\
Q1621-0042 & 3.1065805533 &  3.08282e-05 & 23.48899 &  2.65408 & 13.726607 & 0.062493 \\
Q1621-0042 & 3.1071252216 &   4.3914e-06 &  11.5796 &   0.6346 & 13.192751 & 0.033214 \\
Q1621-0042 & 3.1074610287 &   4.1403e-06 &  9.60536 &   0.5203 &  12.94013 & 0.020344 \\
Q1621-0042 & 3.1078483758 &   1.8378e-06 & 11.53772 &  0.21097 & 13.229401 & 0.005849 \\
Q1621-0042 & 3.1084132807 &   3.1044e-06 &  9.85084 &  0.21009 & 13.596684 &   0.0106 \\
Q1621-0042 & 3.1086531421 &   8.9064e-06 &  6.72933 &   0.7284 & 13.540731 & 0.062748 \\
Q1621-0042 & 3.1088245426 &  2.38638e-05 &  6.24688 &   2.4459 & 13.409847 & 0.430425 \\
Q1621-0042 & 3.1089672074 &  5.28268e-05 &  8.62958 &   4.3338 & 13.525744 & 0.329753 \\
Q1621-0042 & 3.1092576137 &   5.8449e-06 &  9.23257 &  0.92857 & 14.013196 & 0.048196 \\
Q1621-0042 & 3.1094478832 &  2.01567e-05 &  8.42265 &  0.68623 & 13.453503 & 0.126459 \\
Q1621-0042 & 3.1521932281 &  1.33707e-05 & 18.15422 &  1.41772 & 12.285041 & 0.027757 \\
Q1621-0042 & 3.2000297555 &   5.9766e-06 & 17.36085 &  0.67677 & 12.873664 & 0.013454 \\
Q1621-0042 & 3.2009543598 &  2.79514e-05 &  28.4914 &  2.17158 & 12.880799 & 0.037543 \\
Q1621-0042 & 3.2012609092 &   3.4229e-06 &  7.75542 &   0.5938 & 12.695072 & 0.041226 \\
Q1621-0042 &  3.212429793 &  1.79678e-05 & 14.60559 &  1.65625 & 12.531319 &   0.0448 \\
Q1621-0042 &  3.212698733 &  1.10044e-05 &   6.3375 &  1.59889 & 12.168696 & 0.152857 \\
Q1621-0042 &  3.212960841 &  3.19621e-05 & 15.72828 &   3.0136 & 12.402374 & 0.078415 \\
Q1621-0042 & 3.2411022136 &   6.8135e-06 &  6.04755 &  1.33124 & 12.104637 &  0.13014 \\
Q1621-0042 & 3.2411394543 &  1.34001e-05 & 19.35759 &   2.1246 & 12.503914 & 0.048824 \\
Q1621-0042 & 3.3046681229 &   4.5437e-06 &  5.61465 &  0.54849 & 12.179283 & 0.024049 \\
Q1621-0042 &   3.48683067 &  9.14105e-05 & 18.07776 &  6.24838 & 12.739163 & 0.175449 \\
Q1621-0042 & 3.4870499744 &  2.01843e-05 &  5.22384 &  3.75556 & 12.217298 & 0.409601 \\
Q1621-0042 & 3.4873710347 &  1.04503e-05 & 10.63808 &  1.33285 & 13.068555 & 0.046749 \\
Q1621-0042 & 3.4876784768 &  2.44171e-05 &   8.2993 &  1.98165 & 12.544565 &  0.10649 \\
Q1621-0042 & 3.4897682023 &   2.0459e-05 &  9.83864 &  2.05178 &  12.35767 & 0.067929 \\
Q1621-0042 & 3.5438639152 &  1.83891e-05 & 16.72685 &  2.37259 & 12.519013 & 0.045263 \\
Q1621-0042 & 3.5564303329 &  3.76988e-05 &  12.9477 &  2.77518 &  12.59701 & 0.106471 \\
Q1621-0042 & 3.5567142227 &  2.27166e-05 &   8.8303 &    1.509 & 12.497381 & 0.123409 \\
Q1621-0042 & 3.6142126027 &   7.1037e-06 &   9.8769 &  0.69713 & 12.402693 & 0.022883 \\
Q1317-0507 & 2.9004026076 &  1.22946e-05 &  0.87221 &   7.1034 & 11.579673 & 0.156779 \\
Q1317-0507 &  2.900629679 &   6.0066e-06 & 19.08316 &   0.6021 &  13.04434 & 0.012267 \\
Q1317-0507 & 2.9225211952 &   5.5242e-06 &  2.91206 &  1.03071 & 11.931343 & 0.043138 \\
Q1317-0507 & 2.9229121448 &   4.8402e-06 & 10.94707 &  0.55752 & 12.541955 & 0.016817 \\
Q1317-0507 & 2.9331430888 &          0.0 &  6.29257 &  1.68896 & 12.361186 & 0.087427 \\
Q1317-0507 & 2.9333746962 &  2.01536e-05 & 13.63886 &      0.0 & 12.648309 & 0.042141 \\
Q1317-0507 & 2.9341718913 &    8.822e-06 &  9.69636 &      0.0 & 12.690489 & 0.028974 \\
Q1317-0507 &  3.070721429 &  1.32573e-05 &  8.49902 &  1.17226 & 12.535802 & 0.056145 \\
Q1317-0507 & 3.0709091859 &          0.0 & 22.22903 &   4.0068 &  12.80115 & 0.059789 \\
Q1317-0507 & 3.0918008313 &  5.60771e-05 &  28.0323 &  4.59807 & 12.531498 & 0.082297 \\
Q1317-0507 & 3.0921225458 &   1.9864e-06 &  9.55218 &  0.38467 & 12.910705 & 0.026994 \\
Q1317-0507 & 3.1783596002 &    6.818e-06 & 10.08233 &      0.0 & 12.462418 & 0.019891 \\
Q1317-0507 & 3.2768680126 &   4.4351e-06 & 14.46715 &  0.50419 &  12.66785 & 0.011446 \\
Q1317-0507 & 3.2773675181 &   2.0213e-06 & 11.24395 &  0.21711 & 12.885374 & 0.006429 \\
Q1317-0507 & 3.2862554392 &   1.0873e-05 &  8.04702 &  1.48288 & 12.312248 & 0.084819 \\
Q1317-0507 & 3.2864965342 &  2.38539e-05 & 37.05407 &  2.05845 & 13.150894 & 0.018125 \\
Q1317-0507 & 3.2871644918 &  1.21459e-05 &  8.41941 &  1.28285 & 12.542205 & 0.098708 \\
Q1317-0507 &  3.287495904 &    8.418e-06 & 16.56867 &  1.07138 & 13.210506 & 0.023779 \\
Q1317-0507 & 3.2880132387 &   1.4653e-06 &  9.82865 &  0.15702 & 13.746049 & 0.005579 \\
Q1317-0507 & 3.2884039977 &    3.129e-06 & 12.44431 &  0.40661 & 13.685554 & 0.013632 \\
Q1317-0507 & 3.2887573131 &   7.7456e-06 & 13.50743 &  0.49402 & 13.397027 & 0.021033 \\
Q1317-0507 & 3.2939653296 &  2.10001e-05 & 25.12794 &  2.17064 &  12.67825 & 0.031588 \\
Q1317-0507 &        3.302 &          0.0 & 23.22312 &      0.0 & 12.497492 & 0.140011 \\
Q1317-0507 &      3.30214 &          0.0 & 13.64445 &  2.77429 & 12.579384 & 0.171503 \\
Q1317-0507 &  3.302413832 &  1.93748e-05 & 16.87714 &  1.72855 & 12.915588 & 0.048324 \\
Q1317-0507 & 3.3033139972 &  3.11905e-05 & 27.96074 &  3.46743 & 12.605193 & 0.042751 \\
Q1317-0507 & 3.3204245396 &  4.94436e-05 &   7.6654 &  6.39871 & 11.822642 & 0.298012 \\
Q1317-0507 & 3.3206863775 &  2.11024e-05 &  8.39889 &  3.58576 & 12.170275 &  0.14687 \\
Q1317-0507 & 3.3209458047 &  1.38249e-05 &  5.86512 &  1.54782 & 12.079484 & 0.083747 \\
Q1317-0507 &  3.359412477 &   2.6887e-06 &  6.41866 &   0.3087 & 12.578435 & 0.012866 \\
Q1317-0507 & 3.3598496449 &          0.0 &  9.57446 &  0.88721 & 12.478573 & 0.038109 \\
Q1317-0507 &  3.360017569 &    7.598e-06 &  9.05282 &  0.72198 & 12.638404 & 0.026782 \\
Q1317-0507 & 3.3717676227 &  6.18054e-05 & 25.98639 &  3.45339 & 12.542395 & 0.087154 \\
Q1317-0507 & 3.3719703203 &   5.0233e-06 &  9.17464 &  0.69158 &  12.76773 & 0.046227 \\
Q1317-0507 & 3.3770259615 &          0.0 & 17.87803 &   7.0035 & 12.631652 & 0.608004 \\
Q1317-0507 & 3.3770842485 &  2.33734e-05 & 13.22358 &  1.62125 & 13.055206 & 0.247301 \\
Q1317-0507 & 3.3774410587 &  3.83868e-05 &  9.45072 &  3.06732 & 12.018919 & 0.208603 \\
Q1317-0507 & 3.4114309706 &   8.4551e-06 &  7.94305 &  1.09226 & 12.264333 & 0.037119 \\
Q1317-0507 & 3.4464369785 &  6.34562e-05 & 13.56904 &  3.52898 & 12.775405 & 0.255687 \\
Q1317-0507 & 3.4467220231 &  5.51532e-05 & 14.17307 &  2.28212 & 12.908591 & 0.170608 \\
Q1317-0507 & 3.5710945606 & 0.0001155401 & 13.31136 &  5.32564 & 12.838938 & 0.312283 \\
Q1317-0507 &  3.571248586 &   9.5144e-06 &  6.13215 &  2.44734 & 12.633763 & 0.435906 \\
Q1317-0507 & 3.5715573175 &  3.77016e-05 &  8.74919 &  4.93029 & 12.272062 & 0.222558 \\
Q1317-0507 & 3.5717899326 &  3.44592e-05 &  5.12302 &   3.1582 &  11.95238 &  0.25311 \\
Q1317-0507 &       3.5744 &          0.0 & 32.52609 & 19.58588 & 12.429004 & 0.185706 \\
Q1317-0507 &       3.5747 &          0.0 & 10.41755 &  4.18078 & 12.368942 & 0.582417 \\
Q1317-0507 & 3.5750407561 & 0.0001493266 &  21.2707 & 12.72841 & 13.196339 & 0.570069 \\
Q1317-0507 & 3.5755722828 & 0.0002849822 & 25.64414 & 23.14081 & 13.193381 &  0.57588 \\
Q1317-0507 & 3.5762176772 &  8.39378e-05 &   16.288 &  5.25001 & 12.733122 & 0.277566 \\
Q1317-0507 & 3.5812784849 &   6.0982e-06 &  6.37648 &  1.51061 & 12.307989 & 0.144982 \\
Q1317-0507 & 3.5816023499 &   4.7239e-06 &  7.01343 &  0.83526 & 12.556361 & 0.085042 \\
Q1317-0507 & 3.5820333134 &  6.78491e-05 & 37.01685 & 12.08189 & 13.033035 & 0.132476 \\
 Q0055-269 & 2.9138919564 &    9.031e-06 & 20.00192 &  1.08709 & 12.775986 & 0.018586 \\
 Q0055-269 & 2.9452651587 &   5.4567e-06 & 18.43372 &  0.58106 & 12.936683 & 0.011649 \\
 Q0055-269 & 2.9490138136 &  2.62785e-05 & 22.03978 &  2.12674 & 12.938688 & 0.047233 \\
 Q0055-269 & 2.9492146142 &   5.8329e-06 &  4.11516 &  1.25075 & 12.212719 & 0.110483 \\
 Q0055-269 & 2.9495285631 &  1.78516e-05 & 15.05622 &  1.67593 & 12.623985 & 0.060427 \\
 Q0055-269 & 2.9501332271 &   9.1487e-06 & 17.13312 &  1.14666 & 12.768464 & 0.024439 \\
 Q0055-269 & 2.9506014858 &          0.0 &  6.35593 &  0.57517 & 12.600837 & 0.030708 \\
 Q0055-269 & 2.9507414858 &          0.0 & 21.30534 &  0.37134 & 13.532563 & 0.006785 \\
 Q0055-269 & 2.9514090897 &   4.6035e-06 & 18.16122 &    0.546 & 13.069641 & 0.011109 \\
 Q0055-269 & 2.9519279836 &   4.3543e-06 &  8.36994 &  0.49072 & 12.505231 & 0.020431 \\
 Q0055-269 & 3.0048084276 &  2.21313e-05 &  7.98029 &  1.85864 & 12.141798 & 0.119165 \\
 Q0055-269 & 3.0050385337 &  1.24074e-05 &   9.8471 &  1.06813 & 12.544871 & 0.048072 \\
 Q0055-269 &  3.038771664 &   9.1527e-06 & 11.22042 &      0.0 & 12.564649 & 0.023485 \\
 Q0055-269 & 3.0389309016 &          0.0 & 12.95108 &  2.64734 & 12.229224 & 0.070502 \\
 Q0055-269 & 3.0837957057 &  1.14576e-05 &  7.66412 &  1.33925 & 12.102983 & 0.051406 \\
 Q0055-269 & 3.0852831666 &          0.0 &  7.50111 &      0.0 & 11.944129 & 0.065137 \\
 Q0055-269 & 3.0857060954 &          0.0 &  14.6908 &  1.42976 & 12.588531 & 0.033939 \\
 Q0055-269 & 3.0858992208 &   2.5916e-06 &  7.25784 &  0.33765 & 12.984442 & 0.014405 \\
 Q0055-269 & 3.0861637424 &  1.34803e-05 &  8.45035 &  1.51401 & 12.303823 &  0.05816 \\
 Q0055-269 &  3.132319957 &  1.17386e-05 & 10.08214 &  1.83902 & 12.265241 & 0.054215 \\
 Q0055-269 &  3.132619957 &  1.17386e-05 & 10.08214 &  1.83902 & 12.265241 & 0.054215 \\
 Q0055-269 & 3.1791725099 &   7.0695e-06 &  4.24314 &  0.99514 & 12.334079 & 0.044768 \\
 Q0055-269 & 3.1901865016 &          0.0 &  12.6607 &  9.42229 & 12.834432 & 0.272874 \\
 Q0055-269 & 3.1904369536 &   9.6313e-06 &  9.06869 &  0.98742 & 13.884993 & 0.051128 \\
 Q0055-269 &  3.190945713 &  3.03017e-05 &  8.04941 &  1.96469 & 13.867291 & 0.118983 \\
 Q0055-269 & 3.1911780472 &   3.8204e-05 &  7.31322 &  3.96579 & 13.589801 & 0.217565 \\
 Q0055-269 & 3.1915402982 &  2.72933e-05 &  13.2352 &  4.53966 &  13.72595 & 0.107274 \\
 Q0055-269 &   3.19180899 &          0.0 &  6.79959 &  3.07015 & 13.360071 & 0.200784 \\
 Q0055-269 &      3.19207 &          0.0 &     15.0 &      0.0 & 13.706627 & 0.048261 \\
 Q0055-269 &       3.1928 &          0.0 & 25.77681 &      0.0 &      12.7 &      0.0 \\
 Q0055-269 &  3.194175822 &   1.8829e-06 &  5.67797 &  0.23916 &  13.17232 & 0.016562 \\
 Q0055-269 & 3.1942817057 &   1.5857e-05 & 35.74274 &  1.28937 & 13.429946 & 0.024658 \\
 Q0055-269 & 3.1944718545 &      3.8e-06 & 13.06588 &  0.57588 & 13.299972 &  0.02815 \\
 Q0055-269 & 3.1950228482 &   9.1118e-06 &  4.45038 &  1.46555 & 12.054613 & 0.088871 \\
 Q0055-269 & 3.2481234997 &   4.4028e-06 &  4.56674 &  0.58992 &   12.5011 & 0.026912 \\
 Q0055-269 & 3.2484492111 &   9.9425e-06 &  9.90656 &  1.28236 & 12.506088 & 0.037696 \\
 Q0055-269 & 3.2559350025 &  1.46016e-05 & 10.83279 &  1.20701 &  12.87445 & 0.053767 \\
 Q0055-269 & 3.2562110711 &   7.3972e-06 &  9.71797 &  0.69637 & 13.077552 & 0.035503 \\
 Q0055-269 & 3.2566443345 &  1.97907e-05 & 12.01261 &  2.07106 &  12.74184 & 0.075477 \\
 Q0055-269 & 3.2569451922 &  1.13293e-05 & 10.13801 &  1.14031 & 12.859639 & 0.060377 \\
 Q0055-269 & 3.2573653104 &   8.5545e-06 & 16.09103 &  0.87888 & 12.952977 & 0.019552 \\
 Q0055-269 &  3.258330502 &  2.99137e-05 &  10.1537 &  4.63137 & 12.230325 & 0.131515 \\
 Q0055-269 & 3.4233857217 &  1.38934e-05 & 13.37195 &  1.41901 & 12.501856 & 0.038322 \\
 Q0055-269 &  3.423965273 &          0.0 & 19.48252 &  2.88077 & 12.458194 & 0.050852 \\
 Q0055-269 & 3.5262715039 &          0.0 & 30.53252 &      0.0 & 12.829482 & 0.017774 \\
 Q0055-269 & 3.5271105748 &   4.5054e-06 & 16.05429 &  0.43543 & 13.221699 & 0.009864 \\
 Q0055-269 & 3.5276670295 &  1.56912e-05 & 15.77066 &   1.6483 & 12.634992 &  0.03805 \\
 Q0055-269 & 3.5284284857 &  2.08993e-05 & 17.77879 &   2.2101 & 12.435035 & 0.042144 \\
 Q0055-269 & 3.5553839548 &   6.6352e-06 &  7.77835 &  0.50579 & 12.877419 & 0.031155 \\
 Q0055-269 & 3.5556624665 &   7.2942e-06 & 10.34758 &  0.63536 &   13.0187 & 0.024754 \\
 Q0055-269 &  3.556080733 &          0.0 & 24.40416 &  0.90431 & 13.090002 & 0.014113 \\
 Q0055-269 & 3.5570519331 &   9.4411e-06 & 11.95992 &  0.92423 & 12.521328 & 0.026178 \\
 Q0055-269 & 3.5582623832 &    9.981e-06 & 13.23237 &  0.96869 & 12.545528 & 0.025204 \\
 Q0055-269 & 3.5732335317 &  6.47003e-05 & 89.66587 & 11.17765 & 13.486891 &  0.04635 \\
 Q0055-269 & 3.5852692361 &   9.2923e-06 & 21.00809 &   0.8804 & 12.752284 & 0.015146 \\
 Q0055-269 & 3.5910090517 &   4.0061e-06 &   8.9159 &  0.41385 & 12.608968 & 0.014251 \\
 Q0055-269 & 3.5998557416 &   9.4363e-06 & 11.01139 &   0.9426 & 12.628557 & 0.027995 \\
 Q0055-269 & 3.6004740366 &   5.1027e-06 & 10.35066 &  0.47343 & 13.107478 & 0.016063 \\
 Q0055-269 & 3.6008711647 &   3.5986e-06 & 11.10504 &  0.43833 & 13.310441 &  0.01316 \\
 Q0055-269 & 3.6014550748 &   3.1795e-06 & 16.48785 &  0.38397 & 13.732174 & 0.007361 \\
 Q0055-269 & 3.6018846411 &   6.0009e-06 &  9.36386 &  0.48867 & 13.067519 & 0.025569 \\
 Q0055-269 & 3.6031794376 &   8.4266e-06 & 11.38609 &  0.44997 &  13.42118 & 0.035054 \\
 Q0055-269 &   3.60349851 &  2.34995e-05 & 15.18231 &  1.29704 & 13.238627 & 0.053956 \\
  Q1422+23 & 2.9096620176 &    3.437e-06 &   8.9014 &  0.42593 & 12.463322 & 0.014639 \\
  Q1422+23 & 2.9100393993 &  1.42201e-05 & 10.90619 &  1.94748 & 11.954181 & 0.054383 \\
  Q1422+23 & 2.9472198789 &  1.44529e-05 &  6.56098 &  1.96062 & 11.448304 & 0.078736 \\
  Q1422+23 & 2.9475082149 &   1.6626e-06 &  7.72554 &  0.21352 & 12.484069 & 0.007827 \\
  Q1422+23 & 2.9606238419 &   8.5175e-06 & 14.63104 &  0.97622 &  12.55122 & 0.023585 \\
  Q1422+23 &  2.961075968 &   5.4738e-06 & 13.71751 &   0.8695 & 12.703324 & 0.021084 \\
  Q1422+23 & 2.9614361167 &   3.2103e-06 &  8.09866 &  0.41219 & 12.632592 & 0.018124 \\
  Q1422+23 & 2.9619453697 &   4.1217e-06 &  19.1878 &  0.49205 & 13.281684 & 0.009606 \\
  Q1422+23 & 2.9623251176 &   4.5011e-06 & 11.03697 &  0.41941 & 12.835516 & 0.022989 \\
  Q1422+23 & 2.9713556895 &  1.87329e-05 & 20.10835 &   1.6916 & 12.596251 & 0.036874 \\
  Q1422+23 & 2.9716253093 &  2.47317e-05 &  6.54851 &  4.98945 & 11.432409 & 0.440993 \\
  Q1422+23 & 2.9751818552 &  1.18947e-05 &  7.01863 &  1.51066 & 11.659124 & 0.064499 \\
  Q1422+23 & 2.9757996964 &  2.55343e-05 & 24.07431 &   2.8376 & 12.388387 & 0.045724 \\
  Q1422+23 & 2.9761986704 &   1.8949e-06 & 10.25511 &  0.23867 & 12.812102 & 0.014524 \\
  Q1422+23 & 2.9992026699 &   5.1823e-06 & 18.78212 &  0.56228 & 12.706168 & 0.010763 \\
  Q1422+23 & 3.0350231148 &    4.081e-06 &  8.30452 &  0.47815 & 12.120227 & 0.017445 \\
  Q1422+23 &  3.036582361 &  1.41199e-05 & 15.22378 &  1.52057 &  12.17789 & 0.036185 \\
  Q1422+23 & 3.0369090298 &  1.32601e-05 &  8.18178 &   1.6536 & 11.781193 & 0.084737 \\
  Q1422+23 & 3.0633717009 &   8.7451e-06 & 26.22603 &  1.01037 & 12.981046 &  0.01322 \\
  Q1422+23 & 3.0643171111 &   6.4641e-06 & 12.28035 &    0.731 & 12.648215 & 0.020225 \\
  Q1422+23 & 3.0695831092 &  1.56738e-05 &  9.25906 &  1.77955 & 12.094481 & 0.063649 \\
  Q1422+23 & 3.0701552094 &   9.7255e-06 &   2.8343 &   1.8336 & 11.734405 & 0.083725 \\
  Q1422+23 &  3.070995777 &    4.913e-06 &  5.28667 &  0.59031 & 12.242724 & 0.030079 \\
  Q1422+23 &  3.089864459 &   9.6897e-06 & 13.34009 &   0.7628 & 12.943929 & 0.042525 \\
  Q1422+23 & 3.0901724726 &  1.63734e-05 & 20.02859 &      0.0 & 13.172808 & 0.018656 \\
  Q1422+23 &  3.090527212 &          0.0 & 20.17183 &      0.0 & 12.988757 & 0.019337 \\
  Q1422+23 & 3.0910452316 &   2.5449e-06 & 10.66807 &  0.27722 &  13.00579 &  0.00871 \\
  Q1422+23 & 3.0946561227 &   9.3086e-06 & 17.25888 &  1.13249 & 12.225443 & 0.021697 \\
  Q1422+23 & 3.1192916578 &  1.77802e-05 & 13.30337 &  2.19821 & 11.910054 & 0.056997 \\
  Q1422+23 & 3.1196940981 &  1.12849e-05 & 12.00397 &  1.38326 & 12.047441 & 0.040589 \\
  Q1422+23 & 3.1325107233 &  1.74138e-05 & 30.82643 &  1.83635 &  12.56882 & 0.021743 \\
  Q1422+23 & 3.1337678856 &   8.4839e-06 & 17.76769 &  0.87827 & 12.809212 & 0.018089 \\
  Q1422+23 & 3.1340767541 &   6.0331e-06 &  5.62005 &  0.89338 & 12.183141 & 0.065282 \\
  Q1422+23 & 3.1344482855 &   3.2631e-06 & 16.47109 &  0.37761 & 13.001537 & 0.007611 \\
  Q1422+23 & 3.1370896293 &   5.6591e-06 & 13.01771 &  1.35451 & 12.607782 & 0.118792 \\
  Q1422+23 & 3.1371351755 &  2.58488e-05 & 29.04721 &  5.26483 & 12.614788 & 0.103381 \\
  Q1422+23 &  3.138070083 &  2.05467e-05 & 20.63454 &  2.19873 &  12.26817 &   0.0413 \\
  Q1422+23 & 3.1914025827 &   4.2473e-06 &  8.50665 &  0.48495 & 12.126793 &  0.01724 \\
  Q1422+23 & 3.2332935237 &  1.12609e-05 &  8.30242 &  1.29475 & 12.034652 & 0.046362 \\
  Q1422+23 & 3.2425716921 &   7.1162e-06 &    3.669 &  1.07194 & 11.528472 & 0.048661 \\
  Q1422+23 & 3.2567764098 &  1.84377e-05 & 14.59025 &  2.39951 & 12.048606 & 0.069811 \\
  Q1422+23 & 3.2573535221 &  1.82333e-05 & 22.37534 &  3.34309 & 12.300122 & 0.051364 \\
  Q1422+23 & 3.2653979563 &  1.86058e-05 &  9.94755 &  1.77903 & 12.062839 & 0.113134 \\
  Q1422+23 & 3.2657447532 &   1.6705e-05 & 17.68045 &  1.43142 & 12.598546 & 0.034826 \\
  Q1422+23 & 3.2759388643 &   8.9897e-06 & 11.46557 &  0.95684 & 11.893301 & 0.027753 \\
  Q1422+23 & 3.3179607541 &   9.5322e-06 & 13.78775 &  1.03871 & 12.185765 & 0.025066 \\
  Q1422+23 & 3.3799289464 &  1.01908e-05 & 15.29909 &  1.10761 & 12.570812 & 0.024497 \\
  Q1422+23 & 3.3804154538 &   1.1389e-05 & 11.91368 &  1.17175 & 12.367523 & 0.034953 \\
  Q1422+23 & 3.3813358312 &          0.0 &  7.43239 &  1.37775 & 12.032733 & 0.066416 \\
  Q1422+23 & 3.3816400325 &    1.532e-06 & 11.61792 &  0.16856 & 13.275556 & 0.004718 \\
  Q1422+23 & 3.3822046825 &   1.9241e-06 & 14.00338 &  0.22317 & 13.237018 & 0.005081 \\
  Q1422+23 & 3.3826779152 &   4.2982e-06 &   8.5904 &  0.49969 &  12.62269 & 0.022017 \\
  Q1422+23 &  3.383171749 &  4.14237e-05 & 21.22034 &  4.81863 &  12.15999 & 0.076162 \\
  Q1422+23 & 3.4107906207 &  1.22014e-05 & 12.72186 &  1.26799 & 12.253361 &  0.03366 \\
  Q1422+23 & 3.4114545495 &   5.0817e-06 & 17.96023 &  0.51638 & 12.802656 & 0.009995 \\
  Q1422+23 & 3.4468971929 &   5.4188e-06 &  9.69273 &  0.56497 & 13.011219 & 0.018591 \\
  Q1422+23 & 3.4473452748 &   3.2254e-06 & 12.35632 &  0.32989 &  13.41001 & 0.008752 \\
  Q1422+23 & 3.4700205536 &   3.1919e-06 &  5.05533 &  0.37785 & 12.523476 & 0.017192 \\
  Q1422+23 & 3.4711420439 &  1.13389e-05 &  3.32676 &  1.70462 & 11.835061 & 0.075704 \\
  Q1422+23 & 3.4761513199 &  1.99044e-05 &  8.08288 &  2.11739 &    12.082 &   0.0784 \\
  Q1422+23 &  3.479392491 &   3.2598e-05 & 13.92875 &  2.41841 & 12.457468 & 0.087585 \\
  Q1422+23 & 3.4797989707 &  3.96494e-05 & 14.62255 &  3.10079 & 12.402032 & 0.101664 \\
  Q1422+23 & 3.4804984068 &  1.22786e-05 & 13.49009 &   1.2523 &  12.42365 & 0.031397 \\
  Q1422+23 & 3.4948479007 &   6.0135e-06 &   9.9647 &  0.62075 & 12.417679 & 0.019916 \\
  Q1422+23 & 3.5146191149 &   3.2475e-06 &  7.94578 &  0.29139 & 12.833535 & 0.016622 \\
  Q1422+23 & 3.5149473111 &  1.45984e-05 & 13.44553 &  1.34988 &  12.51242 & 0.037462 \\
  Q1422+23 & 3.5349838978 &          0.0 & 18.51921 &      0.0 & 12.884885 & 0.017754 \\
  Q1422+23 & 3.5359678125 &   6.1777e-06 & 23.18871 &  0.47012 & 13.732611 & 0.009063 \\
  Q1422+23 & 3.5366586041 &  2.30045e-05 & 22.05767 &  2.32721 & 13.052911 &  0.04571 \\
  Q1422+23 & 3.5373483321 &   4.7292e-06 & 19.09344 &  0.47329 &  13.24486 & 0.008635 \\
  Q1422+23 & 3.5384595227 & 0.0001013483 & 26.34479 &  4.96295 & 13.346563 & 0.140312 \\
  Q1422+23 & 3.5386978476 &   2.7683e-06 &  9.56292 &  0.61696 &   13.5847 &  0.05679 \\
  Q1422+23 & 3.5393068815 &  1.80434e-05 & 12.17799 &  3.87861 & 13.364441 & 0.719463 \\
  Q1422+23 & 3.5394275025 & 0.0003153286 &  17.2478 &  6.56238 & 13.231333 &  1.02273 \\
  Q1422+23 & 3.5399500832 &  8.07157e-05 & 37.90766 &  5.22007 & 12.944296 & 0.074789 \\

\bottomrule
\end{longtable}


\bsp	
\label{lastpage}
\end{document}